\documentclass[12pt]{article}

\usepackage{latexsym,amsmath,amssymb,theorem,epsfig}

\DeclareFontFamily{OT1}{rsfs10}{}
\DeclareFontShape{OT1}{rsfs10}{m}{n}{ <-> rsfs10 }{}
\DeclareMathAlphabet{\mathscript}{OT1}{rsfs10}{m}{n}

\numberwithin{equation}{section}

\newcommand{\cp}[1]{{\mathbb C}{\mathbb P}^{#1}}

\newcommand{\dslash}[0]{\slash{\hspace{-0.23cm}}\partial}
\newcommand{\deslash}[0]{\slash{\hspace{-0.17cm}}\partial}

\newcommand{\ns}{\normalsize}

\newcommand{\bX}{{\mathbf X}}
\newcommand{\bY}{{\mathbf Y}}
\newcommand{\bZ}{{\mathbf Z}}

\def\cC{{\mathcal C}}
\def\cE{{\mathcal E}}

\def\cL{{\mathcal L}}

\def\cN{{\mathcal N}}
\def\cO{{\mathcal O}}

\def\cS{{\mathcal S}}

\theoremstyle{plain}

{\theorembodyfont{\rmfamily} }

\begin{document}


\begin{titlepage}

\vspace{-5cm}

\title{
   \hfill{\ns UPR-999T} \\[1em]
   {\LARGE Vector Bundle Moduli Superpotentials in Heterotic
Superstrings and M-Theory}
\\
[1em] }
\author{
   Evgeny I. Buchbinder$^1$, Ron Donagi$^2$ and
Burt A.~Ovrut$^1$ \\[0.5em]
   {\ns $^1$Department of Physics, University of Pennsylvania} \\[-0.4em]
   {\ns Philadelphia, PA 19104--6396}\\
   {\ns $^2$Department of Mathematics, University of Pennsylvania}
   \\[-0.4em]
   {\ns Philadelphia, PA 19104--6395, USA}\\}

\date{}

\maketitle

\begin{abstract}

The non-perturbative superpotential generated by a heterotic superstring
wrapped once around a genus-zero holomorphic curve is proportional to the
Pfaffian involving the determinant of a Dirac operator on this curve. We
show
that the space of zero modes of this Dirac operator is the kernel of a
linear
mapping that is dependent on the associated vector bundle moduli. By
explicitly computing the determinant of this map, one can deduce whether
or
not the dimension of the space of zero modes vanishes. It is shown that
this
information is sufficient to completely determine the Pfaffian and, hence,
the
non-perturbative superpotential as explicit holomorphic functions of the
vector bundle moduli. This method is illustrated by a number of
non-trivial
examples.

\end{abstract}

\thispagestyle{empty}

\end{titlepage}


\section{Introduction:}


The non-perturbative superpotentials for the moduli of superstrings
and $M$-theory have been calculated using string worldsheet
conformal field theory \cite{DSWW1,DSWW2} and, alternatively, by
considering the contributions to the partition function
of strings wrapped on holomorphic curves
$z$ in the compactification space
\cite{BBS,Witten1}. This latter approach has been used to compute
non-perturbative superpotentials in both weakly \cite{Witten2} and
strongly \cite{Lima1,Lima2} coupled heterotic superstring theories
on a Calabi-Yau threefold, as well as in other theories
\cite{DGW,HM}. The superpotentials in heterotic string theory were found
to be proportional to a factor involving the Wess-Zumino-Witten term,
which couples the superstring to the background $SO(32)$ and
$E_8 \times E_8$ gauge bundle $V$ \cite{Witten2,Lima1, Lima2}.
Assuming a mild condition on the structure group, this factor
can be shown to be the Pfaffian of a chiral Dirac operator
twisted by the restriction, $V|_{z}$, of the gauge bundle to the
holomorphic
curve. In \cite{Witten2}, it was shown that this Pfaffian and, hence,
the superpotential will vanish if and only if $V|_{z}$ is non-trivial.
Furthermore, it is clear that the Pfaffian must be a holomorphic function
of gauge bundle moduli associated with $V|_{z}$. However, until recently,
neither the vanishing locus of the Pfaffian in vector bundle
moduli space nor its explicit dependence on these moduli was known.
This was rectified in \cite{BDO2}, where we provided solutions to both of
these problems within the framework of weakly and strongly coupled
$E_8 \times E_8$ theories compactified on elliptically fibered
Calabi-Yau threefolds. In \cite{BDO2}, we emphasized concepts, and
did not discuss details of the requisite mathematics. Furthermore, we
applied our method to a single, non-trivial example. It is the
purpose of this paper to greately expand on the content of \cite{BDO2}.
We will present the complete mathematical structure and then, using
this structure, compute many examples of non-perturbative superpotentials.

Specifically, in this paper we do the following. In Section 2, we
review the computation of non-perturbative superpotentials in both
weakly and strongly coupled heterotic string theory following
the approach in \cite{BBS,Witten1}. The superpotentials are shown to
be proportional to the Pfaffian of a chiral Dirac operator, ${\cal
D}_{-}$,
twisted by the vector bundle restricted to the holomorphic curve on which
the string is wrapped. In this paper, we will compute
${\rm Pfaff}({\cal D}_{-})$ algebraically. However, in Appendix A,
an alternative form for the Pfaffian, written in terms of an explicit
gauge one-form solution of the Hermitian Yang-Mills equations
is derived. We discuss how non-perturbative superpotentials
might be computed using this formalism. In Section 3,
the moduli of stable, $SU(n)$ holomorphic vector bundles, $V$, over
elliptically fibered Calabi-Yau threefolds, $X$, are enumerated and their
properties are briefly discussed. Relevant information about such
threefolds, and the construction of holomorphic vector bundles
from spectral data is presented in Appendix B. We then
derive an explicit formula for the number of vector bundle moduli when
the base space of $X$ is any Hirzebruch surface, that is,
$B={\mathbb F}_r$. These results were previously obtained in
\cite{BDO1}. Next, a holomorphic curve $z$ is specified
and the restriction of the vector bundle to the surface
$\pi^* z \subset X$ is discussed. We describe and enumerate the moduli
associated with $V|_{\pi^* z}$, giving an explicit formula
for the number of such moduli when $B={\mathbb F}_r$ and $z={\cal S}$.
Throughout this paper, when specificity is required, we will restrict
$r=0,1,2$.
The global structure of the moduli space is also discussed. The first part
of Section 4 is devoted to presenting these restricted bundle moduli
from the point of view of the curve $z$. It is shown that these moduli
appear as sections of a specific rank $n$ vector bundle over $z$. This
requires expressions for the image on $z$ of certain line bundles
on $\pi^* z$. These expressions are derived in Appendix C.
In the second part of this section, we compute the restriction
of $V$ to $z$ and show that it is the image of the spectral data line
bundle. Thus, on curve $z$, we completely categorize
$V|_z$ and the associated moduli. The main results of this paper
are calculated in Sections 5 and 6. In Section 5, we first discuss
${\rm Pfaff}({\cal D}_{-})$ and show that it will vanish if and only if
the dimension of the space of holomorphic sections
$h^{0}(z, V|_z \otimes  S_{-})$, where $S_{-}$ is the
negative chirality spin bundle on $z$, is non-zero. It is then
demonstrated that $h^{0}(z, V|_z \otimes S_{-})$ will be non-zero,
for all vector bundle moduli, for a specific set of vector bundles.
We then present large classes of theories with $z={\cal S}$ in which,
generically, $h^{0}(z, V|_z \otimes S_{-})$ can vanish over
a subset of moduli space. In these examples, the associated
cohomology class $H^{0}(z, V|_z \otimes  S_{-})$ is contained in an
exact sequence of the form
\begin{equation}
0 \rightarrow H^{0}(z, V|_z \otimes S_{-})
\rightarrow W_1 \stackrel{f}{\rightarrow} W_2
\rightarrow \dots.
\label{intro1}
\end{equation}
Here, $W_1$ and $W_2$ are linear spaces of identical finite
dimension and, hence, linear map $f$ can be represented by a square
matrix. Furthermore, $f$ is an explicit function of the vector bundle
moduli associated with $V|_{z}$. It follows that
$h^{0}(z, V|_z \otimes S_{-})$ will be non-zero if and only if
the determinant of $f$ vanishes. Thus, the Pfaffian and, hence,
the superpotential will vanish if and only if $det f=0$.
In Section 6, we compute the matrix $f$ and its determinant
for a number of explicit examples, all, for simplicity, with
$B={\mathbb F}_r, z={\cal S}$ and structure group
$G=SU(3)$. To begin, we present an example of the type where,
although matrix $f$ is non-trivial, $det f$ vanishes for all
values of the vector bundle moduli. Therefore, for this example,
${\rm Pfaff}({\cal D}_{-})$ and the superpotential always vanish.
However, we then present three explicit examples where
$det f$ are explicit, non-zero functions of the vector bundle moduli.
In one example, a polynomial appearing as a factor in $det f$ is
very lengthy and, hence, is presented in Appendix D. In these
examples $det f$ may vanish, but only on a divisor of the
moduli space. Using $det f$, one can, in each case, explicitly
compute the zero locus of the Pfaffian. Away from the zero locus,
${\rm Pfaff}({\cal D}_{-})$ and, hence, the superpotential are
non-vanishing. Finally, in Section 7, we show how the Pfaffian
can be explicitly computed from the knowledge of its zero locus.
We find that ${\rm Pfaff}({\cal D}_{-})\propto det f$ and,
therefore, that the non-perturbative superpotential $W$ satisfies
$W \propto det f$. Since $det f$ was explicitly computed
as a function of the associated vector bundle moduli in Section 6,
we have determined the vector bundle moduli contribution to the
superpotential.

Although at first sight rather complicated to derive,
the superpotential for vector
bundle
moduli potentially has a number of important physical applications. To
begin
with, it is essential to the study of the stability of the vacuum
structure~\cite{CK,MPS} of
both weakly coupled heterotic string theory and
heterotic $M$-theory
~\cite{hmt1,hmt2,hvb,five,standard1,standard2,standard3,standard4}.
Furthermore, in both theories it allows, for the first time, a discussion
of
the dynamics of the gauge bundles. For example, in heterotic $M$-theory
one
can determine if a bundle is stable or whether it decays, via a small
instanton transition \cite{tony}, into five-branes.
In recent years, there has been
considerable research into the cosmology of superstrings and heterotic
$M$-theory \cite{low4, hos, blo}.
In particular, a completely new approach to early universe
cosmology, Ekpyrotic theory \cite{ekp1,ekp2,ekp3,ekp4,ekp5},
has been introduced within the context of
brane universe theories. The vector bundle superpotentials discussed in
this
paper and \cite{BDO2} allow one to study the dynamics of the small
instanton phase
transitions that occur when a five-brane \cite{ekp1,ekp2} or
an "end-of-the-world" orbifold
plane \cite{ekp3,ekp4,ekp5} collides with our observable brane,
thus producing the Big Bang. These physical applications will be discussed
elsewhere.


\section{Non-Perturbative Superpotentials:}


Let us consider the long wavelength limit of $E_8 \times E_8$
heterotic superstring theory compactified on a Calabi-Yau threefold,
$X$. This four-dimensional, $N=1$ supersymmetric effective theory contains
a number of chiral supermultiplets, some charged with respect to the
low-energy gauge group and some uncharged. Of the uncharged
supermultiplets,
a subset corresponds to the moduli of the compactification. In heterotic
string theory, these moduli break into two types, Calabi-Yau moduli
corresponding to the Kahler and complex structure deformations of $X$ and
the vector bundle moduli. These latter fields are associated with the
integration constants in the vacuum solutions of the Hermitian Yang-Mills
equations for the gauge connection on a fixed $X$.
Here, we will simply denote
any $N=1$ chiral moduli superfield by
\begin{equation}
(Y^{I^{\prime}}, \lambda^{I^{\prime}}),
\label{1.1}
\end{equation}
where $Y^{I^{\prime}}$ is the complex scalar and $\lambda^{I^{\prime}}$
is its spin $\frac{1}{2}$ superpartner. The index $I^{\prime}$
runs over all Calabi-Yau and vector bundle moduli. No matter what their
origin, the dynamics of these moduli is given by a Lagrangian of the
form
\begin{eqnarray}
{\cal L}_{4D}&=&K_{I^{\prime}\bar J^{\prime}}
\partial_u Y^{I^{\prime}} \partial^u \bar Y^{\bar J^{\prime}}
+ e^{\kappa_p^2 K}\left(K^{I^{\prime}\bar J^{\prime}}
D_{I^{\prime}}W \bar D_{\bar J^{\prime}}\bar W
- 3\kappa_p^2 |W|^2\right) \nonumber \\
& & +K_{I^{\prime}\bar J^{\prime}} \lambda^{I^{\prime}}
\dslash \bar\lambda^{\bar J^{\prime}} -
e^{\frac{\kappa_p^2 K}{2}} (D_{I^{\prime}} D_{J^{\prime}}W)
\lambda^{I^{\prime}}\lambda^{ J^{\prime}} +{\rm h.c.} +\dots
\label{1.2}
\end{eqnarray}
Here $\kappa_p^2$ is the four-dimensional Newton's constant,
\begin{equation}
K_{I^{\prime}\bar J^{\prime}}=\partial_{I^{\prime}}
\partial_{\bar J^{\prime}}K
\label{1.3}
\end{equation}
are the Kahler metric and the Kahler potential respectively, and
\begin{equation}
D_{I^{\prime}}W =\partial_{I^{\prime}}W +\kappa_p^2
\frac{\partial K}{\partial Y^{I^{\prime}}} W
\label{1.4}
\end{equation}
is the Kahler covariant derivative acting on the superpotential $W$.
A priori, both $K$ and $W$ are arbitrary. However, since the
supermultiplets $(Y^{I^{\prime}}, \lambda^{I^{\prime}})$ correspond to
the moduli of the compactification of the heterotic string on $X$, one
must have
\begin{equation}
W=0.
\label{1.5}
\end{equation}
The moduli Kahler potential, on the other hand, is non-trivial and
can have a complicated functional structure on $Y^{I^{\prime}}$.
Fortunately, we do not need to know the explicit form of $K$ in this
paper.

Both perturbative and non-perturbative quantum processes can, in
principle,
induce a non-vanishing superpotential for the moduli fields. However, it
is
well known that perturbative radiative corrections to the superpotential
vanish. However, non-zero corrections to $W$ can be generated by
non-perturbative processes involving superstring instantons. Following
the approach of \cite{BBS,Witten1}, we will calculate
the non-perturbative superpotential by computing instanton induced
fermion bilinear interactions and then comparing these to the
fermion bilinear terms in~\eqref{1.2}. In this paper, the instanton
contribution arises from a heterotic superstring wrapped once around
a holomorphic curve $z \subset X$. In order to isolate the vector bundle
moduli contribution to the superpotential, the main objective of this
paper, we find it expedient to calculate the two point functions
of fermions associated with the Calabi-Yau moduli. Denote
these moduli superfields by
\begin{equation}
(T^I, \lambda_T^{I}),
\label{1.6}
\end{equation}
where index $I$ runs over Calabi-Yau moduli only. Then the two-point
function of fermions $\lambda_T^I$ and $\lambda_T^J$, located
at positions $y_1$ and $y_2$, is given by the path integral expression
\begin{eqnarray}
< \lambda^I(y_1) \lambda^J(y_2)> &\propto
&\int {\cal D}\lambda
e^{-\int{\rm d}^4 y \sum_{K}  \lambda^{K}
\deslash \lambda^{K}}
\lambda^I(y_1) \lambda^J(y_2) \nonumber \\
& & \int{\cal D}{\mathbb Z}{\cal D} g
e^{-S_z ({\mathbb Z}, g; {\mathbb E}_{{\mathbb M}}^{{\mathbb A}},
{\mathbb B}_{{\mathbb M}{\mathbb N}}, \phi, {\mathbb A}_{{\mathbb M}})},
\label{1.7}
\end{eqnarray}
where here, and henceforth, we drop the subscript $T$ in
$\lambda_T^I$. ${\mathbb Z}$ and $g$ are the worldsheet fields on the
heterotic superstrings, with
\begin{equation}
{\mathbb Z}=(X, \Theta)
\label{1.8}
\end{equation}
being the boson and fermion coordinates of the target superspace
respectively, and $g$ being the map from $z$ to the gauge group.
In addition, ${\mathbb E}_{{\mathbb M}}^{{\mathbb A}},
{\mathbb B}_{{\mathbb M}{\mathbb N}}, \phi$ and ${\mathbb A}_{{\mathbb
M}}$
are the background superfield vielbein, two-form, dilaton and supergauge
connection in which the superstring propagates.
$S_z$ is the heterotic string action. The result of this calculation
is then compared to the term in~\eqref{1.2} proportional to
$(D_I D_J W)\lambda^I \lambda^J$, and the non-perturbative contribution
to $W$ extracted. Note that $W$ is non-perturbative from the point of view
of
the string worldsheet theory and, hence, must be proportional to
$e^{-\frac{1}{\alpha'}}$, where $\alpha'$ is the worldsheet constant.
However,
it does not contain the dilaton and, therefore, is perturbative in terms
of the string coupling expansion.

To proceed, we must present the structure of the heterotic
string action $S_z$. This action was discussed in
\cite{Lima1,Lima2}, where we refer the reader for details and a complete
description of the notation. Here, we will elucidate notation only if it
is relevant to our calculation. The action is found to be
\begin{equation}
S_z =S_S +S_{WZW},
\label{1.9}
\end{equation}
where
\begin{eqnarray}
S_S({\mathbb Z}; {\mathbb E}_{{\mathbb M}}^{{\mathbb A}},
{\mathbb B}_{{\mathbb M}{\mathbb N}}, \phi)&=&T_S
\int_z {\rm d}\sigma^0 {\rm d}\sigma^1 (\phi
\sqrt{{\rm det}{\mathbb E}_i^{A}{\mathbb E}_j^{B}
\eta_{AB}} \nonumber \\
& & -\frac{1}{2} \epsilon^{ij} {\mathbb E}_{i}^{{\mathbb A}}
{\mathbb E}_{j}^{{\mathbb B}}{\mathbb B}_{{\mathbb B}{\mathbb A}})
\label{1.10}
\end{eqnarray}
with ${\mathbb E}_{i}^{{\mathbb A}}\equiv
(\partial_i {\mathbb Z}^{{\mathbb M}})
{\mathbb E}_{{\mathbb M}}^{{\mathbb A}}$,
couples the superstring to
${\mathbb E}_{{\mathbb M}}^{{\mathbb A}},
{\mathbb B}_{{\mathbb B}{\mathbb A}}$
and $\phi$, and
\begin{eqnarray}
S_{WZW}({\mathbb Z}, g; {\mathbb A}_{{\mathbb M}})&=&-\frac{1}{8\pi}
\int_z {\rm d}\sigma^0 {\rm d}\sigma^1 {\rm tr}
[\frac{1}{2} \sqrt{h}h^{ij}(\omega_i -{\mathbb A}_i)
(\omega_j -{\mathbb A}_j)+i \epsilon^{ij}\omega_i {\mathbb A}_j]
\nonumber \\
& &+ \frac{1}{24} \int_{{\cal B}} {\rm d}^3 \hat{\sigma} i
\epsilon^{\hat{i}\hat{j}\hat{k}} {\rm tr}
\omega^{\prime}_{\hat{k}}
\omega^{\prime}_{\hat{j}}
\omega^{\prime}_{\hat{i}}
\label{1.11}
\end{eqnarray}
is the Wess-Zumino-Witten term coupling the
superstring to the supergauge connection. Here,
${\mathbb A}_i=(\partial_i {\mathbb Z}^{{\mathbb M}})
{\mathbb A}_{{\mathbb M}}$ and $\omega$ is the Lie algebra
valued one-form on the heterotic string worldsheet defined by
\begin{equation}
\omega =g^{-1}dg.
\label{1.17}
\end{equation}
Using~\eqref{1.10} and~\eqref{1.11}, the calculation of the
non-perturbative
contribution to the two-point function~\eqref{1.7} can be carried out.
This calculation is technically tedious and will not be reviewed here.
We refer
the reader to \cite{Lima1,Lima2} for details. Suffice it to say that
one proceeds by expanding each term in the action as a power series
of the fermionic coordinate $\Theta$. That is,
\begin{equation}
S_S=S_0+S_{\Theta}+\dots
\label{1.12}
\end{equation}
and
\begin{equation}
S_{WZW}=S_{WZW0}+\dots,
\label{1.13}
\end{equation}
where we only display relevant terms. In~\eqref{1.12} one finds,
after some rescaling of the fields, that
\begin{equation}
S_0 =\frac{{\cal A}(z)}{2\pi\alpha^{\prime}}-i\int_z B.
\label{1.14}
\end{equation}
Here, ${\cal A}(z)$ is the area of the curve $z$ measured using the
pulled-back target space metric, $\alpha^{\prime}$ is the heterotic string
parameter and $B$ is the pulled-back bosonic two-form. Furthermore, the
$S_{\Theta}$ term can be shown to contain four-dimensional fermions
$\lambda^{I}$ that arise from the dimensional reduction on $X$ of the
ten-dimensional gravitino. After a long calculation
\cite{Lima1,Lima2}, one finds that
\begin{eqnarray}
< \lambda^I(y_1) \lambda^J(y_2)> &\propto
&\int {\cal D}\lambda
e^{-\int{\rm d}^4 y \sum_{K}  \lambda^{K}
\deslash \lambda^{K}}
\lambda^I(y_1) \lambda^J(y_2) \nonumber \\
& & \int {\rm d}^4 x (\partial_L \partial_M
(e^{-S_0} \int {\cal D}g e^{-S_{WZW0}}))
\lambda^{L}\lambda^{M}.
\label{1.15}
\end{eqnarray}
Note that the proportionality symbol includes three determinant
factors that are not relevant to the discussion in this paper.
Comparing these results to the $(D_I D_J W)\lambda^I \lambda^J$
term in~\eqref{1.2}, and using~\eqref{1.14}, we conclude that the
non-perturbative contribution to the superpotential of a heterotic
superstring wrapped once around a holomorphic curve $z$ is given by
\begin{equation}
W \propto \left(\int {\cal D}g e^{-S_{WZW0}}\right)
e^{-\frac{{\cal A}(z)}{2\pi \alpha^{\prime}}+i\int_z B}.
\label{1.16}
\end{equation}
The $e^{-\frac{{\cal A}(z)}{2\pi \alpha^{\prime}}+i\int_z B}$ factor
contains only the Calabi-Yau moduli and is reasonably well-known
\cite{DSWW1,DSWW2,Witten2,Lima1}. In this paper, we turn our
attention to the factor $\int {\cal D}g e^{-S_{WZW0}}$, which depends
on the vector bundle moduli. Although a topological criterion
for the vanishing of this factor was given in \cite{Witten2}, its
functional dependence on the vector bundle moduli has
only recently been computed \cite{BDO2}.

To evaluate $\int {\cal D}g e^{-S_{WZW0}}$, we first note that
$S_{WZW0}$ is precisely the expression~\eqref{1.11} with
${\mathbb A}_i$ replaced by the bosonic gauge connection
$A_i =(\partial_iX^{M})A_M$. Let us split the ten-dimensional vector
field $A_M$ into the four-and six-dimensional components $A_u$ and
$A_U$ respectively. Since the holomorphic curve the superstring
is wrapped on lies entirely inside the Calabi-Yau threefold $X$, we get
\begin{equation}
A_i=(\partial_i X^{U})A_{U}.
\label{1.16.5}
\end{equation}
$A_U$ is valued in the Lie algebra of structure group
$G \subseteq E_8 \times E_8$ and is a solution of the Hermitian
Yang-Mills equations on $X$. Such a solution is called
a $G$-instanton.
At this point, we note that, with some restrictions,
this theory can be written in a simpler form. When $A_i=0$, the theory
described by $S_{WZW0}$ is known \cite{Boson}
to be equivalent to an $SO(16) \times SO(16)$ theory
of free Weyl fermions
on $z$ of the same chirality. Now take $A_i \neq 0$. Since only the
$SO(16) \times SO(16)$ ($\subset E_8 \times E_8$) symmetry is manifest
in the free fermion theory, only gauge connection $A_i$
with structure group
\begin{equation}
G \subseteq SO(16) \times SO(16)
\label{1.18}
\end{equation}
can be coupled to fermions in the usual way. The coupling of gauge
backgrounds with larger structure groups is far from manifest and cannot
be given by a Lagrangian description. For these reason we assume,
henceforth, that~\eqref{1.18} is satisfied. Note that, from a
phenomenological point of view, this restriction is not unreasonable
since, as explained in a number of papers
\cite{hvb,standard1,standard2,standard3,standard4},
realistic heterotic
superstring and $M$-theory models always require the structure group
to be contained in $SO(16) \times SO(16)$. With this is mind, we now write
the action for $SO(16) \times SO(16)$ Weyl fermions coupled to
a background gauge connection $A_i$. Denote the Weyl fermions by
$\psi^a, a=1, \dots, 32$ where each is chosen to be a section
of the negative spin bundle on $z$, $S_{-}$. Then
\begin{equation}
S_{\psi}=\int_z {\rm d}\sigma^0 {\rm d}\sigma^1 {\rm tr}
\bar \psi D_{-} \psi,
\label{1.19}
\end{equation}
where $D_{-}$ is the chiral Dirac operator on $z$ with
gauge connection $A$. The subscript minus implies that it acts on
sections of $S_{-}$. The fermion partition function is easily evaluated
to give
\begin{equation}
\int {\cal D}\psi {\cal D} \bar \psi e^{-S_{\psi}} \propto
det {\cal D}_{-},
\label{1.20}
\end{equation}
where
\begin{equation}
{\cal D}_{-} = \bordermatrix{     & {\ } & {\ }  \cr
                             {\ } & 0 & D_{-} \cr
                             {\ } & i\partial_{+} & 0 \cr}
\label{1.21}
\end{equation}
and $i \partial_{+}$ is the Dirac operator with vanishing gauge
connection that acts on sections of the positive spin bundle, $S_{+}$.
There is one final caveat before we can determine the partition
function
$\int {\cal D}g e^{-S_{WZW0}}$. In computing the above, we have used the
$S_{WZW0}$ action in Euclidean space. Hence, the fermions to which
this theory is equivalent are Weyl fermions. However, the heterotic
superstring actually propagates is Minkowski space, where the
associated fermions are simultaneously Weyl and Majorana spinors. These
have half the number of degrees of freedom of the Euclidean space
Weyl fermions. It follows that, to avoid overcounting degrees
of freedom, one should take the square root of the partition function.
Mindful of this caveat, we see from the above discussion that
\begin{equation}
\int {\cal D}g e^{-S_{WZW0}} \propto
{\rm Pfaff}({\cal D}_{-}),
\label{1.22}
\end{equation}
where
\begin{equation}
{\rm Pfaff}({\cal D}_{-})= \sqrt{det {\cal D}_{-}}
\label{1.23}
\end{equation}
and ${\cal D}_{-}$ is given by~\eqref{1.21}. This expression is the
starting point of the paper.

There are, apparently, at least two ways to evaluate
${\rm Pfaff}({\cal D}_{-})$. One approach is to view $A$ as
a gauge connection on a stable, holomorphic vector bundle $V$ over
$X$ using the theorems of Donaldson \cite{Donald} and
Uhlenbeck and Yau \cite{UY}. ${\rm Pfaff}({\cal D}_{-})$ is then
a global operator on $z$ twisted by the bundle
$V|_{z} \otimes S_{-}$ and $det {\cal D}_{-}$ can be determined
by purely algebraic techniques. This is the method we use in this paper.
A second approach is to return to the original partition function
$\int {\cal D}g e^{-S_{WZW0}}$ and to evaluate it, as completely as
possible, using general properties of solutions of the Hermitian
Yang-Mills equations. This evaluation is carried out in Appendix A.
One would then attempt to find explicit solutions of these equations
on a Calabi-Yau threefold. These will depend, generically, on arbitrary
integration constants, the vector bundle moduli. Inserting any such
solution into $\int {\cal D}g e^{-S_{WZW0}}$ would produce, after
integration, an explicit function of the vector bundle moduli. This
approach is problematic since, at present, no solutions of the
Hermitian Yang-Mills on a Calabi-Yau threefold are known. Be that as it
may,
the importance of computing the superpotentials may provide an
incentive for a re-examination of this question. We emphasize that the
properties of ${\rm Pfaff}({\cal D}_{-})$, such as its vanishing
structure,
that will be discussed in this paper might serve as a non-trivial
guide to the construction of such explicit solutions.


\section{The Moduli of Stable SU(n) Vector Bundles:}


Let $X$ be an elliptically fibered Calabi-Yau threefold with base $B$ and
fiber map $\pi:X \longrightarrow B$. In this paper, we will, when
specificity
is required, take $B$ to be a Hirzebruch surface
\begin{equation}
B={\mathbb F}_{r},
\label{x1}
\end{equation}
where $r$ is any non-negative integer. As discussed in Appendix B,
a stable $SU(n)$ holomorphic vector bundle $V$ is determined
via a Fourier-Mukai transformation from spectral data ${\cal{C}}$ and
${\cal{N}}$, where ${\cal{C}}$ is
a divisor in $X$ which is
an effective, irreducible $n$-fold cover
of
the base $B$, and ${\cal{N}}$ is a line bundle on $X$. That is
\begin{equation}
({\cal{C}},{\cal{N}}) \longleftrightarrow V.
\label{x2}
\end{equation}
It is clear from this construction that the number of moduli of a
stable $SU(n)$ holomorphic
vector bundle $V$ is determined by the number of parameters
specifying its spectral cover $\cC$ and by the dimension of the space of
holomorphic line bundles $\cN$ defined on $\cC$.


\subsection{Parameters of $\cC$ and the Space of Line Bundles ${\cal N}$:}


The spectral cover $\cC$ is a divisor of the Calabi-Yau threefold $X$ and,
hence, uniquely determines a line bundle $\cO_{X}(\cC)$ on $X$. Then
$H^{0}
(X,\cO_{X}(\cC))$ is the space of holomorphic sections of $\cO_{X}(\cC)$.
We denote
its dimension by $h^{0}(X,\cO_{X}(\cC))$. It follows that there must exist
$h^{0}$ holomorphic sections $s_{1},..,s_{h^{0}}$ that span this space.
Note that the zero locus of any non-vanishing element of
$H^{0}(X,\cO_{X}(\cC))$,
\begin{equation}
 s_{\{a_{i}\}}=\Sigma_{i=1}^{h^{0}}a_{i}s_{i},
\label{eq:37}
\end{equation}
for fixed complex coefficients $a_{i}$, determines an effective divisor
$\cC_{\{a_{i}\}}$ of $X$ in the class of $\cC$. Running over all ${a_{i}}$
gives the complete set, $|\cC|$, of effective divisors in the class of
$\cC$.
Clearly
\begin{equation}
 |\cC|={\mathbb P}H^{0}(X,\cO_{X}(\cC)),
\label{eq:38}
\end{equation}
where the right hand side is the projectivization of
$H^{0}(X,\cO_{X}(\cC))$. It
follows that
\begin{equation}
 dim|\cC|=h^{0}(X,\cO_{X}(\cC))-1.
\label{eq:39}
\end{equation}
This quantity counts the number of parameters specifying the
spectral
cover $\cC$. We now consider the line bundles $\cN$ over $\cC$.

The set of holomorphic line bundles $\cN$ over the spectral cover $\cC$
is, by definition, determined by the set of holomorphic transition
functions allowed on $\cC$. These, in turn, are specified as the closed
but not
exact elements of the multiplicative group $C^{1}(\cC,\cO_{\cC}^{*})$ of
non-vanishing holomorphic functions on the intersection of any two open
sets in
the atlas of $\cC$. That is, the group of line bundles of $\cC$ is given
by
\begin{equation}
 Pic(\cC)=H^{1}(\cC,\cO_{\cC}^{*}),
\label{eq:40}
\end{equation}
where $ H^{1}(\cC,\cO_{\cC}^{*})$ is the first \v Cech cohomology group of
$ \cO_{\cC}^{*}$ on $\cC$. Clearly then, the dimension of the space of
line
bundles $\cN$ over $\cC$ is specified by
\begin{equation}
 dimPic(\cC)=h^{1}(\cC,\cO_{\cC}^{*}).
\label{eq:41}
\end{equation}
Putting these results together, we see that the
number of moduli of a stable $SU(n)$ holomorphic vector bundle $V$, which
we will
denote
by $n(V)$, is given by
\begin{equation}
 n(V)= (h^{0}(X,\cO_{X}(\cC))-1) + h^{1}(\cC,\cO_{\cC}^{*}).
\label{eq:42}
\end{equation}
We now turn to the explicit evaluation of each of the terms in this
expression.

Our basic approach to evaluating $h^{0}(X,\cO_{X}(\cC))$ is through the
Riemann-Roch theorem which, in this context, states that
\begin{equation}
\chi_{E}(X,\cO_{X}(\cC))=h^{0}(X,\cO_{X}(\cC))-h^{1}+h^{2}-h^{3},
\label{eq:43}
\end{equation}
where $\chi_{E}(X,\cO_{X}(\cC))$ is the Euler characteristic and $h^{q}$
for
$q=1,2,3$ are the dimensions of the higher cohomology groups of $X$
evaluated in $\cO_{X}(\cC)$. To proceed, we will use the fact that the
stable $SU(n)$ holomorphic vector bundles that we consider in this
paper are positive. This allows us to employ the Kodaira vanishing theorem
which, in turn, allows us to conclude that
\begin{equation}
H^{q}(X,\cO_{X}(\cC))=0
\label{eq:48}
\end{equation}
for $q>0$. It follows that $h^{q}=0$ for $q=1,2,3$ and the Riemann-Roch
theorem simplifies to
\begin{equation}
\chi_{E}(X,\cO_{X}(\cC))=h^{0}(X,\cO_{X}(\cC)).
\label{eq:49}
\end{equation}
Therefore, to evaluate $h^{0}(X,\cO_{X}(\cC))$ we need simply to evaluate
the
Euler characteristic. For the situation at hand, the Euler characteristic
is determined from the Atiyah-Singer index theorem to be
\begin{equation}
\label{eq:50a}
\chi_{E}(X,\cO_{X}(\cC))=
\frac{1}{6} \int_{X}c_{1}^{3}(\cO_{X}(\cC))+
\frac{1}{12}\int_{X} c_{1}(\cO_{X}(\cC)) \wedge c_2(TX).
\end{equation}
The first Chern class $c_1(\cO_{X} (\cC))$ is just given by $\cC$.
The second Chern class of the tangent bundle $c_2(TX)$ has been found in
\cite{FMW1} to be
\begin{equation}
c_2(TX)= \pi^{*}(c_2(B)+ 11 c_1^2 (B))+12 \sigma \cdot \pi^{*}(c_1 (B)).
\label{eq:50b}
\end{equation}
Using equation~\eqref{A3} for the first and second Chern
classes of ${\mathbb F}_{r}$ and the spectral cover given in
equations~\eqref{eq:7} and~\eqref{eq:17},
one can compute $\chi_{E}(X,\cO_{X}(\cC))$.
Of course, there are further restrictions on the integers $a$ and $b$.
These
are 1. the non-negativity conditions~\eqref{eq:18} required to make $\cC$
effective and 2. conditions~\eqref{eq:19},~\eqref{eq:20}
and~\eqref{eq:25a}
necessary to render $\cC$ irreducible and positive.
With these restrictions, equation~\eqref{eq:49} is valid. We find that
\begin{equation}
h^{0}(X,\cO_{X}(\cC))=\frac{n}{3}(4n^{2}-1)+nab-(n^{2}-2)(a+b)
+ar(\frac{n^{2}}{2}-1)-\frac{n}{2}ra^{2}.
\label{eq:54}
\end{equation}
We now proceed to calculate $h^{1}(\cC,\cO_{\cC}^{*})$.

To evaluate $h^{1}(\cC,\cO_{\cC}^{*})$, we note that it is an element in a
long exact sequence of cohomology groups given in part by
\begin{equation}
\rightarrow H^{1}(\cC,\cO_{\cC}) \rightarrow H^{1}(\cC, \cO_{\cC}^{*})
\rightarrow H^{2}(\cC,{\mathbb Z}) \rightarrow,
\label{eq:65}
\end{equation}
where $H^{1}(\cC, \cO_{\cC}^{*}) \rightarrow H^{2}(\cC,{\mathbb Z})$ is
the \v Cech coboundary operation $\delta_{1}$. However, using the
Lefschetz
theorem one can show that
\begin{equation}
H^{1}(\cC,\cO_{\cC})=0.
\label{eq:63}
\end{equation}
Hence, the mapping
\begin{equation}
\delta_{1}: H^{1}(\cC, \cO_{\cC}^{*}) \rightarrow H^{2}(\cC,{\mathbb Z})
\label{eq:66}
\end{equation}
is injective. That is,
\begin{equation}
Pic(\cC)=H^{1}(\cC, \cO_{\cC}^{*}) \subset H^{2}(\cC,{\mathbb Z}).
\label{eq:67}
\end{equation}
Note that $H^{2}(\cC,{\mathbb Z})$ forms a rigid lattice and, hence, there
are no smooth deformations. We conclude that
\begin{equation}
dimH^{1}(\cC, \cO_{\cC}^{*})=0.
\label{eq:68}
\end{equation}
%


\subsection{n(V) for $B={\mathbb F}_{r}$:}


We can now give the final expression for the number of moduli
of a positive, stable $SU(n)$ holomorphic
vector bundle over an elliptically fibered Calabi-Yau threefold with base
$B={\mathbb F}_{r}$. The associated spectral cover is given by
\begin{equation}
\cC=n\sigma+\pi^{*}(a\cS +b\cE),
\label{eq:69}
\end{equation}
where the effectiveness of $\cC$ requires
\begin{equation}
a \geq 0, \qquad b \geq 0,
\label{eq:70}
\end{equation}
the irreducibility of $\cC$ demands that
\begin{equation}
b \geq ar, \qquad a \geq 2n, \qquad b \geq n(r+2)
\label{eq:71}
\end{equation}
and the positivity requirement for $\cC$ implies
\begin{equation}
a > 2n, \qquad b > ar-n(r-2).
\label{eq:72}
\end{equation}
Using equations~\eqref{eq:42},~\eqref{eq:54} and~\eqref{eq:68}, one can
conclude the following.

\begin{itemize}

\item The number of vector bundle moduli is given by
\begin{equation}
n(V)=\frac{n}{3}(4n^{2}-1)+nab-(n^{2}-2)(a+b)
+ar(\frac{n^{2}}{2}-1)-\frac{n}{2}ra^{2}-1.
\label{eq:73}
\end{equation}
\end{itemize}

\noindent This equation will be
essential to the subsequent discussion.


\subsection{Vector Bundles Restricted to $\pi^{*}z$ and their Moduli:}


In this paper, we are interested in a heterotic superstring wrapped once
around
a genus-zero holomorphic curve in the elliptically fibered Calabi-Yau
threefold $X$.
In fact, the curves we consider are of the form
\begin{equation}
\sigma \cdot \pi^{*}z,
\label{x3}
\end{equation}
where $z$ is a holomorphic curve in the base surface $B$ of $X$. We will
frequently
refer to $\sigma \cdot \pi^{*}z$ simply as $z$.
It is clear, furthermore, that the superstring will couple, not to the
complete vector bundle $V$ but, rather, to the restriction of the vector
bundle to the curve $z$, denoted by $V|_{z}$. The properties of this
bundle,
however, are most easily analyzed by first considering $V|_{\pi^{*}z}$,
the
restriction of $V$ to the surface $\pi^{*}z$, to which we now turn.

We begin by noting that the surface $\pi^{*}z$ is elliptically fibered
over
the base curve $z$ and that
\begin{equation}
{\cal{C}}|_{\pi^{*}z} = {\cal C} \cdot \pi^{*}z
\label{smile}
\end{equation}
is an effective, irreducible curve which is an
$n$-fold cover of the base $z$. Similarly, ${\cal{N}}|_{\pi^{*}z}$ is the
restriction of the line bundle ${\cal{N}}$ to $\pi^{*}z$. We will, to
simplify
our notation, denote these quantities by
\begin{equation}
C={\cal{C}}|_{\pi^{*}z}, \qquad N={\cal{N}}|_{\pi^{*}z}.
\label{fun1}
\end{equation}
Given the restricted spectral data $C$ and $N$,
one can construct $V|_{\pi^{*}z}$ via a
Fourier-Mukai
transformation. That is,
\begin{equation}
(C, N) \longleftrightarrow
V|_{\pi^{*}z}.
\label{x4}
\end{equation}
As we did in the previous section for $V$, we can ask how to define and
enumerate the vector bundle moduli associated with $V|_{\pi^{*}z}$
(we want to count the deformations of $V|_{\pi^* z}$ which come as
restrictions of deformations of $V$, not all abstract deformations
of $V|_{\pi^* z}$).
It is
clear from~\eqref{x4} that the number of moduli of $V|_{\pi^{*}z}$ is
determined by the number of parameters specifying the restricted spectral
cover
$C$ and by the size of the space of restricted holomorphic line
bundles $N$. Since the holomorphic line bundle
${\cal N}$ on the threefold does not have any continuous
parameters, neither does its restriction. Therefore, we only need
to consider the parameters of the spectral cover.

The spectral cover $C$ is a divisor of the surface
$\pi^{*}z$
and,
hence, uniquely determines a line bundle $\cO_{\pi^{*}z}(C)$
on $\pi^{*}z$. Then $H^{0}
(\pi^{*}z,\cO_{\pi^{*}z}(C))$ is the space of holomorphic
sections
of $\cO_{\pi^{*}z}(C)$. We
denote
its dimension by $h^{0}(\pi^{*}z,\cO_{\pi^{*}z}(C))$.
For the reasons discussed previously, the set,
$|C|$, of all effective divisors in the class of
$C$ is given by
\begin{equation}
|C|={\mathbb
P}H^{0}(\pi^{*}z,\cO_{\pi^{*}z}(C)),
\label{eq:x6}
\end{equation}
where the right hand side is the projectivization of
$H^{0}(\pi^{*}z,\cO_{\pi^{*}z}(C))$. It
follows that
\begin{equation}
 dim|C|=h^{0}(\pi^{*}z,\cO_{\pi^{*}z}(C))-1.
\label{eq:x7}
\end{equation}
This quantity counts the number of parameters specifying the
spectral
cover $C$.
Putting the results of this subsection together, we see that the
number of moduli of the restricted  vector bundle $V|_{\pi^{*}z}$, which
we will denote
by $n(V|_{\pi^{*}z})$, is given by
\begin{equation}
 n(V|_{\pi^{*}z})= dim|C|
\label{eq:x10a}
\end{equation}
and, hence, using~\eqref{eq:x7} that
\begin{equation}
 n(V|_{\pi^{*}z})= h^{0}(\pi^{*}z,\cO_{\pi^{*}z}
(C))-1.
\label{eq:x10}
\end{equation}
We now turn to the explicit calculation
 of $h^{0}(\pi^{*}z,\cO_{\pi^{*}z}
(C))$.

To evaluate $h^{0}(\pi^{*}z,\cO_{\pi^{*}z}(C))$
we again consider the Riemann-Roch theorem, which now becomes
\begin{equation}
\chi_{E} (\pi^* z, {\cal O}_{\pi^* z}(C))=
h^{0} (\pi^* z, {\cal O}_{\pi^* z}(C))-h^1 +h^2,
\label{star}
\end{equation}
where
$\chi_{E} (\pi^{*} z, {\cal O}_{\pi^* z}(C))$
is the Euler characteristic and $h^{q}, q=1, 2$ are the dimensions
of the higher cohomology groups of $\pi^* z$ with coefficients
in ${\cal O}_{\pi^* z}(C)$. As in the previous
calculation
of $h^{0}(X, {\cal O}_{X}({\cal C}))$, expression~\eqref{star} is only
useful to us if $h^{1}$ and
$h^{2}$ can be shown to vanish.
To do this, we first need to demonstrate that
${\cal O}_{\pi^* z}(C)$ can be decomposed as
\begin{equation}
{\cal O}_{\pi^* z}(C) =
K_{\pi^* z} \otimes L_{\pi^* z},
\label{something}
\end{equation}
where $K_{\pi^* z}$ is the canonical bundle and $L_{\pi^* z}$
is a positive line bundle. If this is the case, one can use the Kodaira
vanishing
theorem which immediately imlies that
\begin{equation}
h^1 =h^2 =0.
\label{doublecheck}
\end{equation}
Decomposition~\eqref{something} can be shown for a wide class of curves
$z$ in ${\mathbb F}_{r}$.
However, the calculations are complicated and unenlightening. Recall from
Appendix B that the effective curves ${\cal{S}}$ and ${\cal{E}}$ form a
basis
of $H_{2}({\mathbb F}_{r}, {\mathbb R})$.
In this paper,
we will restrict our discussion to the curve ${\cal S}$.
The extension of these results to ${\cal E}$ or other effective curves
is straightforward, if tedious. That is, we will always choose the curve
\begin{equation}
z={\cal S},
\label{0.1}
\end{equation}
when specificity is required.
${\cal S}$  is a holomorphic curve of genus zero and, hence, ${\cal
S}={\mathbb
P}^{1}$. Furthermore, note from~\eqref{A2}
that ${\cal S}^2 =-r$ and, therefore, ${\cal S}$ is an isolated
holomorphic curve in ${\mathbb F}_{r}$ for $r=1, 2$. It then
follows that $\sigma \cdot \pi^* {\cal S}$ is an isolated
holomorphic curve of genus zero in $X$. We demonstrated
in \cite{BDO1} that
\begin{equation}
\pi^* {\cal S} = K3, \quad dP_9, \quad {\mathbb P}^{1}\times T^2
\label{0.2}
\end{equation}
for $B={\mathbb F}_0, {\mathbb F}_1$ and ${\mathbb F}_2$ respectively.
As discussed in Appendix B, for $B={\mathbb F}_{r}$ the spectral cover
in $X$ is of the form
\begin{equation}
{\cal C}= n\sigma +\pi^{*} ((a+1){\cal S} +b {\cal E}),
\label{XX}
\end{equation}
where $a+1, b$ are integers satisfying the effectiveness, irreducibility
and positivity conditions~\eqref{eq:18}, ~\eqref{eq:19}, ~\eqref{eq:20}
and~\eqref{eq:25a}. Note that we use $a+1$, rather than
$a$, as the coefficient of ${\cal S}$ in~\eqref{XX} to match
our conventions in \cite{BDO1}. The coefficient $a+1$ signifies the vector
bundle after the small instanton transition, when the data of a single
curve
${\cal S}$ is added to the bundle. Using~\eqref{A2}, it then follows
from~\eqref{smile} and~\eqref{fun1} that
\begin{equation}
C= n\sigma|_{\pi^* {\cal S}} +
(b- (a+1)r)F.
\label{F}
\end{equation}
We showed in \cite{BDO1} that
\begin{equation}
{\cal O}_{\pi^* {\cal S}}(C) =
K_{\pi^* {\cal S}} \otimes L_{\pi^* {\cal{S}}},
\label{0.3}
\end{equation}
where
\begin{equation}
L_{\pi^* {\cal S}} = {\cal O}_{\pi^* {\cal S}}
(n \sigma|_{\pi^{*} {\cal S}} + (b-ar)F),
\label{0.4}
\end{equation}
and that the line bundle $L_{\pi^* {\cal S}}$ is positive if and only if
\begin{equation}
b> ar - n(r-2).
\label{vert}
\end{equation}
However, this is exactly the second positivity condition in~\eqref{A2}.
Therefore, if ${\cal C}$ is chosen to be positive then so is the
line bundle $L_{\pi^* {\cal S}}$, and equation~\eqref{doublecheck} holds.
Therefore, equation~\eqref{star} becomes
\begin{equation}
\chi_{E} (\pi^* {\cal S}, {\cal O}_{\pi^* {\cal S}}
(C))=
h^{0} (\pi^* {\cal S}, {\cal O}_{\pi^* {\cal S}}
(C)) .
\label{0.5}
\end{equation}
The Euler characteristic and, hence,
$h^{0} (\pi^* {\cal S}, {\cal O}_{\pi^* {\cal S}}
(C))$
can be calculated using the Atiyah-Singer index theorem. We showed in
\cite{BDO1}
that
\begin{equation}
h^{0} (\pi^* {\cal S}, {\cal O}_{\pi^* {\cal S}}
(C))=
nb-(n^2 -2) + \frac{rn}{2} (n-1) - r (na +1).
\label{doublecross}
\end{equation}
As was discussed in detail in \cite{BDO1},
$h^{0} (\pi^* {\cal S}, {\cal O}_{\pi^* {\cal S}}
(C))$
has a physical interpretation as the number of ``transition''
moduli associated with the small instanton transition centered
on the curve $z= {\cal S}$. As these transition moduli will not be the
central focus of this paper, we refer the reader to \cite{BDO1}
for details. What is directly relevant to this paper are the number
and properties of the parameters of the projective space
\begin{equation}
|C|=
{\mathbb P}H^{0} (\pi^* {\cal S}, {\cal O}_{\pi^* {\cal S}}
(C))
\label{0.6}
\end{equation}
which specify the vector bundle moduli of $V|_{\pi^{*}{\cal{S}}}$.
From~\eqref{eq:x10} and~\eqref{doublecross} one can conclude the
following.

\begin{itemize}
\item The number of vector bundle moduli of $V|_{\pi^{*}{\cal{S}}}$ is
given by
\begin{equation}
n(V|_{\pi^* {\cal S}})=nb-(n^2 -1) + \frac{rn}{2} (n-1) - r (na
+1).
\label{B}
\end{equation}
\item For each choice of the line bundle ${\cal N}$,
the vector bundle moduli of $V|_{\pi^{*}{\cal{S}}}$
parameterize the projective space
\begin{equation}
{\mathbb P}H^{0} (\pi^* {\cal S}, {\cal O}_{\pi^* {\cal S}}
(C)) \simeq
{\mathbb P}^{nb-(n^2 -1) + \frac{rn}{2} (n-1) - r (na +1)}.
\label{C}
\end{equation}
\end{itemize}

To illustrate these concepts, we now present four examples
that we will explore throughout this paper. In the first two of these
examples, we
will choose
\begin{equation}
B = {\mathbb F}_{1}, \qquad G = SU(3)
\label{0.7}
\end{equation}
and, hence, $r=1$, $n=3$ and
\begin{equation}
\pi^{*}{\cal S} =dP_9.
\label{0.8}
\end{equation}
It follows that~\eqref{B} and~\eqref{C} become
\begin{equation}
n(V|_{dP_{9}}) = 3(b-a)-6
\label{0.9}
\end{equation}
and
\begin{equation}
{\mathbb P}H^{0} (dP_{9}, {\cal O}_{dP_{9}}
(C)) \simeq {\mathbb P}^{3(b-a)-6}
\label{0.10}
\end{equation}
respectively.
Furthermore, in order to assure that ${\cal C}$ is positive,
we will always assume that
\begin{equation}
a>5.
\label{0.11}
\end{equation}
{\bf Example 1 \rm}: In this example, take
\begin{equation}
b-a =5.
\label{0.12}
\end{equation}
Then, from~\eqref{F} we have
\begin{equation}
C= 3 \sigma|_{dP_9} +4F.
\label{alpha}
\end{equation}
Note, that $a+1, b$ satisfy~\eqref{eq:18},~\eqref{eq:19}
and~\eqref{eq:20}. Therefore, $C$ is both
effective and irreducible. In addition, both
conditions in~\eqref{eq:25a} are satisfied, guaranteeing that
${\cal C}$ and, hence, $L_{dP_9}$ are positive.
It follows that
\begin{equation}
n(V|_{dP_9})=9
\label{beta}
\end{equation}
and
\begin{equation}
{\mathbb P}H^{0} (dP_9, {\cal O}_{dP_9} (3 \sigma|_{dP_9} +4F))
\simeq {\mathbb P}^{9}.
\label{0.13}
\end{equation}

\noindent {\bf Example 2 \rm}: In this example, choose
\begin{equation}
b-a =6.
\label{0.14}
\end{equation}
Then~\eqref{F} implies
\begin{equation}
C= 3 \sigma|_{dP_9} +5F.
\label{gamma}
\end{equation}
Integers $a+1, b$ satisfy~\eqref{eq:18},~\eqref{eq:19}
and~\eqref{eq:20} and both conditions in~\eqref{eq:25a}. Therefore
$C$ is effective and irreducible and $L_{dP_{9}}$ is
positive. Then
\begin{equation}
n(V|_{dP_9})=12
\label{delta}
\end{equation}
and
\begin{equation}
{\mathbb P}H^{0} (dP_9, {\cal O}_{dP_9} (3 \sigma|_{dP_9} +5F))
\simeq {\mathbb P}^{12}.
\label{0.15}
\end{equation}

\noindent In the third and the fourth examples, we will take
\begin{equation}
B={\mathbb F}_{2}, \qquad G=SU(3)
\label{zorro1}
\end{equation}
and, hence, $r=2$, $n=3$ and
\begin{equation}
\pi^{*}{\cal{S}}={\mathbb P}^{1} \times T^{2}.
\label{zorro2}
\end{equation}
It follows that~\eqref{B} and~\eqref{C} become
\begin{equation}
n(V|_{{\mathbb P}^{1} \times T^{2}})=3(b-2a)-4
\label{zorro3}
\end{equation}
and
\begin{equation}
{\mathbb P}H^{0}({\mathbb P}^{1} \times T^{2}, {\cal{O}}_{{\mathbb P}^{1}
\times T^{2}}(C)) \simeq
{\mathbb P}^{3(b-2a)-4}
\label{zorro4}
\end{equation}
respectively. Furthermore, we will always take
\begin{equation}
a > 5.
\label{zorro5}
\end{equation}

\noindent {\bf Example 3 \rm}: Choose in this case
\begin{equation}
b-2a =3.
\label{f11}
\end{equation}
Then, we see from~\eqref{F} that
\begin{equation}
C= 3 \sigma|_{{\mathbb P}^{1}
\times T^{2}} +F.
\label{f22}
\end{equation}
Integers $a+1, b$ satisfy~\eqref{eq:18},~\eqref{eq:19}
and~\eqref{eq:20} and both conditions in~\eqref{eq:25a}. Therefore
$C$ is effective and irreducible
and
$L_{{\mathbb P}^{1} \times T^{2}}$ is positive. Then
\begin{equation}
n(V|_{{\mathbb P}^{1} \times T^{2}})=5
\label{f3}
\end{equation}
and
\begin{equation}
{\mathbb P}H^{0} ({\mathbb P}^{1} \times T^{2}, {\cal O}_{{\mathbb P}^{1}
\times T^{2}} (3 \sigma|_{{\mathbb P}^{1} \times T^{2}} +F))
\simeq {\mathbb P}^{5}.
\label{f4}
\end{equation}

\noindent {\bf Example 4 \rm}: Choose in this case
\begin{equation}
b-2a =4.
\label{0.15.5}
\end{equation}
Then, we see from~\eqref{F} that
\begin{equation}
C= 3 \sigma|_{{\mathbb P}^{1}
\times T^{2}} +2F.
\label{ref1}
\end{equation}
Integers $a+1, b$ satisfy~\eqref{eq:18},~\eqref{eq:19}
and~\eqref{eq:20} and both conditions in~\eqref{eq:25a}. Therefore
$C$ is effective and irreducible
and
$L_{{\mathbb P}^{1} \times T^{2}}$ is positive. Then
\begin{equation}
n(V|_{{\mathbb P}^{1} \times T^{2}})=8
\label{ref2}
\end{equation}
and
\begin{equation}
{\mathbb P}H^{0} ({\mathbb P}^{1} \times T^{2}, {\cal O}_{{\mathbb P}^{1}
\times T^{2}} (3 \sigma|_{{\mathbb P}^{1} \times T^{2}} +2F))
\simeq {\mathbb P}^{8}.
\label{ref3}
\end{equation}
%

\section{Vector Bundles on the Curve $z$:}


\subsection{Vector Bundle Moduli on $z$:}


For the calculation of the superpotential, it is very important to know
the direct image of the line bundle
${\cal O}_{\pi^* z} (C)$ on $z$, that is
\begin{equation}
\pi_* {\cal O}_{\pi^* z} (C).
\label{D}
\end{equation}
Clearly, since $C$ is an $n$-fold cover of
$z$,~\eqref{D} will be a rank $n$ vector bundle over $z$.
Since, in all examples in this paper, $z$ will be chosen to be a Riemann
surface of
genus zero, we can always decompose this rank $n$ vector bundle into
the direct sum of $n$ line bundles
\begin{equation}
\bigoplus_{i=1}^{n} {\cal O}_{z}(m_i)
\label{E}
\end{equation}
for some integers $m_i$. To proceed, we once again restrict our attention
to the fundamental curve
\begin{equation}
z={\cal S}.
\label{0.28}
\end{equation}
The extention of
these results to other effective curves $z$ is straightforward.
Then, we showed in~\eqref{F} that $C$
has the form
\begin{equation}
C= n \sigma|_{\pi^* {\cal S}} +fF,
\label{0.29}
\end{equation}
where
\begin{equation}
f = b-(a+1)r.
\label{I}
\end{equation}
It is possible to show, for
\begin{equation}
B={\mathbb F}_0 \Rightarrow \pi^{*}{\cal S} =K3
\label{0.30}
\end{equation}
that
\begin{equation}
\pi_* {\cal O}_{K3} (n \sigma|_{K3} +fF)
={\cal O}_{{\cal S}}(f) \oplus \bigoplus_{i=2}^{n}
{\cal O}_{{\cal S}}(f-2i),
\label{G}
\end{equation}
for
\begin{equation}
B={\mathbb F}_1 \Rightarrow \pi^{*}{\cal S} =dP_9
\label{0.31}
\end{equation}
that
\begin{equation}
\pi_* {\cal O}_{dP_9} (n \sigma|_{dP_9} +fF)
={\cal O}_{{\cal S}}(f) \oplus \bigoplus_{i=2}^{n}
{\cal O}_{{\cal S}}(f-i)
\label{H}
\end{equation}
and for
\begin{equation}
B={\mathbb F}_2 \Rightarrow \pi^{*}{\cal S} ={\mathbb P}^{1} \times T^{2}
\label{zorro7}
\end{equation}
that
\begin{equation}
\pi_* {\cal O}_{{\mathbb P}^{1} \times T^{2}} (n \sigma|_{{\mathbb P}^{1}
\times T^{2}} +fF)=\bigoplus_{i=1}^{n}{\cal O}_{{\cal S}}(f).
\label{ref4}
\end{equation}

\noindent The proofs of these statements are sufficiently complicated
that we present them in Appendix C. Expressions~\eqref{G},~\eqref{H}
and~\eqref{ref4} allow us to compute
$\pi_* {\cal O}_{\pi^* {\cal S}} (C)$
explicitly as the sum of line bundles over ${\cal S}$. We are
particularly
interested in the linear space
\begin{equation}
H^{0} ({\cal S},
\pi_* {\cal O}_{\pi^* {\cal S}} (C))
\label{0.32}
\end{equation}
of holomorphic sections of the rank $n$ vector bundle on ${\cal S}$ and,
in particular, in its dimension
\begin{equation}
h^{0} ({\cal S},
\pi_* {\cal O}_{\pi^* {\cal S}} (C)).
\label{0.33}
\end{equation}

\begin{itemize}

\item Let us first assume that $B ={\mathbb F}_{0}$. Then
\begin{equation}
h^{0} (K3,
\pi_* {\cal O}_{K3} (C))
=h^{0} ({\cal S}, {\cal O}_{\cal S}(f) \oplus \bigoplus_{i=2}^{n}
{\cal O}_{\cal S}(f-2i)).
\label{0.34}
\end{equation}
Using the fact that
\begin{equation}
h^{0} ({\mathbb P}^{1}, {\cal O}_{{\mathbb P}^1} (m))=m+1
\label{L}
\end{equation}
for $m \geq 0$ and
\begin{equation}
h^{0} ({\mathbb P}^{1}, {\cal O}_{{\mathbb P}^1} (m))=0
\label{M}
\end{equation}
for $m < 0$, the summation
\begin{equation}
\sum_{i=2}^{n} i = \frac{n^2+n-2}{2}
\label{J}
\end{equation}
and~\eqref{I} for $r=0$, one can easily show that
\begin{equation}
h^{0} ({\cal S}, {\cal O}_{\cal S}(f) \oplus \bigoplus_{i=2}^{n}
{\cal O}_{\cal S}(f-2i)) = nb -n^2 +2.
\label{0.35}
\end{equation}
Note that this is identical to expression~\eqref{doublecross}
evaluated for $r=0$ and, hence,
$h^{0} ({\cal S}, {\cal O}_{\cal S}(f) \oplus \bigoplus_{i=2}^{n}
{\cal O}_{\cal S}(f-2i))$
gives the number of transition moduli for $z={\cal S}$ and
$B={\mathbb F}_{0}$.

\item Now assume that $B={\mathbb F}_{1}$.
Then
\begin{equation}
h^{0}(dP_{9},\pi_{*}{\cal{O}}_{dP_{9}}(C))=
h^{0} ({\cal S}, {\cal O}_{\cal S}(f) \oplus \bigoplus_{i=2}^{n}
{\cal O}_{\cal S}(f-i)).
\label{0.36}
\end{equation}
Using~\eqref{L},~\eqref{J} and ~\eqref{I} for $r=1$, we find that
\begin{equation}
h^{0} ({\cal S}, {\cal O}_{\cal S}(f) \oplus \bigoplus_{i=2}^{n}
{\cal O}_{\cal S}(f-i))= n(b-a-1) + \frac{n}{2}(1-n) +1,
\label{0.36.5}
\end{equation}
which is identical to expression~\eqref{doublecross} evaluated for $r=1$.
Therefore,
again,
$h^{0} ({\cal S}, {\cal O}_{\cal S}(f) \oplus \bigoplus_{i=2}^{n}
{\cal O}_{\cal S}(f-i))$ gives the number of transition moduli
for $z={\cal S}$ and $B={\mathbb F}_{1}$.

\item Finally, consider $B={\mathbb F}_{2}$. Then
\begin{equation}
h^{0}({\mathbb P}^{1} \times T^{2},\pi_{*}{\cal{O}}_{{\mathbb P}^{1}
\times T^{2}}(C))=
h^{0} ({\cal S}, \bigoplus_{i=1}^{n}{\cal O}_{\cal S}(f)).
\label{0.36aa}
\end{equation}
Using~\eqref{L} and ~\eqref{I} for $r=2$, we find that
\begin{equation}
h^{0} ({\cal S}, \bigoplus_{i=1}^{n}
{\cal O}_{\cal S}(f))= n(b-2a-1),
\label{0.36.5aa}
\end{equation}
which is identical to expression~\eqref{doublecross} evaluated for $r=2$.
Therefore, again,
$h^{0} ({\cal S}, \bigoplus_{i=1}^{n}
{\cal O}_{\cal S}(f))$ gives the number of transition moduli
for $z={\cal S}$ and $B={\mathbb F}_{2}$.

\end{itemize}

The fact that
$h^{0} ({\cal S},
\pi_{*}{\cal O}_{\pi^*{\cal S}}(C))$
is identical to the number of transition moduli is not coincidental.
Using a Leray spectral sequence, one can show that
\begin{equation}
H^{0} ({\cal S},
\pi_{*}{\cal O}_{\pi^*{\cal S}}(C))=
H^{0} (\pi^{*}{\cal S}, {\cal O}_{\pi^{*}{\cal S}}
(C))
\label{0.37}
\end{equation}
for any base $B={\mathbb F}_{r}$.
It follows from this that
$h^{0}({\cal S},
\pi_{*}{\cal O}_{\pi^*{\cal S}}(C))$
is the number of transition moduli.

To illustrate these concepts, we now consider the four
examples
introduced previously. In the first two examples
\begin{equation}
r=1, \quad n=3, \quad a>5.
\label{0.39}
\end{equation}

\noindent {\bf Example 1 \rm}: In this example
\begin{equation}
b-a =5
\label{0.40}
\end{equation}
and we see from~\eqref{alpha} that
\begin{equation}
C= 3 \sigma|_{dP_9} + 4F.
\label{0.41}
\end{equation}
Then, equation~\eqref{H} implies
\begin{equation}
\pi_{*}{\cal O}_{dP_9} (3 \sigma|_{dP_9} +4F)
={\cal O}_{{\cal S}}(4) \oplus {\cal O}_{{\cal S}}(2)
\oplus {\cal O}_{{\cal S}}(1).
\label{0.42}
\end{equation}
Furthermore, using~\eqref{L} we see
\begin{equation}
h^0 ({\cal S}, {\cal O}_{{\cal S}}(4) \oplus {\cal O}_{{\cal S}}(2)
\oplus {\cal O}_{{\cal S}}(1)) =10,
\label{0.43}
\end{equation}
which is consistent with~\eqref{beta}.\\

\noindent {\bf Example 2 \rm}: In this example
\begin{equation}
b-a =6
\label{0.44}
\end{equation}
and from~\eqref{gamma}
\begin{equation}
C= 3 \sigma|_{dP_9} + 5F.
\label{0.45}
\end{equation}
Then, equation~\eqref{H} implies that
\begin{equation}
\pi_{*}{\cal O}_{dP_9} (3 \sigma|_{dP_9} +5F)
={\cal O}_{{\cal S}}(5) \oplus {\cal O}_{{\cal S}}(3)
\oplus {\cal O}_{{\cal S}}(2).
\label{0.46}
\end{equation}
Furthermore, using~\eqref{L} we see
\begin{equation}
h^0 ({\cal S}, {\cal O}_{{\cal S}}(5) \oplus {\cal O}_{{\cal S}}(3)
\oplus {\cal O}_{{\cal S}}(2)) =13
\label{0.47}
\end{equation}
which is consistent with~\eqref{delta}.\\

\noindent In the third and the fourth examples, we choose
\begin{equation}
r=2, \qquad n=3, \qquad a>5.
\label{zorro8}
\end{equation}

\noindent {\bf Example 3 \rm}: In this example
\begin{equation}
b-2a =3
\label{f5}
\end{equation}
and from~\eqref{f22}
\begin{equation}
C= 3 \sigma|_{{\mathbb P}^{1}
\times T^{2}} + F.
\label{f6}
\end{equation}
Then, equation~\eqref{ref4} implies that
\begin{equation}
\pi_{*}{\cal O}_{{\mathbb P}^{1} \times T^{2}} (3 \sigma|_{{\mathbb P}^{1}
\times T^{2}} +F)
={\cal O}_{{\cal S}}(1) \oplus {\cal O}_{{\cal S}}(1)
\oplus {\cal O}_{{\cal S}}(1).
\label{f7}
\end{equation}
Furthermore, using~\eqref{L} we see
\begin{equation}
h^0 ({\cal S}, {\cal O}_{{\cal S}}(1) \oplus {\cal O}_{{\cal S}}(1)
\oplus {\cal O}_{{\cal S}}(1)) =6,
\label{f8}
\end{equation}
which is consistent with~\eqref{f3}.\\

\noindent {\bf Example 4 \rm}: In this example
\begin{equation}
b-2a =4
\label{0.48aa}
\end{equation}
and from~\eqref{ref1}
\begin{equation}
C= 3 \sigma|_{{\mathbb P}^{1}
\times T^{2}} + 2F.
\label{0.49aa}
\end{equation}
Then, equation~\eqref{ref4} implies that
\begin{equation}
\pi_{*}{\cal O}_{{\mathbb P}^{1} \times T^{2}} (3 \sigma|_{{\mathbb P}^{1}
\times T^{2}} +2F)
={\cal O}_{{\cal S}}(2) \oplus {\cal O}_{{\cal S}}(2)
\oplus {\cal O}_{{\cal S}}(2).
\label{0.50aa}
\end{equation}
Furthermore, using~\eqref{L} we see
\begin{equation}
h^0 ({\cal S}, {\cal O}_{{\cal S}}(2) \oplus {\cal O}_{{\cal S}}(2)
\oplus {\cal O}_{{\cal S}}(2)) =9,
\label{0.51}
\end{equation}
which is consistent with~\eqref{ref2}.

Putting everything together, we can conclude the following.

\begin{itemize}
\item The number of transition moduli associated with curve $z$ can be
determined as follows. First consider the spectral cover restricted to
$\pi^{*}z$, $C$. The image
\begin{equation}
\pi_{*}{\cal{O}}_{\pi^{*}z}(C)
\label{tonto1}
\end{equation}
is a rank $n$ vector bundle over $z$. The number of transition moduli
associated with $z$ is given by
\begin{equation}
h^{0}(z,\pi_{*}{\cal{O}}_{\pi^{*}z}(C)).
\label{tonto2}
\end{equation}
\item The number of vector bundle moduli associated with $V|_{\pi^{*}z}$
is
given by the dimension of the projective space, that is
\begin{equation}
n(V|_{\pi^{*}z})=
h^{0}(z,\pi_{*}{\cal{O}}_{\pi^{*}z}(C))-1.
\label{tonto3}
\end{equation}
\end{itemize}


\subsection{The Vector Bundle Restricted to $z$:}


As discussed previously , the superstring will couple
to the restriction of the vector bundle to the curve $z$, $V|_{z}$.
The number of moduli of $V|_{\pi^{*}z}$ was determined above and is
associated on $z$
with the image of the line bundle of $C$, that is,
$\pi_{*}{\cal{O}}_{\pi^{*}z}(C)$. We now want to
construct
$V|_{z}$ itself.
Note, however, that
$z$ is not an elliptic fibration and, hence, $V|_z$ can not be obtained
by the Fourier-Mukai transformation of spectral data. To construct
$V|_{z}$, we need to unwind what the Fourier-Mukai transformation actually
means.
Let ${\cal{P}}$ be the Poincare line bundle on the fiber product
$X \times_{B} X$ and, hence, by restriction on $X \times_{B} {\cal{C}}$.
Let $\pi_1$ and $\pi_2$ be the two projections
of the fiber product $X \times_{B} {\cal C}$ onto the two factors
$X$ and ${\cal C}$ respectively. The fact that $V$ is obtained by
a Fourier-Mukai transformation of $({\cal C}, {\cal N})$ means that
\begin{equation}
V= \pi_{1*} (\pi_{2}^{*} {\cal N}|_{{\cal{C}}} \otimes {\cal{P}}).
\label{puan1}
\end{equation}
Let us now consider the restriction, $V|_{\sigma}$, of $V$ to the zero
section $\sigma \subset X$. Clearly, using the base change theorem, this
can
be carried out as follows. First, restrict the line bundle $\pi_{2}^{*}
{\cal N}|_{{\cal{C}}} \otimes {\cal{P}}$ on $X \times_{B} {\cal{C}}$ to
\begin{equation}
(\pi_{2}^{*} {\cal N}|_{{\cal{C}}})|_{\sigma \times_{B} {\cal{C}}} \otimes
{\cal{P}}|_{\sigma \times_{B} {\cal{C}}}
\label{eng1}
\end{equation}
on $\sigma \times_{B} {\cal{C}} \subset X \times_{B} {\cal{C}}$. Using the
fact that
\begin{equation}
\sigma \times_{B} {\cal{C}}= {\cal{C}},
\label{eng2}
\end{equation}
this line bundle can be written simply as
\begin{equation}
{\cal N}|_{{\cal{C}}} \otimes
{\cal{P}}|_{\sigma \times_{B}{\cal{C}}}.
\label{eng3}
\end{equation}
Finally, recall that
\begin{equation}
\pi_{{\cal{C}}}: {\cal{C}} \rightarrow B
\label{eng4}
\end{equation}
is the $n$-fold covering map of the spectral cover onto the base. It
follows
that
\begin{equation}
V|_{B}=\pi_{{\cal{C}}*}({\cal N}|_{{\cal{C}}} \otimes
{\cal{P}}|_{\sigma \times_{B}{\cal{C}}}).
\label{eng5}
\end{equation}
Identifying $B$ with the zero section $\sigma$, and using the fact that
the
Poincare line bundle becomes trivial when restricted to the zero section,
we
see that
\begin{equation}
V|_{\sigma}=\pi_{{\cal{C}}*}{\cal N}|_{{\cal{C}}} .
\label{eng6}
\end{equation}
We are now free to restrict this result to the curve $z \subset \sigma$.
Recall that we are slightly abusing notation. What we call the curve
$z$ is, in fact, the curve $\sigma \cdot \pi^{*}z$ and, therefore,
it is contained inside the global section $\sigma$. Using the fact
that
\begin{equation}
{\cal{N}}|_{C}=N|_{C}
\label{pray1}
\end{equation}
we obtain the following result.

\begin{itemize}

\item The restriction of the vector bundle to the curve $z$ is given by
\begin{equation}
V|_{z} =\pi_{C*}N|_{C},
\label{puan2}
\end{equation}
where
\begin{equation}
\pi_{C}:C \to z
\label{puan3}
\end{equation}
is the $n$-fold covering map of $C$ onto the base
curve $z$.

\end{itemize}

In all applications in this paper, the curve
$z$ on which the superstring is wrapped will be isomorphic to
${\mathbb P}^{1}$. It is well known that the space of line bundles
over ${\mathbb P}^{1}$ does not have moduli since for a given first
Chern class there is a unique holomorphic line bundle. On the other
hand, equation~\eqref{puan2} depends, by restriction, on the spectral
cover,
which does have continuous parameters. What happens to $V|_{z}$ if we
start changing the parameters of $C$? A naive answer
is that nothing happens.
Since the vector bundle $V|_{z}$ does not have moduli,
it can depend only on the homology class of the spectral cover but
not on the choice of a representative inside this homology class.
This answer turns out to be incorrect. In the next sections,
by studying $h^{0}$ of the vector bundle
$V|_{z} \otimes {\cal O}_{z}(-1)$,
we will discover that a continuous deformation of the spectral cover
inside its homology class can cause a discrete change in the topology
of $V|_{z}$. Exploring how $h^{0}(z, V|_{z} \otimes {\cal O}_{z}(-1))$
changes as we change parameters of the spectral cover will
allow us to compute the superpotential.


\section{General Structure of the Zeros of the Pfaffian:}


\subsection{The Superpotential and the Pfaffian:}


In section $2$, we showed that, in heterotic superstring theory,
the superpotential is proportional to the Pfaffian of the chiral Dirac
operator
${\cal D}_{-}$ restricted to the curve $z$ on which the superstring
is wrapped. That is,
\begin{equation}
W \propto {\rm Pfaff}({\cal D}_{-}),
\label{3.15}
\end{equation}
where
\begin{equation}
{\rm Pfaff}({\cal D}_{-})=\sqrt{det{\cal D}_{-}},
\label{3.15a}
\end{equation}
${\cal D}_{-}$, for an appropriate choice of basis of the Clifford
algebra, is
given by
\begin{equation}
{\cal D}_{-} = \bordermatrix{     & {\ } & {\ }  \cr
                             {\ } & 0 & D_{-} \cr
                             {\ } & i\partial_{+} & 0 \cr}
\label{3.15b}
\end{equation}
and
\begin{equation}
D_{-} : \Gamma (z, V|_{z} \otimes S_{-})
\rightarrow  \Gamma (z, V|_{z} \otimes S_{+}),
\label{3.15c}
\end{equation}
with
\begin{equation}
S=S_{+} \oplus S_{-}
\label{3.15d}
\end{equation}
the spin bundle on the curve $z$. Since ${\cal D}_{-}$ involves $V|_{z}$,
it
follows from the above discussion that Pfaff$({\cal D}_{-})$ will be
dependent
upon the moduli of $V|_{\pi^{*}z}$. As discussed in~\cite{BDO2}, one can
show
that
\begin{equation}
det {\cal D}_{-} \propto \sqrt{detD_{-}D_{+}},
\label{3.15e}
\end{equation}
where
\begin{equation}
D_+ = D_{-}^{\dagger}.
\label{3.15f}
\end{equation}

The key point in the evaluation of ${\rm Pfaff}({\cal D}_{-})$ is to
analyse
under what circumstences it will vanish \cite{Witten2,BDO2}. Clearly,
${\rm Pfaff}({\cal D}_{-})$ vanishes if and only if one or both
of the operators $D_{-}$ and $D_{+}$ have a non-trivial zero mode. Recall
that
\begin{equation}
index D_{+} = dim {\ } ker D_{+} - dim {\ } ker D_{-}.
\label{3.16}
\end{equation}
From the Atiyah-Singer index theorem
\begin{equation}
indexD_{+} = \frac{i}{2 \pi} \int_{z} {\rm tr} {\cal F},
\label{3.17}
\end{equation}
where ${\cal F}$ is the curvature two-form associated with the connection
${\cal A}$ restricted to the curve $z$. Since the structure group of the
holomorphic vector bundle $V$
is contained in the semi-simple group $E_8 \times E_8$, we see that
${\rm tr} {\cal F}$ vanishes. Therefore,
\begin{equation}
index D_{+} =0
\label{3.18}
\end{equation}
and, hence,
\begin{equation}
dim {\ } ker D_{+} = dim {\ } ker D_{-}.
\label{3.19}
\end{equation}
It follows that ${\rm Pfaff}({\cal D}_{-})$ will vanish if and only if
\begin{equation}
dim {\ } ker D_{-} >0.
\label{3.20}
\end{equation}
By using the fact that operator $D_{-}$ depends on a connection
satisfying the Hermitian Yang-Mills equations, one can show
\cite{BDO2} that for a special choice of the gauge one can always set
\begin{equation}
D_{-} = i \bar \partial,
\label{3.21}
\end{equation}
where, for any open neighborhood in $z$ with coordinates
$\sigma$ and $\bar \sigma$
\begin{equation}
\bar \partial = \partial_{\bar \sigma}.
\label{3.22}
\end{equation}
Therefore, ${\rm Pfaff}({\cal D}_{-})$ will vanish if and only if
\begin{equation}
dim {\ } ker \bar \partial >0.
\label{3.23}
\end{equation}
However, it follows from the equation~\eqref{3.21} and~\eqref{3.22} that
the zero modes of operator $\bar \partial$ are precisely the holomorphic
sections of the vector bundle $V|_{z}\otimes S_{-}$. Using the fact that
on $z={\mathbb P}^{1}$
\begin{equation}
S_{-} = {\cal O}_{z}(-1),
\label{3.24}
\end{equation}
and defining
\begin{equation}
V|_{z} (-1) = V|_{z} \otimes {\cal O}_{z}(-1),
\label{3.25}
\end{equation}
we see that
\begin{equation}
dim {\ } ker \bar \partial = h^{0} (z, V|_{z} (-1)).
\label{3.26}
\end{equation}
Hence, the problem of determining the zeros of the Pfaffian of
${\cal D}_{-}$ is reduced to deciding whether or not there are any
non-trivial
global holomorphic sections of the vector bundle $V|_{z} (-1)$ over curve
$z$. That is, we conclude the following.

\begin{itemize}

\item The Pfaffian of ${\cal{D}}_{-}$ vanishes if and only if
\begin{equation}
h^{0}(z,V|_{z}(-1)) >0.
\label{pray2}
\end{equation}
\end{itemize}

An equivalent way of stating the same result is to realize that
the condition for the vanishing of ${\rm Pfaff}({\cal D}_{-})$
is directly related to the non-triviality or triviality of the bundle
$V|_{z}$. To see this, note that any holomorphic $SO(16) \times SO(16)$
vector bundle  $V|_{z}$ over a genus zero curve $z={\mathbb P}^{1}$
is of the form
\begin{equation}
V|_{z} = \bigoplus_{i=1}^{16} {\cal O}_{{\mathbb P}^{1}}(m_i) \oplus
{\cal O}_{{\mathbb P}^{1}}(-m_i)
\label{3.27}
\end{equation}
with non-negative integers $m_i$. Therefore,
\begin{equation}
V|_{z}(-1) = \bigoplus_{i=1}^{16} {\cal O}_{{\mathbb P}^{1}}(m_i-1) \oplus
{\cal O}_{{\mathbb P}^{1}}(-m_i-1).
\label{3.28}
\end{equation}
Using formulas~\eqref{L} and~\eqref{M}
for $h^0({\mathbb P}^1, {\cal O}_{{\mathbb P}^{1}}(m_i))$
from the previous section,
we see that
\begin{equation}
h^{0}(z, V|_{z}(-1)) = \sum_{i=1}^{16} m_i.
\label{3.31}
\end{equation}
Therefore $h^{0}(z, V|_{z}(-1))>0$ if and only if at least one $m_i$
is greater than zero. That is, $h^{0}(z, V|_{z}(-1))>0$ if and only if
the vector bundle $V|_{z}$ is non-trivial.

The rest of this section will be devoted to the general discussion,
in the context of stable, holomorphic vector bundles over elliptically
fibered Calabi-Yau threefolds $X$, of when $V|_{z}$ is trivial
and when it is non-trivial,
that is, equivalently, when $V|_{z}(-1)$ does not have
sections
and when it does. The structure group of the bundle will be assumed to be
an $SU(n)$ subgroup of $SO(16) \times SO(16)$, although all detailed
examples in this paper will, for specificity, be given for an $SU(3)$
structure group.
As we will see later in the paper, generically, the condition of
triviality
or non-triviality of the vector bundle $V|_{z}$ is not topological in the
sense that it is governed not soley by the
choice of the topological type of the spectral cover
or the line bundle but, rather,
by the choice of parameters of the
spectral cover. In other words, for a given
family of
positive spectral covers and,
therefore, a family of stable, holomorphic vector bundles on $X$
parametrized
by its moduli, the answer to the question whether or not $V|_{z}$ is
trivial
depends on the choice of the moduli. In the next section, we will give two
examples of when, for every choice of the moduli, the bundle $V|_{z}$
is non-trivial and the superpotential vanishes identically. We will then
give examples where, for a generic choice of moduli, the bundle $V|_{z}$
is trivial but there is a co-dimension one subspace where $V|_{z}$
becomes non-trivial. This means that, generically,
the Pfaffian and, therefore, the
superpotential is non-vanishing. However, the Pfaffian does vanish for
moduli on the co-dimension one subspace.


\subsection{Vanishing of the Pfaffian:}


We now discuss how the choice of moduli affects the triviality or
non-triviality of $V|_{z}$. In fact, it is technically easier to
study the equivalent problem of whether or not $h^{0}(z, V|_{z}(-1))$
vanishes.
In this paper, we will restrict the discussion to stable,
holomorphic vector bundles
over Calabi-Yau threefolds $X$ elliptically fibered over
Hirzebruch surfaces ${\mathbb F}_{r}$ with $r=0, 1, 2$. The curve $z$ will
be
chosen to be purely horizontal, that is, to lie in the base
$B={\mathbb F}_{r}$. As
discussed previously, for any curve $z$ in the base of $X$, the
restriction,
$V|_{z}$, of the holomorphic vector bundle to $z$ is given by
\begin{equation}
V|_{z} = \pi_{C*}N|_{C},
\label{3.32}
\end{equation}
where $C$ and $N$ denote the restriction of the spectral
cover ${\cal C}$ and line bundle ${\cal{N}}$ to the surface $\pi^{*}z$.
Then
\begin{equation}
H^{0}(z, V|_{z}(-1)) =
H^{0}(z, \pi_{C*}  N|_{C}
\otimes {\cal O}_{z}(-1))
\label{3.32.2}
\end{equation}
and, hence
\begin{equation}
h^{0}(z, V|_{z}(-1)) =
h^{0}(z, \pi_{C*}  N|_{C}
\otimes {\cal O}_{z}(-1)).
\label{3.32.5}
\end{equation}
Using a Leray spectral sequence, one can show that
\begin{equation}
H^{0}(z, \pi_{C*} N|_{C}
\otimes {\cal O}_{z}(-1)) = H^{0} (C,
N (-F)|_{C}),
\label{3.33.0}
\end{equation}
where we define
\begin{equation}
N (-F)= N \otimes
{\cal O}_{\pi^{*}z} (-F)
\label{3.34}
\end{equation}
and $F$ is the class of the elliptic fiber.
Therefore
\begin{equation}
h^{0}(z, \pi_{C*} N|_{C}
\otimes {\cal O}_{z}(-1)) = h^{0} (C,
N (-F)|_{C}).
\label{3.33}
\end{equation}
We conclude from~\eqref{3.32.5} and~\eqref{3.33} that one can determine
when
the bundle $V|_{z}(-1)$ has sections by studying
under what circumstances the line bundle $N(-F)$,
restricted to the Riemann surface $C$, does.
To accomplish this, we consider the short exact sequence
\begin{equation}
0 \rightarrow E \otimes {\cal O}_{\pi^{*}z} (-D)
\stackrel{f_{D}}{\rightarrow} E \stackrel{r}{\rightarrow}
E|_{D} \rightarrow 0,
\label{3.35}
\end{equation}
where $E$ is
any holomorphic vector bundle on $\pi^{*}z$ and $D$ is any effective
divisor
in $\pi^{*}z$. Henceforth, we will choose
\begin{equation}
E = N(-F)
\label{3.36}
\end{equation}
and
\begin{equation}
D = C.
\label{3.37}
\end{equation}
The short exact sequence~\eqref{3.35} can, therefore, be written as
\begin{equation}
0 \rightarrow N(-F- C)
\stackrel{f_{C}}{\rightarrow}
N(-F) \stackrel{r}{\rightarrow}
N(-F)|_{C} \rightarrow 0,
\label{3.38}
\end{equation}
where
\begin{equation}
N(-F-C)=N(-F) \otimes {\cal O}_{\pi^*z}(-C).
\label{3.38.new}
\end{equation}
The map $f_C$ in~\eqref{3.38} is given by
multiplication by a section of the line bundle
${\cal O}_{\pi^{*}z} (C)$ that vanishes precisely on $C$.
The mapping
$r$ is just restriction.
The cohomology long exact sequence corresponding to~\eqref{3.38}
is given by
\begin{eqnarray}
&&0 \rightarrow H^{0}(\pi^*z, N
(-F-C))
\rightarrow  H^{0}(\pi^*z,  N(-F)) \rightarrow
H^{0}(C, N (-F)|_{C})
\rightarrow \nonumber\\
&&H^{1}(\pi^*z, N
(-F-C))
\stackrel{f_C}{\rightarrow}
H^{1}(\pi^*z, N(-F))
\rightarrow
H^{1}(C, N(-F)|_{C}) \rightarrow \nonumber \\
&&H^{2}(\pi^*z, N(-F-C))
\rightarrow  H^{2}(\pi^*z,  N(-F))
\rightarrow 0.
\label{3.39}
\end{eqnarray}
First, we note from~\eqref{3.32.5} and~\eqref{3.33}
that the dimension of the cohomology group $H^{0}(C, N(-F)|_C)$
is precisely the quantity that we want to compute. Furthermore, since
the sequence~\eqref{3.39} is exact, the group $H^{0}(C, N(-F)|_C)$
is the kernel of the
map $f_C$. Mapping $f_C$ in
the long exact sequence is the
linear map from
$H^{1}(\pi^*z, N(-F-C))$ to $H^{1}(\pi^*z, N(-F))$
induced by multiplying by
a section of ${\cal O}_{\pi^{*}z}(C)$ that
vanishes at $C$. As it arises in a similar way to
$f_C$ in
the short exact sequence~\eqref{3.38}, we give this linear mapping the
same symbol.
Finally, note that the cohomology groups
$H^{1}(\pi^*z, N (-F-C))$ and
$H^{1}(\pi^*z, N(-F))$
are finite dimensional vector spaces and, hence,
the map $f_C$ can be represented by
a matrix.

We now arrive at a key point. The restriction of the spectral cover
$C$ represents the homology
class whose elements parametrize a projective space, as discussed above.
As we deform a spectral cover $C$ inside
its homology class, we actually change the map $f_C$
since we change the explicit section that vanishes at $C$.
This implies that the map $f_C$ depends on the
vector bundle moduli. More precisely, it depends not on all vector bundle
moduli but only on those that parametrize $C$.
As we discussed in Section 3,
the set of
parameters of  $C$ forms the projective space
\begin{equation}
{\mathbb P}H^{0}(\pi^{*}z, {\cal O}_{\pi^{*}z}(C)).
\label{3.39.5}
\end{equation}
Therefore, the map $f_C$ depends on the coordinates
of~\eqref{3.39.5}.

Before specifying the curve $z$ and the parameters of the line bundle
${\cal N}$,
let us make a general comment on the cohomology exact
sequence~\eqref{3.39}.
Suppose that
\begin{equation}
h^{0}(\pi^{*}z, N(-F))
>
h^{0}(\pi^{*}z, N (-F-C)).
\label{3.40}
\end{equation}
Then sequence~\eqref{3.39} implies that
\begin{equation}
h^{0}(C, N (-F)|_{C})
\label{3.41}
\end{equation}
can not be zero. Therefore, if condition~\eqref{3.40} is fulfilled the
line bundle $N(-F)|_{C}$ necessarily
has sections and the Pfaffian vanishes identically. We will, shortly,
give an example when this situation indeed happens. However, as will be
amply
demonstrated, in most cases inequality~\eqref{3.40} is not satisfied and
the
Pfaffian need not be zero.

The further analysis is too complicated in the general setting, so
we will
make some specifications. Curve $z$ will be chosen, as previously, to be
\begin{equation}
z={\cal S}
\label{3.41a}
\end{equation}
Then, ${\cal S}={\mathbb P}^{1}$
and, since ${\cal S} \cdot {\cal S}=-r$,
it is an isolated curve in ${\mathbb F}_{r}$ for $r=1, 2$. It follows that
for
$r=1,2$ the curve $\sigma \cdot \pi^{*}{\cal S}$ is ${\mathbb P}^{1}$ and
isolated in $X$.
Furthermore, we need to specify the line bundle
$N(-F)={\cal N}|_{\pi^* {\cal S}}(-F)$.
This involves the restriction of the global line
bundle ${\cal N}$ on $X$ given by equation~\eqref{eq:27} in Appendix B.
That is
\begin{equation}
{\cal N}= {\cal O}_{X}
(n(\frac{1}{2}+\lambda)\sigma+(\frac{1}{2}-\lambda)
\pi^{*}\eta+(\frac{1}{2}+n\lambda)\pi^{*}c_{1}(B)),
\label{3.43.0}
\end{equation}
where $c_1 (B)$ is the first Chern class of the base ${\mathbb F}_{r}$
\begin{equation}
c_1 ({\mathbb F}_{r}) = 2 {\cal S} +(r+2){\cal E}
\label{3.43.1}
\end{equation}
and $\eta$ is the curve on the base specifying the spectral cover.
Following the notation of the previous sections, we write
\begin{equation}
\eta = (a+1){\cal S} + b{\cal E}.
\label{3.43.2}
\end{equation}
The coefficients $a$ and $b$ are chosen in such a way that
the spectral cover
\begin{equation}
{\cal C}=n \sigma + \pi^* \eta
\label{3.43.3}
\end{equation}
satisfies effectiveness, irreducibility and
positivity conditions~\eqref{eq:18},
~\eqref{eq:19},~\eqref{eq:20} and~\eqref{eq:25a}. The restriction of the
spectral cover to $\pi^* {\cal S}$ was given in~\eqref{F} to be
\begin{equation}
C= n \sigma|_{\pi^* {\cal S}} +
(b-(a+1)r)F.
\label{3.43.4}
\end{equation}
Furthermore, the restriction of the line bundle ${\cal N}$ to
$\pi^* {\cal S}$ is of the form
\begin{equation}
N={\cal N}|_{\pi^{*} {\cal S}}= {\cal O}_{\pi^* {\cal S}}
((n(\frac{1}{2}+\lambda)\sigma+(\frac{1}{2}-\lambda)
\pi^{*}\eta+(\frac{1}{2}+n\lambda)\pi^{*}c_{1}(B)) \cdot \pi^{*}{\cal S}).
\label{3.43.5}
\end{equation}
Using equations~\eqref{3.43.5},~\eqref{3.43.1},~\eqref{3.43.2}
and the intersecton properties~\eqref{A2} in Appendix B, we obtain that
\begin{equation}
N(-F) ={\cal O}_{\pi^* {\cal S}}
(n (\lambda  +\frac{1}{2}) \sigma|_{\pi^{*}{\cal S}} +
((\frac{1}{2}-\lambda) (b-(a+1)r) + (\frac{1}{2} +n\lambda)(2-r) -1)F).
\label{3.43}
\end{equation}
We begin our analysis by giving an explicit example
of when ${\rm Pfaff}({\cal D}_{-})$
and, hence, the superpotential vanishes identically.

\begin{itemize}

\item Suppose that the
following conditions on the parameters are fulfilled
\begin{equation}
(\frac{1}{2}-\lambda)(b-(a+1)r) +(\frac{1}{2} +n\lambda)(2-r) -1 =0,
\quad \lambda+\frac{1}{2} \geq 1.
\label{3.45}
\end{equation}
Then it follows that the line bundles
$N(-F)$ and
$N(-F-C)$
are of the form
\begin{equation}
N(-F) =
{\cal O}_{\pi^{*}{\cal S}}(A \sigma|_{\pi^{*}{\cal S}}), \quad A >0
\label{3.46}
\end{equation}
and
\begin{equation}
N(-F-C)
= {\cal O}_{\pi^{*}{\cal S}}
(A^{\prime}\sigma|_{\pi^{*}{\cal S}} - B^{\prime} F), \quad A^{\prime}
\geq 0, B^{\prime} >0
\label{3.46a}
\end{equation}
respectively.
The condition $B^{\prime}>0$ is the consequence of the fact that we
work with positive spectral covers, which implies $b-ar>0$. Using a Leray
spectral sequence, it follows that
\begin{equation}
H^{0}(\pi^* {\cal S}, {\cal O}_{\pi^{*}{\cal S}}
(A^{\prime} \sigma|_{\pi^* {\cal S}} - B^{\prime} F))=
H^{0}({\cal S}, \pi_{*} {\cal O}_{\pi^{*}{\cal S}}
(A^{\prime} \sigma|_{\pi^* {\cal S}} - B^{\prime} F)).
\label{f1}
\end{equation}
Furthermore, equation~\eqref{B24} in Appendix C implies
\begin{equation}
H^{0}(\pi^* {\cal S}, {\cal O}_{\pi^{*}{\cal S}}
(A^{\prime} \sigma|_{\pi^* {\cal S}} - B^{\prime} F))=
H^{0}({\cal S}, {\cal O}_{{\cal S}} (-B^{\prime}) \oplus
\bigoplus_{i=2}^{A^{\prime}} (-B^{\prime} + i(r-2)).
\label{f2}
\end{equation}
Since $B^{\prime}>0$, we see from equation~\eqref{M} that the line
bundle ${\cal O}_{\pi^{*}{\cal S}}
(A^{\prime} \sigma|_{\pi^* {\cal S}} - B^{\prime} F)$
does not have any global holomorphic sections. On the other hand, the same
analysis, using equation~\eqref{B23} in Appendix C,
shows that the line bundle ${\cal O}_{\pi^{*}{\cal S}}
(A \sigma|_{\pi^* {\cal S}})$ has a unique section for $r=0, 1$ and $A$
sections for $r=2$.
Therefore, we have proven
that conditions~\eqref{3.40} are fulfilled whenever equations~\eqref{3.45}
are satisfied. This, in turn, implies that if equations~\eqref{3.45}
are satisfied, the Pfaffian and, therefore, the superpotential
induced by a superstring wrapped on a
holomorphic curve ${\cal S}$ vanish identically.

\end{itemize}

\noindent Parameters
satisfying~\eqref{3.45}, as well as the effectiveness, irreducibility and
positivity
conditions~\eqref{eq:18},~\eqref{eq:19},~\eqref{eq:20} and~\eqref{eq:25a},
are easily found. An explicit example is
\begin{equation}
r=1, \quad n=3, \quad b-a=4, \quad \lambda = \frac{3}{2}.
\label{3.47}
\end{equation}

We now turn to the more interesting situation when there is no obvious
reason for
the Pfaffian
to vanish identically. In fact, several different cases have to be
distinguished. They are
\begin{equation}
\lambda > \frac{1}{2}, \quad  \lambda < -\frac{1}{2}, \quad
\lambda =\frac{1}{2}, \quad \lambda =-\frac{1}{2}.
\label{3.48.0}
\end{equation}
\begin{itemize}

\item Let us first consider the case
\begin{equation}
\lambda > \frac{1}{2}.
\label{3.48}
\end{equation}
Here, of course, we assume that the first condition in~\eqref{3.45}
is not fulfilled since otherwise, as we have just shown, the Pfaffian
would vanish identically.
If condition~\eqref{3.48} is satisfied, one can show, using Leray
spectral
sequences and equations~\eqref{3.43.4} and~\eqref{3.43}, that the second
cohomology groups in~\eqref{3.39} can be related to the second
cohomology groups on the curve ${\cal S}$ and, hence, vanish identically.
That
is
\begin{equation}
H^{2}(\pi^{*}{\cal S}, N(-F -C)) =
H^{2} ({\cal S}, \pi_{*}N(-F -C))=0
\label{3.51}
\end{equation}
and
\begin{equation}
H^{2}(\pi^{*}{\cal S}, N(-F)) =
H^{2} ({\cal S}, \pi_{*}N(-F))=0.
\label{3.51a}
\end{equation}
Here, we have used the fact that
\begin{equation}
R^{1}\pi_* {\cal O}_{\pi^* {\cal S}}
(A \sigma|_{\pi^* {\cal S}} +fF)=0
\label{3.51.1}
\end{equation}
for $A>0$ and any $f$.
To simplify the sequence~\eqref{3.39}, let us assume that
the line bundle
$N(-F)$ will always be of the form
\begin{equation}
{\cal O}_{\pi^* {\cal S}}
(A \sigma|_{\pi^* {\cal S}}- BF), \quad A > 0, B>0.
\label{3.62.5}
\end{equation}
This condition will be satisfied in all the examples we will consider.
As explained above, such line bundles have no global holomorphic
sections and, hence
\begin{equation}
H^{0} (\pi^{*}{\cal S}, N(-F))=0.
\label{3.52}
\end{equation}
Then, the general structure of the exact sequence~\eqref{3.39}
simplifies to
\begin{eqnarray}
&& 0 \rightarrow
H^0 (C, N(-F)|_{C})
\rightarrow
H^{1}(\pi^{*}{\cal S}, N(-F -C))
\stackrel{f_{C}}{\rightarrow} \nonumber \\
&&H^{1}(\pi^{*}{\cal S}, N(-F))\rightarrow
H^1 (C, N(-F)|_{C})
\rightarrow 0.
\label{3.53}
\end{eqnarray}
Now, let us note the following fact.
From equations~\eqref{3.32.2}-\eqref{3.33} we know that
\begin{equation}
H^0 ({\cal S}, V|_{{\cal S}}(-1)) =
H^0 (C, N(-F)|_C)
\label{3.54.0}
\end{equation}
and therefore, that
\begin{equation}
h^0 ({\cal S}, V|_{{\cal S}}(-1)) =
h^0 (C, N(-F)|_C).
\label{3.54}
\end{equation}
On the other hand, using a Leray spectral sequence, it follows that
the first cohomology groups satisfy the same relation, that is
\begin{equation}
H^1 ({\cal S}, V|_{{\cal S}}(-1)) =
H^1 (C, N(-F)|_C)
\label{3.55.0}
\end{equation}
and, hence
\begin{equation}
h^1 ({\cal S}, V|_{{\cal S}}(-1)) =
h^1 (C, N(-F)|_C).
\label{3.55}
\end{equation}
The quantity
\begin{equation}
h^0 ({\cal S}, V|_{{\cal S}}(-1))-h^1 ({\cal S}, V|_{{\cal S}}(-1)),
\label{3.55.1}
\end{equation}
as follows from the Atiyah-Singer theorem and the fact
that ${\cal O}_{{\cal S}}(-1)$ is a spin bundle, is nothing but
the index of the Dirac operator ${\cal D}_{-}$ that has been shown
previously
to vanish. This allows us to conclude that
\begin{equation}
h^0 (C, N(-F)|_C)=
h^1 (C, N(-F)|_C).
\label{3.56}
\end{equation}
It then follows from sequence~\eqref{3.53} that the middle cohomology
groups must have the same dimension. That is,
\begin{equation}
h^{1}(\pi^{*}{\cal S}, N(-F -C))=
h^{1}(\pi^{*}{\cal S}, N(-F)).
\label{3.57}
\end{equation}
This says that $f_C$
in~\eqref{3.53} is a mapping between two linear spaces of the same
dimension
and, therefore, can be represented by a square matrix.
A non-trivial kernel of the matrix
$f_C$ is equivalent to the condition of
the determinant of $f_C$ vanish. We conclude, therefore, that
\begin{equation}
h^0 ({\cal S}, V|_{{\cal S}}(-1))>0
\label{3.57.5}
\end{equation}
if and only if
\begin{equation}
det f_{C}=0.
\label{3.58}
\end{equation}
As we have previously discussed, the matrix
$f_C$
depends on the vector bundle moduli associated with
the restriction of the spectral cover
${\cal C}$ to the surface $\pi^{*}{\cal S}$.
Therefore, $detf_C$
is a function of these moduli. Within
the zero locus of $detf_{C}$
\begin{equation}
h^0 ({\cal S}, V|_{{\cal S}}(-1))>0,
\label{3.60}
\end{equation}
the vector bundle $V|_{{\cal S}}$ is non-trivial and the Pfaffian
vanishes. However,
outside this zero locus
\begin{equation}
h^0 ({\cal S}, V|_{{\cal S}}(-1))=0
\label{3.59}
\end{equation}
and, as a result, the vector bundle $V|_{{\cal S}}$ is trivial
and the Pfaffian does not vanish.

\end{itemize}

To illustrate these concepts, we now consider the examples introduced in
the
previous section. Since
we have restricted $\lambda$ to satisfy~\eqref{3.48}, we need only
consider
Examples 2, 3 and 4 above.
We will continue to refer to them with the same
numbering. The $det f_C$
in these examples will be
explicity computed in the next section.\\

\noindent {\bf Example 2 \rm}: In this example, we take
\begin{equation}
r=1, \quad n=3, \quad a>5
\label{3.60a}
\end{equation}
and, hence, the lift of the curve ${\cal S}$ is the rational elliptic
surface $dP_{9}$. The bundle over it is specified
by
\begin{equation}
b-a=6, \quad \lambda=\frac{3}{2}.
\label{3.60b}
\end{equation}
Note that $\lambda$ satisfies~\eqref{3.48}. Then, from~\eqref{0.45}
\begin{equation}
C= 3 \sigma|_{dP_9} + 5F.
\label{3.62}
\end{equation}
Furthermore, it follows from~\eqref{3.43} that the line bundle
$N(-F)$ is given by
\begin{equation}
N(-F)=
{\cal O}_{dP_9}(6 \sigma|_{dP_9} -F).
\label{3.61}
\end{equation}
The line bundle $N(-F)$ is of the
form~\eqref{3.62.5}
and, therefore,
\begin{equation}
H^0 (dP_9, N(-F))=0.
\label{3.63}
\end{equation}
Since condition~\eqref{3.48} is satisfied, the second cohomology groups
in the exact sequence~\eqref{3.39} vanish.
Therefore,
sequence~\eqref{3.39} is
reduced to~\eqref{3.53} which, using
equations~\eqref{3.54.0},~\eqref{3.55.0}, ~\eqref{3.62}
and~\eqref{3.61}, can be written as
\begin{equation}
0 \rightarrow H^{0}({\cal S}, V|_{{\cal S}}(-1))
\rightarrow W_1
\stackrel{f_C}{\rightarrow} W_2
\rightarrow
H^{1}({\cal S}, V|_{{\cal S}}(-1)) \rightarrow 0,
\label{3.64}
\end{equation}
where
\begin{equation}
W_1 = H^{1}(dP_9, {\cal O}_{dP_9}(3 \sigma|_{dP_9} -6F))
\label{3.65}
\end{equation}
and
\begin{equation}
W_2 = H^{1}(dP_9, {\cal O}_{dP_9}(6 \sigma|_{dP_9} -F)).
\label{3.65a}
\end{equation}
The construction of the map $f_{C}$,
which will be given in the next section,
is based on the fact that one can relate cohomology groups on $dP_9$
to cohomology groups on ${\cal S}={\mathbb P}^1$.
This is achieved by a Leray spectral sequence which implies that
\begin{equation}
H^{1}(dP_9, {\cal O}_{dP_9}(3 \sigma|_{dP_9} -6F))=
H^{1} ({\cal S}, \pi_{*}{\cal O}_{dP_9}(3 \sigma|_{dP_9} -6F))
\label{3.66}
\end{equation}
and
\begin{equation}
H^{1}(dP_9, {\cal O}_{dP_9}(6 \sigma|_{dP_9} -F))=
H^{1} ({\cal S}, \pi_{*}{\cal O}_{dP_9}(6\sigma|_{dP_9} -F)).
\label{3.66a}
\end{equation}
From equations~\eqref{H},~\eqref{3.65} and~\eqref{3.66}, it follows that
\begin{equation}
W_1 \simeq H^{1} ({\cal S}, {\cal O}_{{\cal S}}(-6) \oplus
{\cal O}_{{\cal S}}(-8) \oplus {\cal O}_{{\cal S}}(-9))
\label{3.68}
\end{equation}
and
\begin{equation}
W_2 \simeq H^{1} ({\cal S}, {\cal O}_{{\cal S}}(-1) \oplus
\bigoplus_{i=3}^{7} {\cal O}_{{\cal S}}(-i)).
\label{3.68a}
\end{equation}
Furthermore, by using the Serre duality
\begin{equation}
H^{1} ({\mathbb P}^{1}, {\cal O}_{{\mathbb P}^{1}}(p))
\simeq H^{0} ({\mathbb P}^{1}, {\cal O}_{{\mathbb P}^{1}}(-2-p))^{*}
\label{3.69}
\end{equation}
we can rewrite $W_1$ and $W_2$ as
\begin{equation}
W_1 \simeq H^{0}({\cal S}, {\cal O}_{{\cal S}}(4) \oplus
{\cal O}_{{\cal S}}(6) \oplus {\cal O}_{{\cal S}}(7))^{*}
\label{3.70}
\end{equation}
and
\begin{equation}
W_2 \simeq H^{0} ({\cal S}, {\cal O}_{{\cal S}}(-1) \oplus
\bigoplus_{i=3}^{7} {\cal O}_{{\cal S}}(-2+i))^{*}.
\label{3.70a}
\end{equation}
From equations~\eqref{L} and~\eqref{M}, we conclude that
\begin{equation}
dim W_1 =dim W_2 =20.
\label{3.71}
\end{equation}
Therefore, the linear map $f_{C}$ in this example can be
represented by a $20 \times 20$ matrix.
This matrix depends on the
parameters of the spectral cover restricted to $dP_9$
given in~\eqref{3.62}. From equations~\eqref{0.15},~\eqref{0.46}
and~\eqref{0.47},
we know that the space of these parameters is the projective space
\begin{equation}
{\mathbb P}H^{0}(dP_9, {\cal O}_{dP_9}(3 \sigma|_{dP_9} +5F))
\simeq {\mathbb P}H^{0} ({\cal S}, {\cal O}_{{\cal S}}(5) \oplus
{\cal O}_{{\cal S}}(3) \oplus {\cal O}_{{\cal S}}(2)) \simeq
{\mathbb P}^{12}.
\label{i1}
\end{equation}
This means that the map $f_{C}$ is parametrized by
the coordinates of
the twelve-dimensional projective space~\eqref{i1}.\\

\noindent {\bf Example 3 \rm}: In this example, we take
\begin{equation}
r=2, \quad n=3, \quad a>5
\label{ff8}
\end{equation}
and, hence, the lift of the curve ${\cal S}$ is just the cross product
${\mathbb P}^1 \times T^2$.
The vector bundle on this surface is
specified by
\begin{equation}
b-2a =3, \quad \lambda =\frac{3}{2}.
\label{ff7}
\end{equation}
Note that $\lambda$ satisfies condition~\eqref{3.48}.
Then, from equation~\eqref{f6} it follows that
\begin{equation}
C=
3 \sigma|_{{\mathbb P}^{1}\times T^2} +F.
\label{ff9}
\end{equation}
Furthermore, it follows from~\eqref{3.43} that the line
bundle $N(-F)$ is given by
\begin{equation}
N(-F)= {\cal O}_{{\mathbb P}^{1}\times T^2}
(6\sigma|_{{\mathbb P}^{1}\times T^2} -2F).
\label{ff10}
\end{equation}
As in the previous example,
\begin{equation}
H^{0} ({\mathbb P}^{1}\times T^2, N(-F))=0
\label{ff11}
\end{equation}
and the second cohomology groups vanish
\begin{equation}
H^{2} ({\mathbb P}^{1}\times T^2,
N(-F -C))=
H^{2} ({\mathbb P}^{1}\times T^2, N(-F))=0.
\label{ff12}
\end{equation}
Thus, the corresponding cohomology exact sequence becomes
\begin{equation}
0 \rightarrow H^{0} ({\cal S}, V|_{{\cal S}}(-1)) \rightarrow
W_3
\stackrel{f_C}{\rightarrow}
W_4 \rightarrow H^{1} ({\cal S}, V|_{{\cal S}}(-1))\rightarrow 0,
\label{ff13}
\end{equation}
where
\begin{equation}
W_3 =H^{1} ({\mathbb P}^{1}\times T^2,
{\cal O}_{{\mathbb P}^{1}\times T^2}
(3 \sigma|_{{\mathbb P}^{1}\times T^2} -3F))
\label{ff14}
\end{equation}
and
\begin{equation}
W_4 =H^{1} ({\mathbb P}^{1}\times T^2,
{\cal O}_{{\mathbb P}^{1}\times T^2}
(6 \sigma|_{{\mathbb P}^{1}\times T^2} -2F)).
\label{ff14a}
\end{equation}
Using~\eqref{ref4} and Serre duality~\eqref{3.69}, these cohomology
groups
can be related to those on the curve ${\cal{S}}$ as
\begin{equation}
W_3 \simeq H^{1}({\cal S}, \bigoplus_{i=1}^{3}{\cal O}_{{\cal S}}(-3))
\simeq  H^{0}({\cal S}, \bigoplus_{1}^{3}{\cal O}_{{\cal S}}(1))^{*}
\label{ff15}
\end{equation}
and
\begin{equation}
W_4 \simeq H^{1}({\cal S}, \bigoplus_{i=1}^{6}{\cal O}_{{\cal S}}(-2))
\simeq  H^{0}({\cal S}, \bigoplus_{i=1}^{6}{\cal O}_{{\cal S}})^{*}.
\label{ff15a}
\end{equation}
It follows from~\eqref{L} and~\eqref{M} that
\begin{equation}
dim W_3 = dim W_4=6.
\label{ff17}
\end{equation}
Therefore, the linear map
$f_C$ in this example can be
represented
by a $6 \times 6$ matrix. The restriction of the spectral cover is of the
form~\eqref{ff9} and, as follows from
equations~\eqref{f4},~\eqref{f7} and~\eqref{f8}, its moduli
parametrize the projective space
\begin{equation}
{\mathbb P}H^{0}({\mathbb P}^1 \times T^2,
{\cal O}_{{\mathbb P}^{1} \times T^2}
(3 \sigma|_{{\mathbb P}^{1}\times T^2} +F))
\simeq {\mathbb P}H^{0}({\cal S}, \bigoplus_{i=1}^{3}{\cal O}_{{\cal
S}}(1))
\simeq {\mathbb P}^{5}.
\label{ff16}
\end{equation}
Hence, the map $f_C$ depends
on six homogeneous coordinates.\\

\noindent {\bf Example 4 \rm}: In this example, we take
\begin{equation}
r=2, \quad n=3, \quad a>5
\label{i2a}
\end{equation}
and, hence, the lift of the curve ${\cal S}$ is just the cross product
${\mathbb P}^1 \times T^2$.
The vector bundle on this surface is
specified by
\begin{equation}
b-2a =4, \quad \lambda =\frac{3}{2}.
\label{i2b}
\end{equation}
Note that $\lambda$ satisfies condition~\eqref{3.48}.
Then, from~\eqref{0.49aa}
\begin{equation}
C=3 \sigma|_{{\mathbb P}^{1} \times T^2}
+2F.
\label{n1}
\end{equation}
Furthermore, it follows from~\eqref{3.43} that the line
bundle $N(-F)$ is given by
\begin{equation}
N(-F)=
{\cal O}_{{\mathbb P}^{1}\times T^2}
(6\sigma|_{{\mathbb P}^{1}\times T^2} -3F).
\label{n2}
\end{equation}
As in the previous example,
\begin{equation}
H^{0} ({\mathbb P}^{1}\times T^2, N(-F))=0
\label{n3}
\end{equation}
and the second cohomology groups vanish
\begin{equation}
H^{2} ({\mathbb P}^{1}\times T^2, N(-F -C))=
H^{2} ({\mathbb P}^{1}\times T^2, N(-F))=0.
\label{n4}
\end{equation}
Thus, the corresponding cohomology exact sequence becomes
\begin{equation}
0 \rightarrow H^{0} ({\cal S}, V|_{{\cal S}}(-1)) \rightarrow
W_5
\stackrel{f_C}{\rightarrow}
W_6 \rightarrow H^{1} ({\cal S}, V|_{{\cal S}}(-1))\rightarrow 0,
\label{n5}
\end{equation}
where
\begin{equation}
W_5 =H^{1} ({\mathbb P}^{1}\times T^2,
{\cal O}_{{\mathbb P}^{1}\times T^2}
(3 \sigma|_{{\mathbb P}^{1}\times T^2} -5F))
\label{6}
\end{equation}
and
\begin{equation}
W_6 =H^{1} ({\mathbb P}^{1}\times T^2,
{\cal O}_{{\mathbb P}^{1}\times T^2}
(6 \sigma|_{{\mathbb P}^{1}\times T^2} -3F)).
\label{6a}
\end{equation}
Using~\eqref{ref4} and the Serre duality~\eqref{3.69}, these cohomology
groups
can be related to those on the curve ${\cal{S}}$ as
\begin{equation}
W_5 \simeq H^{1}({\cal S}, \bigoplus_{i=1}^{3}{\cal O}_{{\cal S}}(-5))
\simeq  H^{0}({\cal S}, \bigoplus_{1}^{3}{\cal O}_{{\cal S}}(3))^{*}
\label{n7}
\end{equation}
and
\begin{equation}
W_6 \simeq H^{1}({\cal S}, \bigoplus_{i=1}^{6}{\cal O}_{{\cal S}}(-3))
\simeq  H^{0}({\cal S}, \bigoplus_{i=1}^{6}{\cal O}_{{\cal S}}(1))^{*}.
\label{n7a}
\end{equation}
It follows from~\eqref{L} and~\eqref{M} that
\begin{equation}
dim W_5 = dim W_6=12.
\label{n8}
\end{equation}
Therefore, the linear map
$f_C$ in this example can be
represented
by a $12 \times 12$ matrix. The restriction of the spectral cover is of
the
form~\eqref{n1} and, as follows from
equations~\eqref{ref3},~\eqref{0.50aa} and~\eqref{0.51}, its moduli
parametrize the projective space
\begin{equation}
{\mathbb P}H^{0}({\mathbb P}^1 \times T^2,
{\cal O}_{{\mathbb P}^{1} \times T^2}(3 \sigma|_{{\mathbb P}^{1}\times
T^2} +2F))
\simeq {\mathbb P}H^{0}({\cal S}, \bigoplus_{i=1}^{3}{\cal O}_{{\cal
S}}(2))
\simeq {\mathbb P}^{8}.
\label{n9}
\end{equation}
Hence, the map $f_C$ depends
on nine homogeneous coordinates.

\begin{itemize}

\item We now turn to the second case in~\eqref{3.48.0}, where
\begin{equation}
\lambda < - \frac{1}{2}.
\label{lone1}
\end{equation}
In this case, the procedure presented above has to be slightly modified.
The reason is that
the coefficient in front of
$\sigma|_{\pi^* {\cal S}}$ in~\eqref{3.43} is now negative. We showed
in~\eqref{3.56} that
\begin{equation}
h^0 (C, N(-F)|_C)=
h^1 (C, N(-F)|_C).
\label{rrr4}
\end{equation}
On the other hand, by Serre duality
\begin{equation}
h^{1}(C, N(-F)|_C)=
h^0(C, (N(-F))^{*}|_C \otimes K_C),
\label{rrr5}
\end{equation}
where by $K_C$ we denote the canonical
bundle of $C$.
Thanks to equations~\eqref{rrr4} and~\eqref{rrr5}, instead of studying
the question of when
\begin{equation}
h^0 (C, N(-F)|_C)>0,
\label{rrr6}
\end{equation}
we can study the equivalent question of when
\begin{equation}
h^0(C, (N(-F))^{*}|_C \otimes K_C) >0.
\end{equation}
This observation is helpful for the following reason. By the adjunction
formula,
the canonical bundle $K_C$ can be related to
the
canonical bundle $K_{\pi^{*}{\cal S}}$ by
\begin{equation}
K_C=
(K_{\pi^{*}{\cal S}} \otimes {\cal O}_{\pi^{*}{\cal S}}
(C))|_C.
\label{rrr8}
\end{equation}
The canonical bundle of $\pi^{*}{\cal S}$ was shown
in \cite{BDO1} to be
\begin{equation}
K_{\pi^{*}{\cal S}}= {\cal O}_{\pi^{*}{\cal S}}(-rF).
\label{rrr9}
\end{equation}
Then, using~\eqref{3.43.4}, one finds that
\begin{equation}
K_C=
{\cal O}_{\pi^{*}{\cal S}}(n \sigma|_{\pi^{*}{\cal S}}
+ (b-(a+2)r)F)|_C.
\label{rrr10}
\end{equation}
If condition~\eqref{lone1} is satisfied, it is easy to see that
\begin{equation}
(N(-F))^{*}
\otimes K_{\pi^* {\cal S}}
\otimes {\cal O}_{\pi^* {\cal S}}(C) =
{\cal O}_{\pi^{*}{\cal S}}(A^{\prime \prime} \sigma|_{\pi^* {\cal S}}
+ \dots ),
\label{rrr11}
\end{equation}
where
\begin{equation}
A^{\prime \prime}=n(-\lambda + \frac{1}{2}) >0
\label{rrr12}
\end{equation}
and the ellipses stand for the fiber class term, which is irrelevant to
our
argument. The fact that $A''$ is positive allows us to proceed with our
discussion. Return to the short exact
sequence~\eqref{3.35}. Although we continue to choose $D=C$,
the vector bundle $E$ is now taken to be
\begin{equation}
E=(N(-F))^{*}
\otimes K_{\pi^{*}{\cal{S}}} \otimes {\cal O}_{\pi^* {\cal S}}(C).
\label{rrr13}
\end{equation}
Note that
\begin{equation}
E|_{C}=(N(-F))^{*}|_C \otimes K_C.
\label{rrr13.1}
\end{equation}
The procedure discussed above is now repeated using this different data.

\end{itemize}

Repeating this procedure
in the general setting requires that we introduce rather complicated
notation.
Here, therefore, it is
simplest to work within the context of an explicit example.
Let us consider
Example 1 discussed previously.\\

\noindent {\bf Example 1 \rm}: In this example, we take
\begin{equation}
r=1, \quad n=3, \quad a>5
\label{n9a}
\end{equation}
and, hence, the lift of the curve ${\cal{S}}$ is just $dP_{9}$. The vector
bundle on this surface is specified by
\begin{equation}
b-a=5, \quad \lambda = -\frac{5}{2}.
\label{n9b}
\end{equation}
Note that $\lambda$ satisfies~\eqref{lone1}. Then, from~\eqref{0.41}
\begin{equation}
C= 3 \sigma|_{dP_9} + 4F.
\label{3.73}
\end{equation}
Furthermore, it follows from~\eqref{3.43} that
the line bundle
$N(-F)$ is given by
\begin{equation}
N(-F)=
{\cal O}_{dP_9}(-6 \sigma|_{dP_9} +4F)
\label{3.72}
\end{equation}
The coefficient in front of $\sigma|_{dP_9}$ is negative, so we
introduce the ``Serre dual'' version of
$N(-F)|_C$
\begin{equation}
(N (-F))^{*}|_C
\otimes K_C.
\label{3.72.0}
\end{equation}
In the case under consideration
\begin{equation}
K_C=
{\cal O}_{dP_9}(3 \sigma|_{\pi^{*}{\cal S}}
+ 3F)|_C
\label{3.79}
\end{equation}
and, therefore,
\begin{equation}
(N (-F))^{*}|_C
\otimes K_C
= {\cal O}_{dP_9}(9 \sigma|_{dP_9} -F)|_C.
\label{3.80}
\end{equation}
We see then that an equivalent
condition for the vanishing of the Pfaffian
is given by
\begin{equation}
h^0 (C,
{\cal O}_{dP_9}(9 \sigma|_{dP_9} -F)|_C)>0.
\label{3.81}
\end{equation}
Since the coefficient in front of $\sigma|_{dP_9}$ in~\eqref{3.81}
is positive,
we can now apply the techniques developed previously in this section. The
new expression for $E$ given in~\eqref{rrr13} becomes, in this example,
\begin{equation}
E = {\cal O}_{dP_9}(9 \sigma|_{dP_9}-F).
\label{3.82}
\end{equation}
Then, the exact sequence~\eqref{3.35} is given by
\begin{equation}
0 \rightarrow {\cal O}_{dP_9}(6 \sigma|_{dP_9}-5F)
\stackrel{f_C}{\rightarrow}
{\cal O}_{dP_9}(9 \sigma|_{dP_9}-F) \rightarrow
{\cal O}_{dP_9}(9 \sigma|_{dP_9}-F)|_C
\rightarrow 0.
\label{3.83}
\end{equation}
The cohomology long exact sequence associated with~\eqref{3.83}
is of the form
\begin{eqnarray}
&&0 \rightarrow H^{0} (dP_9,
{\cal O}_{dP_9}(6 \sigma|_{dP_9}-5F)) \rightarrow
H^{0}(dP_9,
{\cal O}_{dP_9}(9 \sigma|_{dP_9}-F)) \rightarrow \nonumber \\
&&H^{0}(C,
{\cal O}_{dP_9}(9 \sigma|_{dP_9}-F)|_C) \rightarrow
H^{1} (dP_9,
{\cal O}_{dP_9}(6 \sigma|_{dP_9}-5F))
\stackrel{f_C}{\rightarrow} \nonumber \\
&&H^{1}(dP_9,
{\cal O}_{dP_9}(9 \sigma|_{dP_9}-F)) \rightarrow
H^{1}(C,
{\cal O}_{dP_9}(9 \sigma|_{dP_9}-F)|_C)
\rightarrow \nonumber \\
&& H^{2} (dP_9,
{\cal O}_{dP_9}(6 \sigma|_{dP_9}-5F)) \rightarrow
H^{2}(dP_9,
{\cal O}_{dP_9}(9 \sigma|_{dP_9}-F)) \rightarrow 0.
\label{3.84}
\end{eqnarray}
The line bundle ${\cal O}_{dP_9}(9 \sigma|_{dP_{9}} -F)$
is of the type~\eqref{3.62.5} and, therefore,
does not have sections. That is
\begin{equation}
H^{0} (dP_9, {\cal O}_{dP_9}(9 \sigma|_{dP_{9}} -F))=0.
\label{3.85}
\end{equation}
By using a Leray spectral sequence, it is easy to see that second
cohomology
groups in~\eqref{3.84} vanish
\begin{equation}
H^{2} (dP_9,
{\cal O}_{dP_9}(6 \sigma|_{dP_9}-5F)) =
H^{2}(dP_9,
{\cal O}_{dP_9}(9 \sigma|_{dP_9}-F))=0.
\label{3.86}
\end{equation}
As a result, sequence~\eqref{3.84} is reduced to
\begin{equation}
0 \rightarrow H^{0}(C,
{\cal O}_{dP_9}(9 \sigma|_{dP_9}-F)|_C) \rightarrow
W_7
\stackrel{f_C}{\rightarrow}
W_8 \rightarrow
H^{1}(C,
{\cal O}_{dP_9}(9 \sigma|_{dP_9}-F)|_C)
\rightarrow 0,
\label{3.87}
\end{equation}
where
\begin{equation}
W_7 = H^{1} (dP_9, {\cal O}_{dP_9}(6 \sigma|_{dP_9}-5F))
\label{3.88}
\end{equation}
and
\begin{equation}
W_8 = H^{1} (dP_9, {\cal O}_{dP_9}(9 \sigma|_{dP_9}-F)).
\label{3.88a}
\end{equation}
Using equation~\eqref{H} and a Leray spectral sequence, we find that
\begin{equation}
W_7 \simeq H^{1} ({\cal S}, {\cal O}_{\cal S}(-5) \oplus
\bigoplus_{i=7}^{11}{\cal O}_{\cal S}(-i))
\label{3.89}
\end{equation}
and
\begin{equation}
W_8 \simeq H^{1} ({\cal S}, {\cal O}_{\cal S}(-1) \oplus
\bigoplus_{i=3}^{10}{\cal O}_{\cal S}(-i)).
\label{3.89a}
\end{equation}
By Serre duality~\eqref{3.69}, we can also rewrite $W_7$ and $W_8$ in the
form
\begin{equation}
W_{7} \simeq H^{0}({\cal S}, {\cal O}_{\cal S}(3) \oplus
\bigoplus_{i=5}^{9}{\cal O}_{\cal S}(i))^{*}
\label{3.90}
\end{equation}
and
\begin{equation}
W_8 \simeq H^{0} ({\cal S}, {\cal O}_{\cal S}(-1) \oplus
\bigoplus_{i=1}^{8}{\cal O}_{\cal S}(i))^{*}.
\label{3.90a}
\end{equation}
It then follows from equations~\eqref{L},~\eqref{M} that
\begin{equation}
dim W_7= dim W_8=44.
\label{3.91}
\end{equation}
Therefore, the linear map $f_C$ in this example can be
represented by a $44 \times 44$ matrix. It follows from
equations~\eqref{0.13},~\eqref{0.42} and~\eqref{0.43} that
the restriction of the spectral cover to
$dP_9$ given in~\eqref{3.73} has moduli which parameterize the
projective space
\begin{equation}
{\mathbb P}H^{0}(dP_9, {\cal O}_{dP_9}(3 \sigma|_{dP_9} +4F))
\simeq {\mathbb P}H^{0} ({\cal S}, {\cal O}_{{\cal S}}(4) \oplus
{\cal O}_{{\cal S}}(2) \oplus {\cal O}_{{\cal S}}(1)) \simeq
{\mathbb P}^{9}.
\label{i2}
\end{equation}
Hence, the matrix $f_C$ is parametrized by the coordinates
of the nine-dimensional projective space~\eqref{i2}.

The remaining two cases in~\eqref{3.48.0},
\begin{equation}
\lambda =\frac{1}{2}
\label{no1}
\end{equation}
and
\begin{equation}
\lambda =-\frac{1}{2}
\label{no2}
\end{equation}
require a substantial modification of the procedure discussed above
and will not be considered in this paper.


\section{Computing $detf_C$:}


In this section, for the four examples involving isolated curve ${\cal
S}$
presented in the previous sections, we will show how to explicitly find
the matrix
$f_C$ whose kernel is $h^{0} ({\cal S}, V|_{{\cal
S}}(-1))$.
We then compute the determinant $detf_C$ and
give it a global interpretation. This will allow us to obtain an explicit
expression for the superpotential. The preliminary set-up
for constructing $f_C$ was given in the previous section.

Let us begin with \\

\noindent
{\bf Example 2 \rm}: In this example, we take
\begin{equation}
r=1, \quad n=3, \quad a>5
\label{russ1}
\end{equation}
and, hence, $\pi^{*}{\cal{S}}=dP_{9}$. The bundle over this surface is
specified by
\begin{equation}
b-a=6, \quad \lambda=\frac{3}{2}.
\label{russ2}
\end{equation}
Let us recall the basic set-up from the previous section. The sequence
that we are going to study was given in~\eqref{3.64} by
\begin{equation}
0 \rightarrow H^{0}({\cal S}, V|_{{\cal S}}(-1))
\rightarrow W_1
\stackrel{f_C}{\rightarrow} W_2
\rightarrow
H^{1}({\cal S}, V|_{{\cal S}}(-1)) \rightarrow 0,
\label{4.1}
\end{equation}
where
\begin{equation}
W_1 = H^{1}(dP_9, {\cal O}_{dP_9}(3 \sigma|_{dP_9} -6F))
\label{4.2}
\end{equation}
and
\begin{equation}
W_2 = H^{1}(dP_9, {\cal O}_{dP_9}(6 \sigma|_{dP_9} -F)).
\label{4.2a}
\end{equation}
The dimension of each of the linear spaces linear spaces $W_1$ and $W_2$
is $20$
and, hence, the map
$f_C$ can be represented by a $20 \times 20$ matrix.
This matrix depends on the
parameters of the spectral cover restricted to $dP_9$, that is, on
the homogeneous
coordinates of
\begin{equation}
{\mathbb P}H^{0} (dP_9, {\cal O}_{dP_9}(3 \sigma|_{dP_9} +5F)) \simeq
{\mathbb P}H^{0} ({\cal S}, {\cal O}_{{\cal S}}(5)
\oplus {\cal O}_{{\cal S}}(3) \oplus {\cal O}_{{\cal S}}(2))
\simeq {\mathbb P}^{12}.
\label{4.3}
\end{equation}
We will extensively use the relation of $W_1$ and $W_2$ to the cohomology
groups
on the curve ${\cal S}$
(equations~\eqref{3.68},~\eqref{3.68a},~\eqref{3.70} and~\eqref{3.70a}),
given by
\begin{equation}
W_1 \simeq H^{1} ({\cal S}, {\cal O}_{{\cal S}}(-6) \oplus
{\cal O}_{{\cal S}}(-8) \oplus {\cal O}_{{\cal S}}(-9))
\simeq
H^{0}({\cal S}, {\cal O}_{{\cal S}}(4) \oplus
{\cal O}_{{\cal S}}(6) \oplus {\cal O}_{{\cal S}}(7))^{*}
\label{4.4}
\end{equation}
and
\begin{equation}
W_2 \simeq H^{1} ({\cal S}, {\cal O}_{{\cal S}}(-1) \oplus
\bigoplus_{i=3}^{7} {\cal O}_{{\cal S}}(-i)) \simeq
H^{0} ({\cal S}, {\cal O}_{{\cal S}}(-1) \oplus
\bigoplus_{i=3}^{7} {\cal O}_{{\cal S}}(-2+i))^{*}.
\label{4.4a}
\end{equation}
The main idea in constructing the matrix $f_C$
is to express all data on $dP_9$ in terms of data on the curve ${\cal S}$
using the isomorphisms~\eqref{4.4} and~\eqref{4.4a}. Let us denote by
\begin{equation}
\tilde{w}_1 = B_{-6} \oplus  B_{-8} \oplus B_{-9}
\label{4.5}
\end{equation}
an arbitrary element of $H^{1}({\cal S}, {\cal O}_{{\cal S}}(-6)
\oplus {\cal O}_{{\cal S}}(-8) \oplus
{\cal O}_{{\cal S}}(-9))$,
where $B_{-i}, i=6, 8, 9,$ represents an arbitrary differential in
$H^{1}({\cal S}, {\cal O}_{{\cal S}}(-i))$. We see from the Serre duality
relation~\eqref{3.69} and~\eqref{L} that
\begin{equation}
h^{1}({\cal S}, {\cal O}_{{\cal S}}(-i)) =i-1.
\label{4.5.5}
\end{equation}
Let us now lift $\tilde{w}_{1}$
to an element $w_1$ in $H^{1}(dP_9, {\cal O}_{dP_9}
(3 \sigma|_{dP_9} -6F))$ and denote by $b_i=\pi^{*}B_i$ the lift of $B_i$.
Clearly, $w_1$ is given by a sum of $b_i$ multiplied by global sections
$s_1, s_2, s_3$ of some line bundles on $dP_9$. That is
\begin{equation}
w_1 = b_{-6} s_1 + b_{-8} s_2 + b_{-9} s_3.
\label{4.6}
\end{equation}
It is easy to see that
\begin{eqnarray}
&&s_1 \in H^{0}(dP_9, {\cal O}_{dP_9}(3 \sigma|_{dP_9})), \nonumber \\
&&s_2 \in H^{0}(dP_9, {\cal O}_{dP_9}(3 \sigma|_{dP_9}+2F)), \nonumber \\
&&s_3 \in H^{0}(dP_9, {\cal O}_{dP_9}(3 \sigma|_{dP_9}+3F)).
\label{4.7}
\end{eqnarray}
Note that the line bundle ${\cal O}_{dP_9}(3 \sigma|_{dP_9})$ has only
one section vanishing along $\sigma|_{dP_9}$ with multiplicity $3$.
This section is identical~\cite{BDO1,FMW1} to $z$ in the Weierstrass
equation
for $dP_9$
\begin{equation}
y^2 z = 4x^3 -g_2 xz^2 -g_3 z^3.
\label{4.8}
\end{equation}
That is,
\begin{equation}
s_{1}=z.
\label{germ1}
\end{equation}
Furthermore, it was shown in \cite{BDO1} that
\begin{equation}
x \in H^{0}(dP_9, {\cal O}_{dP_9}(3 \sigma|_{dP_9}+2F)), \quad
y \in H^{0}(dP_9, {\cal O}_{dP_9}(3 \sigma|_{dP_9}+3F))
\label{4.9}
\end{equation}
and
\begin{equation}
g_2 \in H^{0}(dP_9, {\cal O}_{dP_9}(4F)), \quad
g_3 \in H^{0}(dP_9, {\cal O}_{dP_9}(6F)).
\label{4.10}
\end{equation}
We see, therefore, that the sections $s_2$ and $s_3$ can be taken to be
\begin{equation}
s_2 =x, \quad s_3 =y.
\label{4.11}
\end{equation}
It follows that
\begin{equation}
w_1= b_{-6} z + b_{-8} x + b_{-9} y.
\label{4.12}
\end{equation}
Expression~\eqref{4.12} completely characterizes an element
$w_1 \in W_1$.

In a similar way, any element
$w_2 \in W_2= H^{1}(dP_9, {\cal O}_{dP_9} (6 \sigma|_{dP_9} -F))$
can be written
as
\begin{equation}
w_{2} = c_{-3} zx + c_{-4} zy + c_{-5} x^2 + c_{-6}xy + c_{-7} y^2
\label{4.13}
\end{equation}
where for $j=3, \cdots ,7$, $c_{-j}=\pi^{*}C_{-j}$ is an element of
$H^{1}(dP_9, {\cal O}_{dP_9} (-jF))$ and $C_{-j}$ is a section in the
$j-1$-dimensional space $H^{1}({\cal{S}}, {\cal{O}}_{{\cal{S}}}(-j))$.
The map $f_C$ is can also be expressed in terms of
data
on curve ${\cal S}$. By using~\eqref{4.3} and~\eqref{4.11},
$f_C$ can be written as
\begin{equation}
f_C = m_5 z + m_3 x + m_2 y,
\label{4.14}
\end{equation}
where $m_k=\pi^{*}M_{k}$, $k=2, 3, 5$, is an element in
$H^{0}(dP_9, {\cal O}_{dP_9} (kF))$ and $M_{k}$ is a section in the
$k+1$-dimensional space $H^{0}({\cal{S}},{\cal{O}}_{{\cal{S}}}(k))$.
Although
there are thirteen parameters in $m_{k}$, $k=2,3,5$, it must be remembered
that
they are homogeneous coordinates for the twelve dimensional projective
space
${\mathbb P}H^{0}(dP_9, {\cal O}_{dP_9} (3 \sigma|_{dP_9} +5F))$.

Putting this all together, we can completely specify the linear mapping
$W_1 \stackrel{f_C}{\rightarrow} W_2$.
First, note that with respect to fixed basis vectors of $W_{1}$ and
$W_{2}$,
the linear map $f_C$ is a $20 \times 20$ matrix. In
order
to find this matrix explicitly, we have to study its action on these basis
vectors. This action is generated through
multiplcation by a section $f_C$
of the form~\eqref{4.14}. Suppressing, for
the time being, the vector coefficients $b_{-i}$ and $c_{-j}$,
we see from~\eqref{4.6} that the linear space $W_1$ is spanned by the
basis vector blocks
\begin{equation}
z, \quad x, \quad y
\label{4.15}
\end{equation}
whereas it follows from~\eqref{4.13} that the linear space $W_2$ is
spanned by basis vector blocks
\begin{equation}
zx, \quad zy, \quad x^2, \quad xy, \quad y^2.
\label{4.16}
\end{equation}
The explicit matrix $M_{IJ}$ representing
$f_C$ is determined by multiplying the
basis vectors~\eqref{4.15} of $W_{1}$ by $f_C$
in~\eqref{4.14}. Expanding the
resulting vectors in $W_{2}$ in the
basis~\eqref{4.16} yields the matrix.  We find that $M_{IJ}$ is given by
\begin{equation}
\bordermatrix{    & z   & x   & y  \cr
              xz  & m_3 & m_5 & 0  \cr
              yz  & m_2 & 0   & m_5 \cr
              x^2 & 0   & m_3 & 0   \cr
              xy  & 0   & m_2 & m_3 \cr
              y^2 & 0   & 0   & m_2 \cr}.
\label{4.17}
\end{equation}
Of course, $M_{IJ}$ is a $20 \times 20$ matrix, so each of the elements
of~\eqref{4.17} represents
a $(j-1) \times (i-1)$ matrix for the corresponding $j=3, 4, 5, 6, 7$
and $i=6, 8, 9$.
Note, that if we multiply $m_5 z$ by a basis vector block $z$ we
get $m_5 z^2$ which is not in the space $W_2$. Therefore, it does not
represent a holomorphic differential and should be put to zero.
For example, let us compute $M_{11}$. This
corresponds to the $xz - z$ component of~\eqref{4.17} where
\begin{equation}
H^{1} (dP_9, {\cal O}_{dP_9} (3 \sigma|_{dP_9} -6F))|_{b_{-6}}
\stackrel{m_3}{\rightarrow}
H^{1} (dP_9, {\cal O}_{dP_9} (6 \sigma|_{dP_9} -F))|_{c_{-3}}.
\label{4.18}
\end{equation}
Note, from~\eqref{4.5.5}, that
\begin{equation}
h^{1} (dP_9, {\cal O}_{dP_9} (3 \sigma|_{dP_9} -6F))|_{b_{-6}} =5
\label{4.19}
\end{equation}
and
\begin{equation}
h^{1} (dP_9, {\cal O}_{dP_9} (6 \sigma|_{dP_9} -F))|_{c_{-3}} =2.
\label{4.20}
\end{equation}
An explicit matrix for $m_3$ is most easily obtained if we use
equations~\eqref{4.2},~\eqref{4.4} and ~\eqref{4.2a},~\eqref{4.4a} to
identify
\begin{equation}
H^{1} (dP_9, {\cal O}_{dP_9} (3 \sigma|_{dP_9} -6F))|_{b_{-6}} =
H^{0} ({\cal S}, {\cal O}_{{\cal S}} (4))^{*}
\label{4.21}
\end{equation}
and
\begin{equation}
H^{1} (dP_9, {\cal O}_{dP_9} (6 \sigma|_{dP_9} -F))|_{c_{-3}}=
H^{0} ({\cal S}, {\cal O}_{{\cal S}} (1))^{*}.
\label{4.22}
\end{equation}
If we define the two dimensional linear space
\begin{equation}
\hat{V}= H^{0}({\cal S}, {\cal O}(1)),
\label{4.23}
\end{equation}
then we see that
\begin{equation}
H^{1} (dP_9, {\cal O}_{dP_9} (3 \sigma|_{dP_9} -6F))|_{b_{-6}}=
Sym^{4}\hat{V}^{*}
\label{4.24}
\end{equation}
and
\begin{equation}
H^{1} (dP_9, {\cal O}_{dP_9} (6 \sigma|_{dP_9} -F))|_{c_{-3}}=
\hat{V}^{*},
\label{4.25}
\end{equation}
where by $Sym^{k}\hat{V}^{*}$ we denote the $k$-th symmetrized tensor
product
of the dual vector space $\hat{V}^{*}$ of $\hat{V}$.
Similarly, it follows from~\eqref{4.3} and~\eqref{4.23} that $m_3$
is an element in
\begin{equation}
H^{0}(dP_9, {\cal O}_{dP_9} (3 \sigma|_{dP_9} +5F))|_{m_3} =
Sym^{3}\hat{V}.
\label{4.26}
\end{equation}
Let us now introduce a basis
\begin{equation}
\{u, v\} \in \hat{V}
\label{4.27}
\end{equation}
and the dual basis
\begin{equation}
\{u^{*}, v^{*}\} \in \hat{V}^{*},
\label{4.28}
\end{equation}
where
\begin{equation}
u^{*} u =v^{*} v =1, \qquad u^{*}v =v^{*} u =0.
\label{4.29}
\end{equation}
Then the space $Sym^{k}\hat{V}^{*}$ is spanned by all possible homogeneous
polynomials in $u^*, v^{*}$ of degree $k$. Specifically,
\begin{equation}
\{u^{*4}, u^{*3}v^{*}, u^{*2}v^{*2}, u^{*}v^{*3}, v^{*4} \}\in
Sym^{4}\hat{V}^{*}
\label{4.30}
\end{equation}
is a basis of $Sym^{4}\hat{V}^{*}$ and
\begin{equation}
\{u^{3}, u^{2}v, uv^{2}, v^{3} \}\in Sym^{3}\hat{V}.
\label{4.31}
\end{equation}
is a basis of $Sym^{3}\hat{V}$.
Clearly, any section $m_3$ can be written in the basis~\eqref{4.31} as
\begin{equation}
m_3 = \phi_1 u^3 +\phi_2 u^2v + \phi_3 uv^2 + \phi_4 v^3,
\label{4.32}
\end{equation}
where $\phi_{a}, a=1, \cdots 4$ represent the associated moduli. Now, by
using
the multiplication rules~\eqref{4.29}, we find that the explicit $2 \times
5$
matrix representation of $m_3$ in this basis, that is, the
$M_{11}$ submatrix of $M_{IJ}$, is given by
\begin{equation}
\bordermatrix{    & u^{*4}  & u^{*3}v^{*} & u^{*2}v^{*2}  & u^{*}v^{*3} &
v^{*4} \cr
              u^* & \phi_1  & \phi_2      & \phi_3        & \phi_4      &
0      \cr
              v^* & 0       & \phi_1      & \phi_2        & \phi_3      &
\phi_4  \cr}.
\label{4.33}
\end{equation}
Continuing in this manner, we can fill out the complete $20 \times 20$
matrix $M_{IJ}$.
It is not particularly enlightening, so we will not present the matrix
$M_{IJ}$
in this paper. What is important is the determinant of $M_{IJ}$. Let us
parametrize the
sections $m_2$ and $m_5$ as
\begin{eqnarray}
&&m_2 = \chi_1 u^2 + \chi_2 uv + \chi_3 v^2 \nonumber \\
&&m_5 = \psi_1 u^5 + \psi_2 u^4v + \psi_3 u^3v^2 + \psi_4 u^2v^3
+ \psi_5 uv^4 + \psi_6 v^5
\label{4.34}
\end{eqnarray}
where $\chi_b, b=1, 2, 3$ and $\psi_c, c=1, \cdots 6$ represent the
associated
moduli. It is then straightforward to compute the determinant of
$M_{IJ}$ using the programs Maple or MATHEMATICA. We find that
it has two rather remarkable properties.
First, it factors in a simple way.
And second, it is independent of some of the variables.
Specifically, we find the following.

\begin{itemize}

\item The determinant of $f_{C}$ is given by
\begin{equation}
det f_C = det M_{IJ} = {\cal{P}}^{4},
\label{4.35}
\end{equation}
where
\begin{eqnarray}
&&{\cal{P}} =
\chi_1^2 \chi_3 \phi_3^2 -
\chi_1^2 \chi_2 \phi_3 \phi_4 -
2\chi_1 \chi_3^2  \phi_3 \phi_1 - \nonumber \\
&&\chi_1 \chi_2 \chi_3  \phi_3 \phi_2 +
\chi_2^2 \chi_3  \phi_1 \phi_3 +
\phi_4^2 \chi_1^3 -              \nonumber \\
&&2 \phi_2 \phi_4 \chi_3 \chi_1^2  +
\chi_1 \chi_3^2 \phi_2^2 +
3 \phi_1 \phi_4 \chi_1 \chi_2 \chi_3 + \nonumber \\
&&\phi_2 \chi_1 \phi_4 \chi_2^2 +
\phi_1^2 \chi_3^3 -
\phi_2 \chi_2 \phi_1 \chi_3^2-
\phi_4 \phi_1 \chi_2^3
\label{4.36}
\end{eqnarray}
is a homogeneous polynomial of degree 5 in
the seven projective parameters
$\phi_a$ and $\chi_b$. Note that none of the remaining six moduli $\psi_a$
appear in ${\cal{P}}$.

\end{itemize}

We conclude this example by pointing out that $detf_C$
will vanish precisely on the zero locus of the homogeneous
polynomial ${\cal P}$. Globally, ${\cal P}$ must be a holomorphic
section of some complex line bundle over
${\mathbb P}^{12}$. Therefore, there must exist a divisor
$D_{{\cal P}} \subset {\mathbb P}^{12}$ such that ${\cal P}$
is a section of
\begin{equation}
{\cal O}_{{\mathbb P}^{12}}(D_{{\cal P}})
\label{div1.1}
\end{equation}
and vanishes on the co-dimension one submanifold
$D_{{\cal P}}$ of ${\mathbb P}^{12}$. This categorizes the space of zeroes
of ${\cal P}$ and, hence, $detf_C$. The exact
eleven dimensional submanifold
\begin{equation}
D_{{\cal P}} \subset
{\mathbb P}H^{0} ({\cal S}, {\cal O}_{{\cal S}}(5) \oplus
{\cal O}_{{\cal S}}(3) \oplus {\cal O}_{{\cal S}}(2))
\label{div1.2}
\end{equation}
can be determined by solving the equation ${\cal P}=0$
using~\eqref{4.36}.

As a second example, let us consider\\

\noindent {\bf Example 1 \rm}: In this example, we take
\begin{equation}
r=1, \quad n=3, \quad a>5
\label{germ2}
\end{equation}
and, hence, $\pi^{*}{\cal{S}}=dP_{9}$.
The vector bundle over $dP_{9}$ is specified by
\begin{equation}
b-a=5, \quad \lambda=-\frac{5}{2}.
\label{germ3}
\end{equation}
According to the previous section, the cohomology
exact sequence that we need to study is
\begin{equation}
0 \rightarrow H^{0}({\cal S}, V|_{{\cal S}}(-1))
\rightarrow W_7
\stackrel{f_C}{\rightarrow} W_8
\rightarrow
H^{1}({\cal S}, V|_{{\cal S}}(-1)) \rightarrow 0,
\label{4.37}
\end{equation}
where
\begin{equation}
W_7 = H^{1}(dP_9, {\cal O}_{dP_9}(6 \sigma|_{dP_9} -5F))
\label{4.38}
\end{equation}
and
\begin{equation}
W_8 = H^{1}(dP_9, {\cal O}_{dP_9}(9 \sigma|_{dP_9} -F)).
\label{4.38a}
\end{equation}
The relation of $W_7$ and $W_8$ to the cohomology groups on the curve
${\cal
S}$
is the following
\begin{equation}
W_7 \simeq H^{1} ({\cal S}, {\cal O}_{{\cal S}}(-5) \oplus
\bigoplus_{i=7}^{11}{\cal O}_{{\cal S}}(-1))
\simeq
H^{0}({\cal S}, {\cal O}_{{\cal S}}(3) \oplus \bigoplus_{i=5}^{9}
{\cal O}_{{\cal S}}(i))^{*}
\label{4.39}
\end{equation}
and
\begin{equation}
W_8 \simeq H^{1} ({\cal S}, {\cal O}_{{\cal S}}(-1) \oplus
\bigoplus_{i=3}^{10} {\cal O}_{{\cal S}}(-i)) \simeq
H^{0} ({\cal S}, {\cal O}_{{\cal S}}(-1) \oplus
\bigoplus_{i=1}^{8} {\cal O}_{{\cal S}}(i))^{*}.
\label{4.39a}
\end{equation}
The dimension of each of the linear spaces $W_7$ and $W_8$ is $44$ and,
hence,
the map $f_C$ can be represented by a $44 \times 44$
matrix.
This matrix depends on the
parameters of the spectral cover restricted to $dP_9$, that is, on
the homogeneous
coordinates of
\begin{equation}
{\mathbb P}H^{0} (dP_9, {\cal O}_{dP_9}(3 \sigma|_{dP_9} +4F)) \simeq
{\mathbb P}H^{0} ({\cal S}, {\cal O}_{{\cal S}}(4)
\oplus {\cal O}_{{\cal S}}(2) {\cal O}_{{\cal S}}(1))
\simeq {\mathbb P}^{9}.
\label{4.40}
\end{equation}
An arbitrary element $\tilde{w}_{7} \in
H^{1} ({\cal S}, {\cal O}_{{\cal S}}(-5) \oplus
\bigoplus_{i=7}^{11}{\cal O}_{{\cal S}}(-1))$  can be written as
\begin{equation}
\tilde{w}_7 = B_{-5} \oplus  \bigoplus_{i=7}^{11} B_{-i},
\label{4.41}
\end{equation}
where $B_{-i}, i=5, 7, \cdots  11,$ denote an arbitrary element of
the $i-1$-dimensional space
$H^{1}({\cal S}, {\cal O}_{{\cal S}}(-i))$.
Let $b_i=\pi^{*}B_i$ represent the lift of $B_i$.
We can now write  an element
$w_7 \in H^{1}(dP_9, {\cal O}_{dP_9}(6 \sigma|_{dP_9} -5F))$
as a sum of $b_{i}$ multiplied by global
sections of appropriate line bundles on $dP_9$. These global sections can
be
chosen to be $z^2, xz, yz, x^2, xy, x^3z^{-1}$. We find that
\begin{equation}
w_7= b_{-5} z^2 + b_{-7} xz + b_{-8} yz + b_{-9} x^2 + b_{-10} xy +
b_{-11}x^3 z^{-1}.
\label{4.42}
\end{equation}
Note that a section $x^3 z^{-1}$ is holomorphic since $x$ vanishes along
$\sigma|_{dP_9}$ with multiplicity one \cite{FMW1}. This cancels the
pole
coming from $z^{-1}$. Using~\eqref{4.9}, it is easy to see that
that every term in~\eqref{4.42} is an element in
$H^{1}(dP_9, {\cal O}_{dP_9}(6 \sigma|_{dP_9} -5F))$. In a similar way,
any element $w_8 \in W_8$ can be expressed as
\begin{equation}
w_{8} = c_{-3} xz^2 + c_{-4} yz^2 + c_{-5} x^2z + c_{-6}xyz + c_{-7} x^3
+ c_{-8} x^2 y + c_{-9}x^4 z^{-1} c_{-10} x^3 y z^{-1},
\label{4.43}
\end{equation}
where for $j=3, \cdots 10$, $c_{-j}=\pi^{*}C_{-j}$ is an element of
$H^{1}(dP_9, {\cal O}_{dP_9} (-jF))$ and $C_{-j}$ is a section in the
$j-1$-dimensional space $H^{1}({\cal{S}}, {\cal{O}}_{{\cal{S}}}(-j))$.
We now need to express the map $f_C$ in terms of data
on the curve ${\cal S}$. By using~\eqref{4.40}, $f_C$
can be written as
\begin{equation}
f_C = m_4 z + m_2 x + m_1 y,
\label{4.45}
\end{equation}
where $m_k=\pi^{*}M_{k}$, $k=1, 2, 4$, is an element of
$H^{0}(dP_9, {\cal O}_{dP_9} (kF))$ and $M_{k}$ is a section in the
$k+1$-dimensional space $H^{0}({\cal{S}},{\cal{O}}_{{\cal{S}}}(k))$.
Although
there are ten parameters in $m_{k}$, $k=1, 2, 4$,
they are homogeneous coordinates for the nine dimensional projective
space
${\mathbb P}H^{0}(dP_9, {\cal O}_{dP_9} (3 \sigma|_{dP_9} +4F))$.

Putting this all together, we can completely specify the linear mapping
$W_{7}
\stackrel{f_C}{\rightarrow} W_{8}$. First, note that
with
respect to fixed basis vectors of $W_{7}$ and $W_{8}$, linear map
$f_C$ is a $44 \times 44$ matrix. To find this matrix
explicitly, we have to study its action on these basis vectors. This
action is
generated through multiplication by a section $f_C$ of
the
form~\eqref{4.45}. Suppressing, for the time being, the vector
coefficients
$b_{-i}$ and $c_{-i}$, we see from~\eqref{4.42} that
the linear space $W_7$
is spanned by the following basis vector blocks
\begin{equation}
z^2, \quad xz, \quad yz, \quad x^2, \quad xy, \quad x^3 z^{-1}
\label{4.46}
\end{equation}
whereas it follows from~\eqref{4.43} that the linear space $W_8$ is
spanned by basis vector blocks
\begin{equation}
xz^2, \quad yz^2, \quad x^2z, \quad xyz, \quad x^3, \quad x^2y, \quad
x^4z^{-1}, \quad x^3 y z^{-1}.
\label{4.47}
\end{equation}
The explicit matrix $M_{IJ}$ representing $f_C$ is
obtained by multiplying the basis vectors~\eqref{4.46} by
$f_C$ in~\eqref{4.45}. Expanding the resulting vectors
in
$W_{8}$ in the basis~\eqref{4.47} yields the matrix. We find that $M_{IJ}$
is
given by
\begin{equation}
\bordermatrix{           & z^2 & xz  & yz       & x^2 & xy       &
x^3z^{-1}\cr
              xz^2       & m_2 & m_4 & -m_1 g_2 & 0   & -m_1 g_3 & 0
\cr
              yz^2       & m_1 & 0   & m_4      & 0   & 0        & 0
\cr
              x^2z       & 0   & m_2 & 0        & m_4 & -m_1g_2  & 0   \cr
              xyz        & 0   & m_1 & m_2      & 0   & m_4      & 0   \cr
              x^3        & 0   & 0   & 4m_1     & m_2 & 0        & m_4
\cr
              x^2y       & 0   & 0   & 0        & m_1 & m_2      & 0
\cr
              x^4z^{-1}  & 0   & 0   & 0        & 0   & 4m_1     & m_2
\cr
              x^3yz^{-1} & 0   & 0   & 0        & 0   & 0        & m_1
\cr }.
\label{4.48}
\end{equation}
This matrix has a more complicated structure than the one in the previous
example and requires some explanation.
First, note that if we multiply $z^2$ by $m_4 z$ we obtain $m_4z^3$, which
is not in the space $W_8$. This implies that we must set this element to
zero.
Second, if we multiply $yz$ by $m_1 y$ we get $m_1 y^2 z$, which is also
not
among the basis vector blocks in~\eqref{4.47}. On the other hand, from
the Weierstass equation~\eqref{4.8}, it follows that
\begin{equation}
m_1 z y^2 = 4 m_1 x^3 -m_1 g_2 x z^2 - m_1 g_3 z^3.
\label{4.49}
\end{equation}
Therefore, we get non-vanishing coefficients in the basis
vector blocks $x^3, xz^2$ and
$z^3$. This is the reason for the appearance of the entries
$4m_1, -m_1g_2, -m_1 g_3$ in~\eqref{4.48}. Third, the matrix~\eqref{4.48}
depends not only on the vector bundle moduli but also on the
moduli of the underlying Calabi-Yau space
contained in $g_2$ and $g_3$. Now, $M_{IJ}$ is a $44 \times 44$ matrix so
every entry in~\eqref{4.48}
represents a matrix block of dimension $(j-1) \times (i-1)$ for the
corresponding $j=3, \dots, 10$ and $i=5,7, \dots ,11$. We will not
write the whole matrix but, rather, show how to explicitly compute two
blocks,
one with and one without Calabi-Yau moduli.
For example, let us compute $M_{11}$. This
corresponds to the $xz^2 - z^2$ component of~\eqref{4.48} where
\begin{equation}
H^{1} (dP_9, {\cal O}_{dP_9} (6 \sigma|_{dP_9} -5F))|_{b_{-5}}
\stackrel{m_2}{\rightarrow}
H^{1} (dP_9, {\cal O}_{dP_9} (9 \sigma|_{dP_9} -F))|_{c_{-3}}.
\label{4.50}
\end{equation}
Note from~\eqref{4.5.5} that
\begin{equation}
h^{1} (dP_9, {\cal O}_{dP_9} (6 \sigma|_{dP_9} -F))|_{b_{-5}} =4
\label{4.51}
\end{equation}
and
\begin{equation}
h^{1} (dP_9, {\cal O}_{dP_9} (9 \sigma|_{dP_9} -F))|_{c_{-3}} =2.
\label{4.52}
\end{equation}
We now relate the data on $dP_9$ to data on curve ${\cal S}$.
Using~\eqref{4.38}-\eqref{4.39a}, we can identify
\begin{equation}
H^{1} (dP_9, {\cal O}_{dP_9} (6 \sigma|_{dP_9} -F))|_{b_{-5}} =
H^{0} ({\cal S}, {\cal O}_{{\cal S}} (3))^{*}
\label{4.53}
\end{equation}
and
\begin{equation}
H^{1} (dP_9, {\cal O}_{dP_9} (9 \sigma|_{dP_9} -F))|_{c_{-3}}=
H^{0} ({\cal S}, {\cal O}_{{\cal S}} (1))^{*}.
\label{4.53.5}
\end{equation}
In terms of the two dimensional linear space~\eqref{4.23}, we see that
\begin{equation}
H^{1} (dP_9, {\cal O}_{dP_9} (6 \sigma|_{dP_9} -5F))|_{b_{-5}}=
Sym^{3}\hat{V}^{*}
\label{4.55}
\end{equation}
and
\begin{equation}
H^{1} (dP_9, {\cal O}_{dP_9} (9 \sigma|_{dP_9} -F))|_{c_{-3}}=
\hat{V}^{*}.
\label{4.56}
\end{equation}
Similarly, it follows from~\eqref{4.23} and~\eqref{4.40} that $m_2$
is an element in
\begin{equation}
H^{0}(dP_9, {\cal O}_{dP_9} (3 \sigma|_{dP_9} +4F))|_{m_2} =
Sym^{3}\hat{V}.
\label{4.57}
\end{equation}
Now introduce the basis' defined in~\eqref{4.27},~\eqref{4.28} and
~\eqref{4.29}.
Then
\begin{equation}
\{u^{*3}, u^{*2}v^{*}, u^{*}v^{*2}, v^{*3} \}\in
Sym^{3}\hat{V}^{*}
\label{4.61}
\end{equation}
is a basis for $Sym^{3}\hat{V}^{*}$ and
\begin{equation}
\{u^{2}, uv, v^{2} \}\in Sym^{2}\hat{V}.
\label{4.62}
\end{equation}
is a basis for $Sym^{2}\hat{V}$.
Furthermore, any section $m_2$ can be written as
\begin{equation}
m_2 = \lambda_1 u^2 +\lambda_2 uv + \lambda_3 v^2 ,
\label{4.63}
\end{equation}
where $\lambda_{a}, a=1, 2, 3,$ represent the associated moduli.
Using
the multiplication rules~\eqref{4.29}, we find that the explicit $2 \times
4$
matrix representation of $m_2$ in this basis, that is, the
$M_{11}$ submatrix of $M_{IJ}$, is given by
\begin{equation}
\bordermatrix{    & u^{*3}     & u^{*2}v^{*} & uv^{*2}   & v^{*2}   \cr
              u^* & \lambda_1  & \lambda_2   & \lambda_3 & 0       \cr
              v^* & 0           & \lambda_1  & \lambda_2 & \lambda_3
\cr}.
\label{4.64}
\end{equation}
Let us now compute a matrix block containing Calabi-Yau moduli, for
example, $M_{31}$.
This corresponds to the $xz^2 - yz$ component of~\eqref{4.48}, where
\begin{equation}
H^{1}(dP_9, {\cal O}_{dP_9}(6 \sigma|_{dP_9} - 5F))|_{b_{-8}}
\stackrel{-m_1 g_2}{\rightarrow}
H^{1}(dP_9, {\cal O}_{dP_9}(9 \sigma|_{dP_9} - F))|_{c_{-3}}.
\label{4.65}
\end{equation}
Note from~\eqref{4.5.5} that
\begin{equation}
h^{1}(dP_9, {\cal O}_{dP_9}(6 \sigma|_{dP_9} - 5F))|_{b_{-8}} =7
\label{4.65a}
\end{equation}
and
\begin{equation}
h^{1}(dP_{9}, {\cal{O}}_{dP_{9}}(9\sigma|_{dP_{9}}-F))|_{c_{-3}} =2.
\label{4.65b}
\end{equation}
Using equations~\eqref{4.38} and~\eqref{4.39}, we can identify
\begin{equation}
H^{1}(dP_9, {\cal O}_{dP_9}(6 \sigma|_{dP_9} - 5F))|_{b_{-8}}
= H^{0}({\cal S}, {\cal O}_{{\cal S}}(6))^{*} =Sym^{6}\hat{V}^{*}.
\label{4.66}
\end{equation}
The basis for $Sym^{6}V^{*}$
is given by
\begin{equation}
\{u^{*6}, u^{*5}v^{*}, u^{4*}v^{*2}, u^{*3}v^{*3}, u^{*2}v^{*4},
u^{*}v^{*5}, v^{*6} \}\in
Sym^{6}\hat{V}^{*}.
\label{4.67}
\end{equation}
A section $m_1$ is an element in
\begin{equation}
H^{0}(dP_9, {\cal O}_{dP_9}(3 \sigma|_{dP_9} +4F))|_{m_1}=
H^{0}({\cal S}, {\cal O}_{{\cal S}}(1)) = \hat{V}
\label{4.68}
\end{equation}
and can be written as
\begin{equation}
m_1= \rho_1 u + \rho_2 v,
\label{4.69}
\end{equation}
where $\rho_{b}, b=1, 2$, are the associated moduli.
Recall from~\eqref{4.10} that $g_2$ is a section of
${\cal O}_{dP_9}(4F)$. Since
\begin{equation}
H^{0}(dP_9, {\cal O}_{dP_9}(4F)) \simeq
H^{0}({\cal S}, {\cal O}_{{\cal S}}(4))
\label{4.70}
\end{equation}
we have
\begin{equation}
g_2 =h_1 u^4 +h_2 u^3 v +h_3 u^2 v^2 +h_4 uv^3 +h_5  v^4,
\label{4.71}
\end{equation}
where $h_{d}, d=1, \cdots , 5$ are the associated Calabi-Yau moduli.
Using the multiplication rules ~\eqref{4.29}, we find that the $2 \times
7$
matrix representation of $m_1g_2$ is given by
\begin{equation}
\bordermatrix{    & u^{*6}      & u^{*5}v^{*} & u^{*4}v^{*2} &
u^{*3}v^{*3} & u^{*2}v^{*4} & u^{*}v^{*5} & v^{*6}  \cr
              u^{*} & \rho_1 h_1 & \rho_1 h_2 + \rho_2 h_1 & \rho_1 h_3 +
\rho_2 h_2  & \rho_1 h_4 +\rho_2 h_3 & \rho_1 h_5 +\rho_2 h_4  & \rho_2h_5
& 0      \cr
              v^{*} & 0 & \rho_1 h_1 & \rho_1 h_2 +\rho_2 h_1 & \rho_1 h_3
+ \rho_2 h_2 & \rho_1 h_4 +\rho_2 h_3 & \rho_1 h_5 +\rho_2 h_4  &
\rho_2h_5  \cr}.
\label{4.72}
\end{equation}
Continuing in this manner, we can fill out the complete $44 \times 44$
matrix $M_{IJ}$.
Let us parametrize
the remaining
sections $m_4$ as
\begin{equation}
m_4 = \mu_1 u^4 + \mu_2 u^3v + \mu_3 u^2v^2 + \mu_4 u v^3
+ \mu_5 v^4 ,
\label{4.73}
\end{equation}
where $\mu_c, c=1, \cdots, 5$ represent the
associated moduli. Also parametrize
\begin{equation}
g_3 = t_1 u^6 + t_2 u^5v + t_3 u^4v^2 + t_4 u^3v^3
+ t_5 u^2 v^4 + t_6 u v^5 +t_7 v^6,
\label{4.74}
\end{equation}
where $t_{e}, e=1, \cdots, 7$ are the associated Calabi-Yau moduli.
It is then straightforward to compute the determinant of
$M_{IJ}$ using the programs Maple or MATHEMATICA.
The result is the following.

\begin{itemize}

\item The determinant of $f_{C}$ is
\begin{equation}
det f_C = det M_{IJ} = 2^{20}{\cal Q}^{11}\tilde{{\cal Q}},
\label{4.75}
\end{equation}
where
\begin{equation}
{\cal Q} = \rho_1 \rho_2 \lambda_2 - \lambda_1 \rho_2^2 - \lambda_3
\rho_1^2
\label{4.76}
\end{equation}
and
\begin{equation}
\tilde{{\cal Q}}=\tilde{{\cal Q}}(\lambda_a, \rho_b, \mu_c, h_d)
\label{4.77}
\end{equation}
is a polynomial of degree $11$ depending on all the vector bundle moduli
$\lambda_a, \rho_b, \mu_c$
and on the Calabi-Yau moduli $h_d$ but is independent of the Calabi-Yau
moduli
$t_e$. Polynomial ${\cal{Q}}$ has a very complicated form and, hence, we
present it in Appendix D.

\end{itemize}

\noindent Let us emphasize the main differences
between this example and the previous one.
\begin{itemize}

\item
In this example $det f_C$ contains two factors, one of
them
appearing with multiplicity one.
This is a very important difference since, later on, we will argue
that the vector bundle moduli superpotential precisely is equal to
$det f_C$.

\item
Unlike in the first example, the zero locus of
$det f_C$
depends not just on vector bundle moduli but also on some of the
moduli of the underlying Calabi-Yau threefold.
\end{itemize}

We conclude this example by pointing out that
$det f_C$ will vanish precisely on the union
of the zero loci of the homogeneous polynomials ${\cal Q}$
and $\tilde{\cal Q}$. Globally, ${\cal Q}\tilde{\cal Q}$ must be
a holomorphic section of some complex line bundle over
${\mathbb P}^{9}$. Therefore, there must exist a divisor
$D_{{\cal Q}\tilde{\cal Q}} \subset {\mathbb P}^{9}$ such that
${\cal Q}\tilde{\cal Q}$ is a section of
\begin{equation}
{\cal O}_{{\mathbb P}^9}(D_{{\cal Q}\tilde{\cal Q}})
\label{div2.1}
\end{equation}
and vanishes on the co-dimension one submanifold
$D_{{\cal Q}\tilde{\cal Q}}$ of ${\mathbb P}^{9}$. This categorizes
the space of zeroes of the product ${\cal Q}\tilde{\cal Q}$ and, hence,
$det f_C$. The exact eight-dimensional submanifold
\begin{equation}
D_{{\cal Q}\tilde{\cal Q}} \subset
{\mathbb P}H^{0} ({\cal S}, {\cal O}_{{\cal S}} (4) \oplus
{\cal O}_{{\cal S}} (2) \oplus
{\cal O}_{{\cal S}} (1))
\label{div2.2}
\end{equation}
can be determined by solving the equation
${\cal Q}\tilde{\cal Q}=0$ using~\eqref{4.76} and~\eqref{C1} from
Appendix D.

As a third example, consider\\

\noindent {\bf Example 3 \rm}: Here we take
\begin{equation}
r=2, \quad n=3, \quad a>5
\label{qqq1}
\end{equation}
and, hence, $\pi^{*}{\cal{S}}={\mathbb P}^{1} \times T^{2}$. The vector
bundle
on this surface is
specified by
\begin{equation}
b-2a =3, \quad \lambda=\frac{3}{2}.
\label{qqq2}
\end{equation}
Recall from the previous
section, that the cohomology exact sequence under consideration is
\begin{equation}
0 \rightarrow H^{0} ({\cal S}, V|_{{\cal S}}(-1)) \rightarrow
W_3
\stackrel{f_C}{\rightarrow}
W_4 \rightarrow H^{1} ({\cal S}, V|_{{\cal S}}(-1))\rightarrow 0,
\label{ff17.5}
\end{equation}
where
\begin{equation}
W_3 =H^{1} ({\mathbb P}^{1}\times T^2,
{\cal O}_{{\mathbb P}^{1}\times T^2}
(3 \sigma|_{{\mathbb P}^{1}\times T^2} -3F))
\label{ff18}
\end{equation}
and
\begin{equation}
W_4 =H^{1} ({\mathbb P}^{1}\times T^2,
{\cal O}_{{\mathbb P}^{1}\times T^2}
(6 \sigma|_{{\mathbb P}^{1}\times T^2} -2F)).
\label{ff18a}
\end{equation}
The relation of $W_3$ and $W_4$ to the cohomology groups on ${\cal S}$ is
the
following
\begin{equation}
W_3 \simeq H^{1}({\cal S}, \bigoplus_{i=1}^{3}{\cal O}_{{\cal S}}(-3))
\simeq  H^{0}({\cal S}, \bigoplus_{1}^{3}{\cal O}_{{\cal S}}(1))^{*}
\label{ff19}
\end{equation}
and
\begin{equation}
W_4 \simeq H^{1}({\cal S}, \bigoplus_{i=1}^{6}{\cal O}_{{\cal S}}(-2))
\simeq  H^{0}({\cal S}, \bigoplus_{i=1}^{6}{\cal O}_{{\cal S}})^{*}.
\label{ff19a}
\end{equation}
The dimension of each of the linear spaces $W_{3}$ and $W_{4}$ is 6 and,
hence, the linear map $f_C$ can
be
represented by a $6 \times 6$ matrix. This matrix
depends on the parameters of the spectral cover restricted to ${\mathbb
P}^{1}
\times T^{2}$, that is, on the six homogeneous coordinates of the
projective space
\begin{equation}
{\mathbb P}H^{0}({\mathbb P}^1 \times T^2,
{\cal O}_{{\mathbb P}^{1} \times T^2}(3 \sigma|_{{\mathbb P}^{1}\times
T^2} +F))
\simeq {\mathbb P}H^{0}({\cal S}, \bigoplus_{i=1}^{3}{\cal O}_{{\cal
S}}(1))
\simeq {\mathbb P}^{5}.
\label{ff21}
\end{equation}
An arbitrary element $\tilde{w}_3 \in
H^{1}({\cal S}, \bigoplus_{i=1}^{3}{\cal O}_{{\cal S}}(-3)) $
can be written as
\begin{equation}
\tilde{w}_3= B_{-3}^{(1)} \oplus B_{-3}^{(2)}
\oplus B_{-3}^{(3)},
\label{ff22}
\end{equation}
where $B_{-3}^{(i)}, i=1, 2, 3$, are elements of the two-dimensional
vector
space
$H^{1}({\cal S}, {\cal O}_{{\cal S}}(-3))$. Let
$b_{-3}^{(i)} =\pi^* B_{-3}^{(i)}$ be the lift of $B_{-3}^{(i)}$. We can
now
write any element
$w_3 \in
H^{1} ({\mathbb P}^{1}\times T^2,
{\cal O}_{{\mathbb P}^{1}\times T^2}
(3 \sigma|_{{\mathbb P}^{1}\times T^2} -3F))$ as
\begin{equation}
w_3 =b_{-3}^{(1)} z +b_{-3}^{(2)} x +b_{-3}^{(3)} y.
\label{ff23}
\end{equation}
In a similar way, any element $w_4 \in
H^{1} ({\mathbb P}^{1}\times T^2,
{\cal O}_{{\mathbb P}^{1}\times T^2}
(6 \sigma|_{{\mathbb P}^{1}\times T^2} -2F))$
can be expressed as
\begin{equation}
w_4 = c_{-2}^{(1)} z^2 +c_{-2}^{(2)} xz + c_{-2}^{(3)} yz +
c_{-2}^{(4)} x^2 + c_{-2}^{(5)} xy + c_{-2}^{(6)} y^2,
\label{ff24}
\end{equation}
where for $j=1, \dots, 6, c_{-2}^{(j)}=\pi^* C_{-2}^{(j)}$ are
elements of the one-dimensional vector space
$H^{1} ({\mathbb P}^{1}\times T^2,
{\cal O}_{{\mathbb P}^{1}\times T^2}(-2F))$
and $C_{-2}^{(j)}$ are elements of
$H^{1} ({\cal S},
{\cal O}_{{\cal S}}(-2))$. We now need to express the map
$f_C$ in terms of the data on the
curve ${\cal S}$. By using~\eqref{ff21},
$f_C$ can be written as
\begin{equation}
f_{C}=m_1 z+ m_2 x +m_3 y,
\label{ff25}
\end{equation}
where $m_k=\pi^* M_k, k=1, 2, 3,$ are elements of
$H^{0} ({\mathbb P}^1 \times T^2,
{\cal O}_{{\mathbb P}^1 \times T^2}(F))$
and $M_k$ are elements of the two-dimensional vector space
$H^{0}({\cal S}, {\cal O}_{{\cal S}}(1))$. Although there are six
parameters
$m_k$, they are homogeneous coordinates for the five-dimensional
projective
space ${\mathbb P}H^{0}({\mathbb P}^{1}\times T^2,
{\cal O}_{{\mathbb P}^1 \times T^2}
(3 \sigma|_{{\mathbb P}^1 \times T^2} + F))$.

Putting this all together, we can specify the linear mapping
$W_3 \stackrel{f_C}{\rightarrow}W_4$.
With respect to fixed basis vectors of $W_3$ and $W_4$, the linear
map $f_C$ is a $6 \times 6$ matrix.
To find this matrix, we have to study its action on these basis
vectors. The action is generated through multiplication by a section
$f_C$ of the form~\eqref{ff25}.
Suppressing, for the time being, the vector coefficients $b_{-3}^{(i)}$
and
$c_{-2}^{(i)}$, we see from~\eqref{ff23} that the linear space $W_3$
is spanned by the following basis vector blocks
\begin{equation}
z, \quad x, \quad y
\label{ff25.0}
\end{equation}
whereas it follows from~\eqref{ff24} that the linear space $W_4$ is
spanned
by basis vector blocks
\begin{equation}
z^2, \quad xz, \quad yz, \quad x^2, \quad xy, \quad y^2.
\label{ff25.1}
\end{equation}
The explicit matrix $M_{IJ}$ is obtained by multiplying the basis
vector blocks~\eqref{ff25.0} by
$f_C$ in~\eqref{ff25}.
Expanding the resulting vectors in $W_4$ in the basis~\eqref{ff25.1}
yields the matrix. We find that $M_{IJ}$ is given by
\begin{equation}
\bordermatrix{    & z   & x   & y  \cr
              z^2 & m_1 & 0   & 0  \cr
              xz  & m_2 & m_1 & 0  \cr
              yz  & m_3 & 0   & m_1 \cr
              x^2 & 0   & m_2 & 0   \cr
              xy  & 0   & m_3 & m_2 \cr
              y^2 & 0   & 0   & m_3 \cr}.
\label{ff25.2}
\end{equation}
The matrix $M_{IJ}$ is a $6 \times 6$ matrix, with every entry
in~\eqref{ff25.2} representing a matrix block of dimension
$1 \times 2$. Let us, for example, find the $z^2 - z$ component
of~\eqref{ff25.2} where
\begin{equation}
H^{1} ({\mathbb P}^1 \times T^2,
{\cal O}_{{\mathbb P}^1 \times T^2}
(3 \sigma|_{{\mathbb P}^1 \times T^2} -3F ))|_{b_{-3}^{(1)}}
\stackrel{m_1}{\rightarrow}
H^{1} ({\mathbb P}^1 \times T^2,
{\cal O}_{{\mathbb P}^1 \times T^2}
(6 \sigma|_{{\mathbb P}^1 \times T^2} -2F ))|_{c_{-1}^{(1)}}.
\label{ff25.3}
\end{equation}
Now, we relate data on ${\mathbb P}^1 \times T^2$ to data
on the curve ${\cal S}$. Using equations~\eqref{ff18}-\eqref{ff19a},
we can identify
\begin{equation}
H^{1}({\mathbb P}^1 \times T^2,
{\cal O}_{{\mathbb P}^1 \times T^2}
(3 \sigma|_{{\mathbb P}^1 \times T^2} -3F ))|_{b_{-3}^{(1)}} =
H^{0}({\cal S}, {\cal O}_{{\cal S}}(1))^{*}
\label{ff25.4}
\end{equation}
and
\begin{equation}
H^{1} ({\mathbb P}^1 \times T^2,
{\cal O}_{{\mathbb P}^1 \times T^2}
(6 \sigma|_{{\mathbb P}^1 \times T^2} -2F ))|_{c_{-2}^{(1)}} =
H^{0}({\cal S}, {\cal O}_{{\cal S}})^{*}.
\label{ff25.5}
\end{equation}
In terms of the two dimensional linear space~\eqref{4.23},
we see that
\begin{equation}
H^{1}({\mathbb P}^1 \times T^2,
{\cal O}_{{\mathbb P}^1 \times T^2}
(3 \sigma|_{{\mathbb P}^1 \times T^2} -3F ))|_{b_{-3}^{(1)}} =
\hat{V}^{*}
\label{ff25.7}
\end{equation}
and
\begin{equation}
H^{1} ({\mathbb P}^1 \times T^2,
{\cal O}_{{\mathbb P}^1 \times T^2}
(6 \sigma|_{{\mathbb P}^1 \times T^2} -2F ))|_{c_{-2}^{(1)}} ={\mathbb C}.
\label{ff25.8}
\end{equation}
Similarly, it follows from~\eqref{ff21} and~\eqref{4.23} that $m_1$ is
an
element of
\begin{equation}
H^{0} ({\mathbb P}^1 \times T^2,
{\cal O}_{{\mathbb P}^1 \times T^2}
(3 \sigma|_{{\mathbb P}^1 \times T^2} +F ))|_{m_1} =
\hat{V}.
\label{ff25.9}
\end{equation}
Now consider the basis' defined in~\eqref{4.27},~\eqref{4.28}
and~\eqref{4.29}.
Then, any section $m_1$ can be written as
\begin{equation}
m_1 = \xi_1 u +\xi_2 v,
\label{ff25.15}
\end{equation}
where $\xi_a, a=1, 2,$ represent the associated moduli. Using the
multiplication rules~\eqref{4.29}, we find the explicit $1 \times 2$
matrix representation of $m_1$ in this basis, that is, the $M_{11}$
submatrix of $M_{IJ}$ is given by
\begin{equation}
\bordermatrix{    & u^{*}   & v^{*}    \cr
              1 & \xi_1  & \xi_2        \cr}.
\label{ff25.16}
\end{equation}
Continuing in this way, we can construct the entire matrix $M_{IJ}$. The
answer
is
\begin{equation}
M_{IJ}=\bordermatrix{      & {\ } &{\ }& {\ }& {\ }& {\ } &{\ } \cr
 {\ } & \xi_1 & \xi_2 & 0 & 0 & 0 & 0 \cr
 {\ } & \tau_1  & \tau_2   & \xi_1 & \xi_2 & 0 & 0 \cr
 {\ } & \zeta_1 & \zeta_2 & 0 & 0 & \xi_1 & \xi_2 \cr
 {\ } & 0 & 0 & \tau_1 & \tau_2 & 0 & 0 & \cr
 {\ } & 0 & 0 & \zeta_1 & \zeta_2 & \tau_1 & \tau_2  \cr
 {\ } & 0 & 0 & 0 & 0 & \zeta_1 & \zeta_2 \cr},
\label{ff26}
\end{equation}
where we have used the following parametrization of $m_2$ and $m_3$
in terms of the moduli $\tau_b, b=1,2$, and $\zeta_c, c=1, 2$
\begin{eqnarray}
&&m_2 = \tau_1 u + \tau_2 v,
\nonumber \\
&&m_3 = \zeta_1 u + \zeta_2 v.
\label{ff27}
\end{eqnarray}
Surprisingly, the determinant of the
matrix~\eqref{ff26} turns out to be identically zero for all values of the
moduli. We conclude the following.

\begin{itemize}

\item The determinant of $f_{C}$ is given by
\begin{equation}
det f_C= 0
\label{ff28}
\end{equation}
for all values of the moduli.

\end{itemize}

\noindent Therefore, in this example, unlike in the
previous two cases, $det f_C$ vanishes identically.
Note that the data~\eqref{qqq1} and~\eqref{qqq2} does not satisfy the
first
condition in~\eqref{3.45}. Therefore, in the present example, the
conditions~\eqref{3.45} are not responsible for the vanishing of
$detf_{C}$.

As a fourth example, consider\\

\noindent {\bf Example 4 \rm}: Here we take
\begin{equation}
r=2, \quad n=3, \quad a>5
\label{fra1}
\end{equation}
and, hence, $\pi^{*}{\cal{S}}={\mathbb P}^{1} \times T^{2}$. The vector
bundle
on this surface is
specified by
\begin{equation}
b-2a =4, \quad \lambda=\frac{3}{2}.
\label{fra1.1}
\end{equation}

Recall from the previous
section that the cohomology exact sequence under consideration is
\begin{equation}
0 \rightarrow H^{0} ({\cal S}, V|_{{\cal S}}(-1)) \rightarrow
W_5
\stackrel{f_C}{\rightarrow}
W_6 \rightarrow H^{1} ({\cal S}, V|_{{\cal S}}(-1))\rightarrow 0,
\label{n10}
\end{equation}
where
\begin{equation}
W_5 =H^{1} ({\mathbb P}^{1}\times T^2,
{\cal O}_{{\mathbb P}^{1}\times T^2}
(3 \sigma|_{{\mathbb P}^{1}\times T^2} -5F))
\label{n11}
\end{equation}
and
\begin{equation}
W_6 =H^{1} ({\mathbb P}^{1}\times T^2,
{\cal O}_{{\mathbb P}^{1}\times T^2}
(6 \sigma|_{{\mathbb P}^{1}\times T^2} -3F)).
\label{n11a}
\end{equation}
The relation of $W_5$ and $W_6$ to the cohomology groups on ${\cal S}$ is
the
following
\begin{equation}
W_5 \simeq H^{1}({\cal S}, \bigoplus_{i=1}^{3}{\cal O}_{{\cal S}}(-5))
\simeq  H^{0}({\cal S}, \bigoplus_{1}^{3}{\cal O}_{{\cal S}}(3))^{*}
\label{n12}
\end{equation}
and
\begin{equation}
W_6 \simeq H^{1}({\cal S}, \bigoplus_{i=1}^{6}{\cal O}_{{\cal S}}(-3))
\simeq  H^{0}({\cal S}, \bigoplus_{i=1}^{6}{\cal O}_{{\cal S}}(1))^{*}.
\label{n12a}
\end{equation}
The dimension of each of the linear spaces $W_{5}$ and $W_{6}$ is 12 and,
hence, the linear map $f_{C}$ can
be
represented by a $12 \times 12$ matrix. This matrix
depends on the parameters of the spectral cover restricted to ${\mathbb
P}^{1}
\times T^{2}$, that is, on the nine homogeneous coordinates of the
projective space
\begin{equation}
{\mathbb P}H^{0}({\mathbb P}^1 \times T^2,
{\cal O}_{{\mathbb P}^{1}T^2}(3 \sigma|_{{\mathbb P}^{1}\times T^2} +2F))
\simeq {\mathbb P}H^{0}({\cal S}, \bigoplus_{i=1}^{3}{\cal O}_{{\cal
S}}(2))
\simeq {\mathbb P}^{8}.
\label{n13}
\end{equation}
An arbitrary element $\tilde{w}_5 \in
H^{1}({\cal S}, \bigoplus_{i=1}^{3}{\cal O}_{{\cal S}}(-5)) $
can be written as
\begin{equation}
\tilde{w}_5= B_{-5}^{(1)} \oplus B_{-5}^{(2)}
\oplus B_{-5}^{(3)},
\label{n14}
\end{equation}
where $B_{-5}^{(i)}, i=1, 2, 3$, are elements of the four-dimensional
vector space
$H^{1}({\cal S}, {\cal O}_{{\cal S}}(-5))$. Let
$b_{-5}^{(i)} =\pi^* B_{-5}^{(i)}$ be the lift of $B_{-5}^{(i)}$. We can
now
write any element
$w_5 \in
H^{1} ({\mathbb P}^{1}\times T^2,
{\cal O}_{{\mathbb P}^{1}\times T^2}
(3 \sigma|_{{\mathbb P}^{1}\times T^2} -5F))$ as
\begin{equation}
w_5 =b_{-5}^{(1)} z +b_{-5}^{(2)} x +b_{-5}^{(3)} y.
\label{n15}
\end{equation}
In a similar way, any element $w_6 \in
H^{1} ({\mathbb P}^{1}\times T^2,
{\cal O}_{{\mathbb P}^{1}\times T^2}
(6 \sigma|_{{\mathbb P}^{1}\times T^2} -3F))$
can be expressed as
\begin{equation}
w_6 = c_{-3}^{(1)} z^2 +c_{-3}^{(2)} xz + c_{-3}^{(3)} yz +
c_{-3}^{(4)} x^2 + c_{-3}^{(5)} xy + c_{-3}^{(6)} y^2,
\label{n16}
\end{equation}
where for $j=1, \dots, 6, c_{-3}^{(j)}=\pi^* C_{-3}^{(j)}$ are
elements of the two-dimensional vector space
$H^{1} ({\mathbb P}^{1}\times T^2,
{\cal O}_{{\mathbb P}^{1}\times T^2}(-3F))$
and $C_{-3}^{(j)}$ are elements of
$H^{1} ({\cal S},
{\cal O}_{{\cal S}}(-3))$. We now need to express the map
$f_C$ in terms of the data on the
curve ${\cal S}$. By using~\eqref{n13},
$f_C$ can be written as
\begin{equation}
f_C=m_1 z+ m_2 x +m_3 y,
\label{n17}
\end{equation}
where $m_k=\pi^* M_k, k=1, 2, 3,$ are elements of
$H^{0} ({\mathbb P}^1 \times T^2,
{\cal O}_{{\mathbb P}^1 \times T^2}(2F))$
and $M_k$ are elements of the three-dimensional vector space
$H^{0}({\cal S}, {\cal O}_{{\cal S}}(2))$. Although there are nine
parameters
$m_k$, they are homogeneous coordinates for the eight-dimensional
projective
space ${\mathbb P}H^{0}({\mathbb P}^{1}\times T^2,
{\cal O}_{{\mathbb P}^1 \times T^2}
(3 \sigma|_{{\mathbb P}^1 \times T^2} + 2F))$.

Putting this all together, we can specify the linear mapping
$W_5 \stackrel{f_C}{\rightarrow}W_6$.
With respect to fixed basis vectors of $W_5$ and $W_6$ the linear
map $f_C$ is a $12 \times 12$ matrix.
To find this matrix, we have to study its action on these basis
vectors. The action is generated through multiplication by a section
$f_C$ of the form~\eqref{n17}.
Suppressing, for the time being, the vector coefficients $b_{-5}^{(i)}$
and
$c_{-3}^{(i)}$ we see from~\eqref{n15} that the linear space $W_5$
is spanned by the following basis vector blocks
\begin{equation}
z, \quad x, \quad y
\label{n17.0}
\end{equation}
whereas it follows from~\eqref{n16} that the linear space $W_6$ is spanned
by basis vector blocks
\begin{equation}
z^2, \quad xz, \quad yz, \quad x^2, \quad xy, \quad y^2.
\label{n17.1}
\end{equation}
The explicit matrix $M_{IJ}$ is obtained by multiplying the basis
vector blocks~\eqref{n17.0} by
$f_C$ in~\eqref{n17}.
Expanding the resulting vectors in $W_6$ in the basis~\eqref{n17.1}
yields the matrix. We find that $M_{IJ}$ is given by
\begin{equation}
\bordermatrix{    & z   & x   & y  \cr
              z^2 & m_1 & 0   & 0  \cr
              xz  & m_2 & m_1 & 0  \cr
              yz  & m_3 & 0   & m_1 \cr
              x^2 & 0   & m_2 & 0   \cr
              xy  & 0   & m_3 & m_2 \cr
              y^2 & 0   & 0   & m_3 \cr}.
\label{n17.2}
\end{equation}
The matrix $M_{IJ}$ is a $12 \times 12$ matrix with every entry
in~\eqref{n17.2} representing a matrix block of dimension
$2 \times 4$. Let us, for example, find the $z^2 - z$ component
of~\eqref{n17.2} where
\begin{equation}
H^{1} ({\mathbb P}^1 \times T^2,
{\cal O}_{{\mathbb P}^1 \times T^2}
(3 \sigma|_{{\mathbb P}^1 \times T^2} -5F ))|_{b_{-5}^{(1)}}
\stackrel{m_1}{\rightarrow}
H^{1} ({\mathbb P}^1 \times T^2,
{\cal O}_{{\mathbb P}^1 \times T^2}
(6 \sigma|_{{\mathbb P}^1 \times T^2} -3F ))|_{c_{-3}^{(1)}}.
\label{n17.3}
\end{equation}
Now we relate data on ${\mathbb P}^1 \times T^2$ to data
on curve ${\cal S}$. Using equations~\eqref{n11}-\eqref{n12a},
we can identify
\begin{equation}
H^{1}({\mathbb P}^1 \times T^2,
{\cal O}_{{\mathbb P}^1 \times T^2}
(3 \sigma|_{{\mathbb P}^1 \times T^2} -5F ))|_{b_{-5}^{(1)}} =
H^{0}({\cal S}, {\cal O}_{{\cal S}}(3))^{*}
\label{n17.4}
\end{equation}
and
\begin{equation}
H^{1} ({\mathbb P}^1 \times T^2,
{\cal O}_{{\mathbb P}^1 \times T^2}
(6 \sigma|_{{\mathbb P}^1 \times T^2} -3F ))|_{c_{-3}^{(1)}} =
H^{0}({\cal S}, {\cal O}_{{\cal S}}(1))^{*}.
\label{n17.5}
\end{equation}
Defining the two dimensional linear space~\eqref{4.23},
we see that
\begin{equation}
H^{1}({\mathbb P}^1 \times T^2,
{\cal O}_{{\mathbb P}^1 \times T^2}
(3 \sigma|_{{\mathbb P}^1 \times T^2} -5F ))|_{b_{-5}^{(1)}} =
Sym^3 \hat{V}^{*}
\label{n17.7}
\end{equation}
and
\begin{equation}
H^{1} ({\mathbb P}^1 \times T^2,
{\cal O}_{{\mathbb P}^1 \times T^2}
(6 \sigma|_{{\mathbb P}^1 \times T^2} -3F ))|_{c_{-3}^{(1)}} =
\hat{V}^{*}.
\label{n17.8}
\end{equation}
Similarly, it follows from~\eqref{n13},~\eqref{4.23} that $m_1$ is an
element of
\begin{equation}
H^{0} ({\mathbb P}^1 \times T^2,
{\cal O}_{{\mathbb P}^1 \times T^2}
(3 \sigma|_{{\mathbb P}^1 \times T^2} +2F ))|_{m_1} =
Sym^2 \hat{V}.
\label{n17.9}
\end{equation}
Now consider the basis' defined in~\eqref{4.27},~\eqref{4.28}
and~\eqref{4.29}.
Then
\begin{equation}
\{ u^{*3}, u^{*2}v^*, u^{*}v^{*2}, v^{*3} \} \in Sym^3 \hat{V}^{*}
\label{n17.13}
\end{equation}
is a basis for $Sym^3 \hat{V}^{*}$ and
\begin{equation}
\{ u^{2}, u v, v^{2} \} \in Sym^2 \hat{V}.
\label{n17.14}
\end{equation}
Furthermore, any section $m_1$ can be written as
\begin{equation}
m_1 = \alpha_1 u^2 +\alpha_2 uv +\alpha_3 v^2 ,
\label{n17.15}
\end{equation}
where $\alpha_a, a=1, 2, 3,$ represent the associated moduli. Using the
multiplication rules~\eqref{4.29}, we can find the explicit $2 \times 4
$
matrix representation of $m_1$ in this basis, that is, the $M_{11}$
submatrix of $M_{IJ}$. It is given by
\begin{equation}
\bordermatrix{    & u^{*3}     & u^{*2}v^{*} & u^*v^{*2}   & v^{*3}   \cr
              u^* & \alpha_1  & \alpha_2   & \alpha_3 & 0       \cr
              v^* & 0           & \alpha_1  & \alpha_2 & \alpha_3
\cr}.
\label{n17.16}
\end{equation}
Continuing in this way, we can construct the entire matrix $M_{IJ}$. The
answer
is
\begin{equation}
M_{IJ}=\bordermatrix{      & {\ } &{\ }& {\ }& {\ }& {\ } &{\ } &{\ } & {\
} & {\
} & {\ } & {\ } &{\ }\cr
 {\ } & \alpha_1 & \alpha_2 & \alpha_3 & 0 & 0 & 0 & 0 & 0 & 0 & 0 & 0 & 0
\cr
 {\ } & 0 & \alpha_1 & \alpha_2 & \alpha_3 & 0 & 0 & 0 & 0 & 0 & 0 & 0 & 0
\cr
 {\ } & \beta_1 & \beta_2 & \beta_3 & 0 & \alpha_1 & \alpha_2 & \alpha_3 &
0 & 0 & 0 & 0& 0 \cr
 {\ } & 0 & \beta_1 & \beta_2 & \beta_3 & 0 & \alpha_1 & \alpha_2 &
\alpha_3 & 0 & 0 & 0 & 0 \cr
 {\ } & \gamma_1 & \gamma_2 & \gamma_3 & 0 & 0 & 0 & 0 & 0 & \alpha_1 &
\alpha_2 & \alpha_3 & 0 \cr
 {\ } & 0 & \gamma_1 & \gamma_2 & \gamma_3 & 0 & 0 & 0 & 0 & 0 & \alpha_1
& \alpha_2 & \alpha_3  \cr
 {\ } & 0 & 0 & 0 & 0 & \beta_1 & \beta_2 & \beta_3 & 0 & 0 & 0 & 0 & 0
\cr
 {\ } & 0 & 0 & 0 & 0 & 0 & \beta_1 & \beta_2 & \beta_3 & 0 & 0 & 0 & 0
\cr
 {\ } & 0 & 0 & 0 & 0 & \gamma_1 & \gamma_2 & \gamma_3 & 0 & \beta_1 &
\beta_2 & \beta_3 & 0 \cr
 {\ } & 0 & 0 & 0 & 0 & 0 & \gamma_1 & \gamma_2 & \gamma_3 & 0 & \beta_1 &
\beta_2 & \beta_3 \cr
 {\ } & 0 & 0 & 0 & 0 & 0 & 0 & 0 & 0 & \gamma_1 & \gamma_2 & \gamma_3 & 0
\cr
{\ } & 0 & 0 & 0 & 0 & 0 & 0 & 0 & 0 & 0 & \gamma_1 & \gamma_2 & \gamma_3
\cr}.
\label{n19}
\end{equation}
where we have used the following parametrization of $m_2$ and $m_3$
in terms of the moduli $\beta_b, b=1, 2, 3$, and $\gamma_c, c=1, 2, 3$
\begin{eqnarray}
&&m_2 = \beta_1 u^2 + \beta_2 uv + \beta_3 v^2,
\nonumber \\
&&m_3 = \gamma_1 u^2 + \gamma_2 uv + \gamma_3 v^2.
\label{n22}
\end{eqnarray}
We now can compute the determinant and find the following.

\begin{itemize}

\item
The determinant of the matrix $M_{IJ}$ is given by
\begin{equation}
det f_C= -{\cal R}^4,
\label{n20}
\end{equation}
where
\begin{equation}
{\cal R}=
\alpha_1 \beta_2 \gamma_3 -\alpha_1 \beta_3 \gamma_2 +
\alpha_2 \beta_3 \gamma_1 -\alpha_2 \beta_1 \gamma_3 +
\alpha_3 \beta_1 \gamma_2 -\alpha_3 \beta_2 \gamma_1.
\label{n21}
\end{equation}

\end{itemize}

We conclude this example by pointing out that
$det f_C$ will vanish precisely
on the zero locus of the homogeneous polynomial
${\cal R}$. Globally, ${\cal R}$ must be a holomorphic section
of some complex line bundle over ${\mathbb P}^{8}$.
Therefore, there must exist a divisor
$D_{{\cal R}} \subset {\mathbb P}^{8}$ such that ${\cal R}$ is a section
of
\begin{equation}
{\cal O}_{{\mathbb P}^{8}}(D_{{\cal R}})
\label{div3.1}
\end{equation}
and vanishes on the co-dimension one submanifold
$D_{{\cal R}}$ of ${\mathbb P}^{8}$. This categorizes the zeroes
of ${\cal R}$ and, hence, $det f_C$.
The exact co-dimension one submanifold
\begin{equation}
D_{{\cal R}} \subset {\mathbb P}H^{0}
({\cal S}, {\cal O}_{{\cal S}}(2) \oplus {\cal O}_{{\cal S}}(2)
\oplus {\cal O}_{{\cal S}}(2))
\label{div3.2}
\end{equation}
can be determined by solving the equation ${\cal R}=0$ using~\eqref{n21}.


\section{Expressions for the Superpotential:}


In the previous section, we computed, for several non-trivial
examples, the explicit expressions for
$det f_C$ in terms of the associated vector bundle
moduli. In each case, $h^{0}(z, V|_{z}(-1))$ will be non-zero and, hence,
the Pfaffian  will vanish if and only if
\begin{equation}
det f_C=0.
\label{insert1}
\end{equation}
In special cases, such as those described by
conditions~\eqref{3.45} or Example 3, $det f_C$
may vanish everywhere in moduli space. In these cases, the conclusions
are straightforward. The Pfaffian and, hence, the superpotential
$W$ vanish for all values of the moduli. That is,
\begin{equation}
W=0.
\label{insert2}
\end{equation}
Generically, however, $det f_C$ will not vanish, as
was shown explicitly in Examples 2, 1 and 4. However, even in this case,
there
will always be a divisor $D$ of the vector bundle moduli space
where~\eqref{insert1} is satisfied. For Examples 2, 1 and 4, this
divisor is given in equations~\eqref{div1.2},~\eqref{div2.2}
and~\eqref{div3.2} respectively. Therefore, on $D$,
$det f_C$ and, hence, the Pfaffian vanish. This will
allow us to explicitly compute the Pfaffian.

To proceed, note
that the Pfaffian of the chiral Dirac operator in $8k+2$ dimensions
also represents a global section of a line bundle over the space of
parameters (see \cite{Freed} for a review). That it is a section,
rather than a function, is a reflection of the fact that the Pfaffian
is not gauge invariant. In our case, it is a global section of a line
bundle over ${\mathbb P}^{N}$. Any global section of a line bundle
over ${\mathbb P}^{N}$ is known to be a homogeneous polynomial.
This means that
the Pfaffian must be a polynomial. Furthemore, we have shown that
this polynomial vanishes precisely along the same zero locus
as $det f_C$. Therefore, since
${\mathbb P}^{N}$ is compact,
the Pfaffian
and $det f_C$ must be equal to each other up to a
possible power.
One can define a purely algebraic analogue of
${\rm Pfaff}({\cal D}_{-})$. A more careful analysis of our argument
shows that, quite generically, the algebraic version of
${\rm Pfaff}({\cal D}_{-})$ equals $det f_C$,
up to a constant. Furthermore, it was shown in
\cite{Bismut1,Bismut2,Bismut3} that the algebraic and analytic
notions of ${\rm Pfaff}({\cal D}_{-})$ agree. Therefore,
\begin{equation}
{\rm Pfaff}({\cal D}_{-}) \propto det f_C.
\label{s0}
\end{equation}
One can now present the final answer for the vector bundle moduli
contribution to the non-perturbative superpotential. Since the
superpotential
is proportional to the Pfaffian, we conclude that
\begin{equation}
W \propto det f_C.
\label{s1}
\end{equation}
Let us now illustrate this explicitly for the
three examples considered above. The ordering of the examples will be the
same as
in the previous subsection. \\

\noindent
{\bf Example 2 \rm}:
In this example, $det f_C$ is a polynomial of
degree
$20$, so it is a global section of the line bundle of degree $20$ over
${\mathbb P}^{12}$.

\begin{itemize}

\item The superpotential equals
\begin{equation}
W \propto {\cal P}^{4},
\label{s2}
\end{equation}
where
\begin{eqnarray}
&&{\cal{P}} =
\chi_1^2 \chi_3 \phi_3^2 -
\chi_1^2 \chi_2 \phi_3 \phi_4 -
2\chi_1 \chi_3^2  \phi_3 \phi_1 - \nonumber \\
&&\chi_1 \chi_2 \chi_3  \phi_3 \phi_2 +
\chi_2^2 \chi_3  \phi_1 \phi_3 +
\phi_4^2 \chi_1^3 -              \nonumber \\
&&2 \phi_2 \phi_4 \chi_3 \chi_1^2  +
\chi_1 \chi_3^2 \phi_2^2 +
3 \phi_1 \phi_4 \chi_1 \chi_2 \chi_3 + \nonumber \\
&&\phi_2 \chi_1 \phi_4 \chi_2^2 +
\phi_1^2 \chi_3^3 -
\phi_2 \chi_2 \phi_1 \chi_3^2-
\phi_4 \phi_1 \chi_2^3.
\label{s3}
\end{eqnarray}
\end{itemize}

Note that $W$ depends on only seven of the thirteen projective vector
bundle
moduli which parameterize ${\mathbb P}^{12}$.\\

\noindent
{\bf Example 1 \rm}: In this example, $det f_C$
is a polynomial
of degree $44$ on the projective space ${\mathbb P}^{9}$.

\begin{itemize}

\item The superpotential is given by
\begin{equation}
W \propto {\cal Q}^{11}\tilde{{\cal Q}},
\label{s7}
\end{equation}
where
\begin{equation}
{\cal Q} = \rho_1 \rho_2 \lambda_2 - \lambda_1 \rho_2^2 - \lambda_3
\rho_1^2
\label{s8}
\end{equation}
and
$\tilde{{\cal Q}}=\tilde{{\cal Q}}(\lambda_a, \rho_b, \mu_c, h_d)$
is given in the Appendix D.
\end{itemize}
Note that the superpotential depends on all ten projective vector bundle
moduli which parameterize ${\mathbb P}^{9}$ and, in addition,
contains non-trivial interactions
between vector bundle moduli and some moduli of the
underlying Calabi-Yau space. \\

\noindent {\bf Example 4 \rm}:
In this example, $det f_C$
is a polynomial of degree $12$ depending on nine homogeneous variables
and,
therefore, it is a global section of the line bundle of degree $12$
over ${\mathbb P}^{8}$.

\begin{itemize}

\item The superpotential is
\begin{equation}
W \propto {\cal R}^4,
\label{n23}
\end{equation}
where
\begin{equation}
{\cal R}=
\alpha_1 \beta_2 \gamma_3 -\alpha_1 \beta_3 \gamma_2 +
\alpha_2 \beta_3 \gamma_1 -\alpha_2 \beta_1 \gamma_3 +
\alpha_3 \beta_1 \gamma_2 -\alpha_3 \beta_2 \gamma_1.
\label{n24}
\end{equation}
\end{itemize}

Note that $W$ depends on all nine projective vector bundle moduli which
parameterize ${\mathbb P}^{8}$.
As in Example 2, $W$ is given by a
perfect power of a polynomial.

\section{Appendix A: Evaluation of $\int {\cal D}g e^{-S_{WZW0}}$}


In this Appendix we will evaluate the path integral
\begin{equation}
\int {\cal D}g e^{-S_{WZW0}},
\label{WZW1}
\end{equation}
where
\begin{eqnarray}
S_{WZW0}&=&-\frac{1}{8\pi}
\int_z {\rm d}\sigma^0 {\rm d}\sigma^1 {\rm tr}
[\frac{1}{2} \sqrt{h}h^{ij}(\omega_i - A_i)
(\omega_j - A_j)+i \epsilon^{ij}\omega_i  A_j)]
\nonumber \\
& &+ \frac{1}{24} \int_{{\cal B}} {\rm d}^3 \hat{\sigma} i
\epsilon^{\hat{i}\hat{j}\hat{k}}{\rm tr}
\omega^{\prime}_{\hat{k}}
\omega^{\prime}_{\hat{j}}
\omega^{\prime}_{\hat{i}},
\label{WZW2}
\end{eqnarray}
the Lie algebra valued one form $\omega$ is given by
\begin{equation}
\omega =g^{-1}dg
\label{WZW3}
\end{equation}
and $A_i$ is the pullback of the Hermitian connection on the
Calabi-Yau threefold $X$ to the holomorphic curve $z$. If we
introduce the complex coordinates on $z$
\begin{equation}
\sigma =\sigma^0 +i\sigma^1, \quad \bar \sigma =\sigma^0 -i\sigma^1,
\label{WZW3.5}
\end{equation}
the action $S_{WZW0}$ can be written as
\begin{equation}
S_{WZW0}= S(g)
+\frac{1}{4\pi}
\int_z {\rm d}^2 \sigma {\rm tr}
\omega_{\bar \sigma}  A_{\sigma} -
\frac{1}{8\pi}
\int_z {\rm d}^2 \sigma {\rm tr}
A_{\bar \sigma}  A_{\sigma},
\label{WZW4}
\end{equation}
where
\begin{equation}
{\rm d}^2\sigma ={\rm d}\sigma {\rm d}\bar \sigma
\label{WZW5}
\end{equation}
and
\begin{equation}
S(g)=-\frac{1}{8\pi}
\int_z {\rm d}^2\sigma {\rm tr}
(g^{-1}\partial_{\sigma}g)  (g^{-1}\partial_{\bar \sigma}g)+
\frac{1}{24\pi} \int_{{\cal B}} {\rm d}^3 \hat{\sigma} i
\epsilon^{\hat{i}\hat{j}\hat{k}} {\rm tr}
(g^{\prime-1}\partial_{\hat{k}}g^{\prime})
(g^{\prime-1}\partial_{\hat{j}}g^{\prime})
(g^{\prime-1}\partial_{\hat{i}}g^{\prime})
\label{WZW6}
\end{equation}
is the WZW action with zero background gauge field.
Since $A_i$ is a connection on a holomorphic vector bundle
and, hence, satisfies the Hermitian Yang-Mills equations,
in any
neighborhood ${\cal U} \subset z$ it is of the form
\begin{equation}
A_{\sigma}=\partial_{\sigma} a \cdot a^{-1}, \quad
A_{\bar \sigma}=\partial_{\bar \sigma} a^{\dagger -1} \cdot a^{\dagger},
\label{WZW7}
\end{equation}
where $a$ is a map from ${\cal U}$ to the complexification
of the structure group $G$. Using~\eqref{WZW7} and the
Polyakov-Wiegmann identity \cite{Polyakov}
\begin{equation}
S(a_1a_2)=S(a_1) +S(a_2) +
\frac{1}{4\pi}\int_z {\rm d}^2 \sigma {\rm tr}(a_1^{-1}
\partial_{\bar \sigma}a_1)
(a_2\partial_{ \sigma}a_2^{-1}),
\label{WZW8}
\end{equation}
the action~\eqref{WZW4} can be re-written as
\begin{equation}
S_{WZW0} = S(ga) - S(a) -
\frac{1}{8\pi} \int_z {\rm d}^2\sigma
(\partial_{\sigma} a \cdot a^{-1})
(\partial_{\bar\sigma} a^{\dagger -1} \cdot a^{\dagger}).
\label{WZW9}
\end{equation}
It is now easy to evaluate the integral~\eqref{WZW1}. Since
${\cal D}g$ is the right-invariant group measure, the integral
\begin{equation}
\int {\cal D}g e^{-S(ga)}
\label{WZW10}
\end{equation}
is a constant, independent of $a$ and, hence, of the vector bundle
moduli. Therefore, we have the following result for the
integral~\eqref{WZW1}
\begin{equation}
\int {\cal D}g e^{-S_{WZW0}} \propto
exp \left( S(a) +\frac{1}{8 \pi}\int_z {\rm d}^2 \sigma
{\rm tr} (\partial_{\sigma}a \cdot a^{-1})
(\partial_{\bar\sigma}a^{\dagger -1} \cdot a^{\dagger}) \right),
\label{WZW11}
\end{equation}
where the functional $S$ is defined in~\eqref{WZW6} with $g$ replaced by
$a$ and
the relation between $a$ and the gauge connection $A$ is given
in~\eqref{WZW7}. Note that the result is expressed in terms of
$a$ rather than the gauge connection $A$. It does not appear to be
possible to present~\eqref{WZW11} soley in terms of the gauge
connection. Also, note that $a$ is globally defined since
$A$ is.

Equation~\eqref{WZW11} represents the general formula for
the vector bundle moduli superpotential and, also, for the Pfaffian
of the chiral Dirac operator twisted by a holomorphic vector bundle
whose structure group is contained in $SO(16) \times SO(16)$.
To find the explicit result, one has to find an exact solution
of the Hermitian Yang-Mills equations on a Calabi-Yau threefold
together with the integration constants, pull it back to the
curve $z$ and perform the integration over the curve.
As we discussed
in Section 2, this procedure is problematic since no solutions
of the Hermitian Yang-Mills equations on a Calabi-Yau threefold are known.

It is straightforward to check that the expression in the exponent
in~\eqref{WZW11} transforms under a gauge transformation with
parameter $\epsilon$ as
\begin{equation}
-\frac{i}{8 \pi}\int_z {\rm d}^2 \sigma{\rm tr} (\epsilon
dA).
\label{WZW12}
\end{equation}
This anomaly is cancelled by the variation of the
$B$-field \cite{Witten2}
\begin{equation}
\delta B=\frac{1}{8 \pi}{\rm tr} (\epsilon dA),
\label{WZW13}
\end{equation}
leaving the superpotential~\eqref{1.16}
gauge invariant.\footnote{The coefficients in~\eqref{WZW12}
and~\eqref{WZW13} are different
from those in \cite{Witten2} since we are following the conventions
of \cite{Lima1,Lima2} for the normalization of the
Wess-Zumino-Witten term.}
Note that expression~\eqref{WZW12} precisely
equals the anomaly of the square root of the chiral Dirac operator.


\section{Appendix B: Spectral Data and $SU(n)$ Vector Bundles}


\subsection{Elliptically Fibered Calabi-Yau Threefolds:}

In this Appendix, we will consider Calabi--Yau threefolds, $X$, that are
structured as elliptic curves fibered over a base surface,
$B$. We denote the natural projection as
$\pi: X\to B$  and by  $\sigma : B\to X$ the analytic map that defines the
zero section.

A simple representation of an elliptic curve is given in the
projective space $\cp{2}$ by the Weierstrass equation
\begin{equation}
zy^2=4x^3-g_2xz^2-g_3z^3,
\label{eq:1}
\end{equation}
where $(x,y,z)$ are the homogeneous coordinates of $\cp{2}$ and $g_2$,
$g_3$ are constants. This same equation can represent the elliptic
fibration, $X$, if the coefficients $g_2$, $g_3$ in the Weierstrass
equation
are functions over the base surface, $B$. The correct way to express this
globally is to replace the projective plane
$\cp{2}$ by a $\cp{2}$-bundle $P \to B$ and then
require that $g_2$, $g_3$ be sections of appropriate line
bundles over the base. If we denote the conormal bundle to the
zero section $\sigma(B)$ by $\cL$, then $P = {\mathbb P}({\mathcal
O}_{B}\oplus \cL^{2} \oplus \cL^{3})$, where ${\mathbb P}$ stands
for the projectivization of the vector bundle. There is a hyperplane
line
bundle ${\mathcal O}_{P}(1)$ on $P$ which corresponds to the divisor
${\mathbb P}(\cL^{2}\oplus \cL^{3}) \subset P$ and the
coordinates $x,y,z$ are sections of
$\cO_{P}(1)\otimes\cL^2, \cO_{P}(1)\otimes
\cL^3$ and $\cO_{P}(1)$ respectively.
It then follows from \eqref{eq:1} that the coefficients $g_2$ and $g_3$
are
sections of $\cL^4$ and $ \cL^6$.

It is useful to define new coordinates, $\bX,\bY,\bZ$ on $X$ by
$x=\bX\bZ$,
$y=\bY$ and $z=\bZ^3$. It follows that $\bX,\bY,\bZ$ are now sections of
line bundles
\begin{equation}
\bX \sim \cO_{X}(2\sigma)\otimes\cL^2 , \qquad
\bY \sim \cO_{X}(3\sigma)\otimes \cL^3 , \qquad
\bZ \sim \cO_{X}(\sigma)
\label{eq:3}
\end{equation}
respectively, where we use the fact that
$\cO_{X}(3\sigma)=\cO_{P}(1)|_{X}$.
The coefficients $g_2$ and $g_3$ remain
sections of line bundles
\begin{equation}
g_2 \sim \cL^4, \qquad
g_3 \sim \cL^6.
\label{eq:4}
\end{equation}
The symbol ``$\sim$'' simply means ``section of''.

The requirement that elliptically fibered threefold, $X$, be a
Calabi--Yau space
constrains the first Chern class of the tangent bundle, $TX$, to
vanish. That is,
\begin{equation}
c_1(TX)=0.
\label{eq:5}
\end{equation}
It follows from this that
\begin{equation}
\cL=K_B^{-1},
\label{eq:6}
\end{equation}
where $K_B$ is the canonical bundle on the base, $B$.
Condition~\eqref{eq:6} is
rather strong and restricts the allowed base spaces of an elliptically
fibered Calabi--Yau threefold to be del Pezzo, Hirzebruch and Enriques
surfaces, as well as certain blow--ups of Hirzebruch surfaces \cite{vafa}.
The definitions and properties of these surfaces have been reviewed
in~\cite{hvb}. In this paper, we will primarily be concerned with the case
\begin{equation}
B={\mathbb F}_{r}
\label{A1}
\end{equation}
where $r$ is any non-negative integer. Recall from~\cite{hvb} that the
homology group
$H_{2}({\mathbb F}_{r}, {\mathbb Z})$ has as a basis of effective classes
${\cal{S}}$ and ${\cal{E}}$ with the intersection numbers
\begin{equation}
{\cal{S}}^{2}=-r, \qquad {\cal{S}}\cdot{\cal{E}}=1, \qquad {\cal{E}}^{2}=0
\label{A2}
\end{equation}
and that the Chern classes are given by
\begin{equation}
c_{1}({\mathbb F}_{r})=2{\cal{S}}+(r+2){\cal{E}}, \qquad c_{2}({\mathbb
F}_{r})=4.
\label{A3}
\end{equation}

\subsection*{Spectral cover Description of $SU(n)$ Vector Bundles:}

As discussed in \cite{FMW1,AJ}, $SU(n)$ vector bundles over an
elliptically
fibered Calabi--Yau threefold can be explicitly constructed from two
mathematical objects, a divisor $\cC$ of $X$, called the spectral cover,
and a
line bundle $\cN$ on $\cC$. In this paper, we will
consider only stable $SU(n)$ vector bundles
constructed from irreducible spectral covers. In addition, we will impose
the condition that the spectral cover be a ``positive'' divisor.

\subsection*{Spectral Cover}

A spectral cover, $\cC$, is a surface in $X$ that is an $n$-fold cover of
the
base $B$. That is, $\pi_{\cC}: \cC\to B$. The general form for a
spectral cover is given by
\begin{equation}
\cC=n\sigma + \pi^*\eta,
\label{eq:7}
\end{equation}
where $\sigma$ is the zero section and $\eta$ is some curve in the base
$B$.
The terms in~\eqref{eq:7} can be considered either as
elements of the homology group $H_{4}(X, {\mathbb Z})$ or, by
Poincare duality, as elements of cohomology $H^{2}(X, {\mathbb Z})$.

In terms of the coordinates $\bX$, $\bY$, $\bZ$ introduced above,
it can be shown that
the spectral cover can be represented as the zero set of the polynomial
\begin{equation}
s=a_{0}\bZ^{n} + a_{2}\bX\bZ^{n-2} + a_{3}\bY\bZ^{n-3} + \ldots +
a_{n}\bX^{\frac{n}{2}}
\label{eq:8}
\end{equation}
for $n$ even and ending in $a_{n}\bX^{\frac{n-3}{2}}\bY$ if $n$ is odd,
along with the relations~\eqref{eq:3}. This tells us that the polynomial
$s$ must be a holomorphic section of the line bundle of the spectral
cover,
$\cO_{X}(\cC)$. That is,
\begin{equation}
s \sim \cO_{X}(n\sigma + \pi^{*}\eta).
\label{eq:9}
\end{equation}
It follows from this and equation~\eqref{eq:3}, that the
coefficients $a_{i}$ in the polynomial $s$ must be sections of the line
bundles
\begin{equation}
a_{i} \sim \pi^*K_{B}^{i} \otimes \cO_{X}(\pi^{*}\eta)
\label{eq:10}
\end{equation}
for $i=1,\ldots ,n$ where we have used expression~\eqref{eq:6}.

In order to describe vector bundles most simply, there are two properties
that we require the spectral cover to possess. The first, which is shared
by
all spectral covers, is that
\begin{itemize}
\item $\cC$ must be an effective class in $H_{4}(X,{\mathbb Z})$.
\end{itemize}
This property is simply an expression of the fact the spectral cover must
be
an actual surface in $X$. It can easily be shown that
\begin{equation}
\cC \subset X \text{ is effective } \Longleftrightarrow \eta
\text{ is an effective class in } H_{2}(B, {\mathbb Z}).
\label{eq:12}
\end{equation}
The second property that we require for the spectral cover is that

\begin{itemize}

\item $\cC$ is an irreducible surface.

\end{itemize}
This condition is imposed because it guarantees that the associated vector
bundle is stable.

In this paper, we require a third condition on the spectral cover,
namely that

\begin{itemize}

\item $\cC$ is positive.

\end{itemize}
This condition is imposed because it allows one to use
the Kodaira vanishing theorem and the Lefschetz theorem when evaluating
the number
of vector bundle moduli. These theorems greatly simplify this calculation.
By definition, $\cC$ is positive (or ample) if and only if the first Chern
class of the associated line bundle $\cO_{X}(\cC)$ is a positive class in
$H^{2}_{DR}(X, {\mathbb R})$. An equivalent definition of positive is the
following. $\cC$ is a positive divisor if and only if
\begin{equation}
\cC \cdot c >0
\label{eq:20a}
\end{equation}
for every effective curve $c$ in $X$.

In order to make these concepts more concrete, we take the
base surface to be
\begin{equation}
B={\mathbb F}_{r}
\label{eq:15}
\end{equation}
and write the conditions under
which the three properties stated above are satisfied.
Then, in general, $\cC$ is given by expression~\eqref{eq:7} where
\begin{equation}
\eta= a\cS+b\cE
\label{eq:17}
\end{equation}
and $a$,$b$ are integers. One can easily check that $\eta$ is an effective
class in ${\mathbb F}_{r}$ and, hence, that $\cC$ is an effective class
in
$X$, if and only if
\begin{equation}
a \geq 0, \qquad b \geq 0.
\label{eq:18}
\end{equation}
It was shown in \cite{tony} that ${\cal C}$ is irreducible if and only
if
\begin{equation}
b \geq ar
\label{eq:19}
\end{equation}
and
\begin{equation}
a \geq 2n, \qquad b \geq n(r+2).
\label{eq:20}
\end{equation}
Finally, it was shown in~\cite{BDO1} using
expressions~\eqref{A2} and~\eqref{A3} that spectral cover ${\cal{C}}$ will
be
positive if and only if
\begin{equation}
a>2n, \qquad b>ar-n(r-2).
\label{eq:25a}
\end {equation}
\subsection*{The Line Bundle $\cN$}

As discussed in \cite{FMW1,AJ}, in addition to the
spectral cover it is necessary to
specify a line bundle, $\cN$, over $\cC$. For $SU(n)$ vector bundles, this
line bundle must be a restriction of a global line bundle on $X$
(which we will again denote by ${\mathcal N}$), satisfying the condition
\begin{equation}
c_{1}(\cN)=n(\frac{1}{2}+\lambda)\sigma+(\frac{1}{2}-\lambda)
\pi^{*}\eta+(\frac{1}{2}+n\lambda)\pi^{*}c_{1}(B),
\label{eq:27}
\end{equation}
where $c_{1}(\cN)$, $c_{1}(B)$ are the first Chern classes of $\cN$ and
$B$
respectively and $\lambda$ is, a priori, a rational number. Since
$c_{1}(\cN)$
must be an integer class, it follows that either
\begin{equation}
n \quad \mbox{is odd}, \qquad \lambda=m+\frac{1}{2}
\label{eq:28}
\end{equation}
or
\begin{equation}
n \quad \mbox{is even}, \qquad \lambda=m, \qquad \eta=c_{1}(B) \mod 2,
\label{eq:29}
\end{equation}
where $m \in {\mathbb Z}$.

\subsection*{$SU(n)$ Vector Bundle}

Given a spectral cover, $\cC$, and a line bundle, $\cN$, satisfying the
above
properties, one can now uniquely construct an $SU(n)$ vector bundle, $V$.
This
can be accomplished in two ways. First, as discussed in \cite{AJ}, the
vector
bundle
can be directly constructed using the associated Poincare bundle,
${\mathcal P}$.
The result is that
\begin{equation}
V=\pi_{1*}(\pi^{*}_{2}\cN\otimes {\mathcal P}),
\label{eq:32}
\end{equation}
where $\pi_{1}$ and $\pi_{2}$ are the two projections of the fiber
product $X \times_{B} \cC$ onto the two factors $X$ and $\cC$. We refer
the reader to \cite{AJ,gub} for a detailed discussion. Equivalently, $V$
can be
constructed directly from $\cC$ and $\cN$ using the Fourier-Mukai
transformation, as discussed in \cite{FMW1,AJ}. Both of these
constructions work in
reverse, yielding the spectral data $(\cC, \cN)$ up to the overall factor
of $K_B$ given the vector bundle $V$.
Throughout this paper we will indicate this relationship between the
spectral data and the vector bundle by writing
\begin{equation}
(\cC,\cN) \longleftrightarrow V.
\label{eq:33}
\end{equation}
The Chern classes for the $SU(n)$ vector bundle $V$ have been computed in
\cite{FMW1} and \cite{gub,smb}. The results are
\begin{equation}
c_{1}(V)=0
\label{eq:34}
\end{equation}
since $\operatorname{tr} F=0$ for the structure group $SU(n)$,
\begin{equation}
c_2(V)=\pi^{*}\eta\cdot\sigma-\frac{1}{24}\pi^{*}c_1(B)^2(n^3-n)
+\frac{1}{2}(\lambda^2-\frac{1}{4})n\pi^{*}(\eta\cdot(\eta-nc_1(B)))
\label{eq:35}
\end{equation}
and
\begin{equation}
c_3(V)= 2\lambda \sigma\cdot \pi^{*}(\eta\cdot(\eta-nc_1(B))).
\label{eq:36}
\end{equation}
Finally, we note that it was shown in \cite{gub} that
\begin{equation}
N_{gen}=\frac{c_{3}(V)}{2}
\label{eq:36a}
\end{equation}
gives the number of quark and lepton generations.


\section{Appendix C: The Direct Image
$\pi_{*}{\cal O}_{\pi^* {\cal S}}(A \sigma|_{\pi^{*}{\cal S}}+fF)$}


The goal of this Appendix is to show how to find the direct image of the
line bundle
\begin{equation}
{\cal O}_{\pi^* {\cal S}}(A \sigma|_{\pi^{*}{\cal S}}+fF), \quad A>0
\label{B0}
\end{equation}
on the curve ${\cal S} \subset {\mathbb F}_{r}$ used
throughout the paper.

First, note that
\begin{equation}
{\cal O}_{\pi^* {\cal S}}(A \sigma|_{\pi^{*}{\cal S}}+fF)=
{\cal O}_{\pi^* {\cal S}}(A \sigma|_{\pi^{*}{\cal S}}) \otimes
{\cal O}_{\pi^* {\cal S}}(fF)=
{\cal O}_{\pi^* {\cal S}}(A \sigma|_{\pi^{*}{\cal S}}) \otimes
\pi^* {\cal O}_{{\cal S}}(f).
\label{B1}
\end{equation}
From the projection formula
\begin{equation}
\pi_{*}({\cal F}\otimes \pi^* {\cal G})=(\pi_* {\cal F})\otimes {\cal G}
\label{B1.5}
\end{equation}
for arbitrary bundles ${\cal F}$ and ${\cal G}$,
it follows that
\begin{equation}
\pi_{*} {\cal O}_{\pi^{*}{\cal S}} (A \sigma|_{\pi^{*}{\cal S}} +fF)=
\pi_{*}{\cal O}_{\pi^* {\cal S}}(A \sigma|_{\pi^{*}{\cal S}})
\otimes {\cal O}_{{\cal S}}(f).
\label{B2}
\end{equation}
This implies that it is enough to find
$\pi_{*}{\cal O}_{\pi^* {\cal S}}(A \sigma|_{\pi^{*}{\cal S}})$,
which can be done by induction.
To begin with, consider the short
exact sequence
\begin{equation}
0 \rightarrow {\cal O}_{\pi^* {\cal S}} \rightarrow
{\cal O}_{\pi^* {\cal S}}(\sigma|_{\pi^* {\cal S}})
\rightarrow
{\cal O}_{\pi^* {\cal S}}
(\sigma|_{\pi^* {\cal S}})|_{\sigma|_{\pi^* {\cal S}}}
\rightarrow 0.
\label{B3}
\end{equation}
The inclusion
\begin{equation}
{\cal O}_{\pi^* {\cal S}} \hookrightarrow
{\cal O}_{\pi^* {\cal S}}(\sigma|_{\pi^* {\cal S}})
\label{B3.1}
\end{equation}
induces a sheaf map
\begin{equation}
i:\pi_{*}{\cal O}_{\pi^* {\cal S}} \hookrightarrow
\pi_*{\cal O}_{\pi^* {\cal S}}(\sigma|_{\pi^* {\cal S}}).
\label{B3.2}
\end{equation}
For each point $p \in {\cal S}$, consider a map
\begin{equation}
i_{p} : H^{0}(F_p, {\cal O}_{F_p}) \rightarrow
H^{0}(F_p, {\cal O}_{F_p}((\sigma|_{\pi^* {\cal S}})|_{F_p})),
\label{B3.3}
\end{equation}
where $F_p$ is the elliptic fiber above the point $p$.
Since $\sigma|_{\pi^* {\cal S}}$ intersects $F$ at one point,
it follows that the degree of the line bundle
${\cal O}_{F_p}((\sigma|_{\pi^* {\cal S}})|_{F_p})$ is one and, hence,
it has a unique global holomorphic section. Therefore, the map
$i_p$ is an isomorphism, so
\begin{equation}
\pi_*{\cal O}_{\pi^* {\cal S}}(\sigma|_{\pi^* {\cal S}})\simeq
\pi_{*} {\cal O}_{\pi^* {\cal S}} \simeq
{\cal O}_{{\cal S}}.
\label{B18}
\end{equation}
Thus, we have found the direct image of the line bundle
${\cal O}_{\pi^* {\cal S}}(A \sigma|_{\pi^* {\cal S}})$ for $A=1$.
Now, consider the sequence
\begin{equation}
0 \rightarrow
{\cal O}_{\pi^* {\cal S}}(\sigma|_{\pi^* {\cal S}})
\rightarrow
{\cal O}_{\pi^* {\cal S}}(2\sigma|_{\pi^* {\cal S}})
\rightarrow
{\cal O}_{\pi^* {\cal S}}
(2\sigma|_{\pi^* {\cal S}})|_{\sigma|_{\pi^* {\cal S}}}
\rightarrow 0
\label{B19}
\end{equation}
and take its direct image, remembering that the direct image of the first
term was already found in~\eqref{B18}.
By using the facts that the
last term in the sequence can be written as
$\pi^{*}{\cal O}_{{\cal S}}(2\sigma|_{\pi^*{\cal S}} \cdot
\sigma|_{\pi^*{\cal S}})|_{\sigma|_{\pi^*{\cal S}}}$
and all higher direct images
vanish by the Kodaira vanishing theorem, we obtain
\begin{equation}
0 \rightarrow {\cal O}_{{\cal S}} \rightarrow
\pi_*{\cal O}_{\pi^* {\cal S}}(2\sigma|_{\pi^* {\cal S}})
\rightarrow {\cal O}_{{\cal S}}(2(r-2)) \rightarrow 0,
\label{B20}
\end{equation}
where the intersection property \cite{BDO1}
\begin{equation}
\sigma|_{\pi^* {\cal S}} \cdot \sigma|_{\pi^* {\cal S}}=-(2-r)
\label{B20.1}
\end{equation}
has been used.
Since the extention group
\begin{equation}
Ext({\cal O}_{{\cal S}}(2 (r-2)), {\cal O}_{{\cal S}})
\simeq H^{1} ({\cal S}, {\cal O}_{{\cal S}}(2 (r-2))^{*} \otimes
{\cal O}_{{\cal S}})
\simeq H^{0} ({\cal S}, {\cal O}_{{\cal S}}(2r-6))^{*}
\label{B21}
\end{equation}
vanishes for $r=0, 1, 2$, we have
\begin{equation}
\pi_*{\cal O}_{\pi^* {\cal S}}(2\sigma|_{\pi^* {\cal S}})
\simeq {\cal O}_{{\cal S}} \oplus {\cal O}_{{\cal S}}(2 (r-2)).
\label{B22}
\end{equation}
Therefore, we have found the direct image of the line bundle
$\pi_*{\cal O}_{\pi^* {\cal S}}(A\sigma|_{\pi^* {\cal S}})$
for $A=2$. Continuing this way we obtain
\begin{equation}
\pi_*{\cal O}_{\pi^* {\cal S}}(A\sigma|_{\pi^* {\cal S}}) \simeq
{\cal O}_{{\cal S}} \oplus \bigoplus_{i=2}^{A} {\cal O}_{{\cal S}}
(i (r-2))
\label{B23}
\end{equation}
for arbitrary $A \geq 1$.
Tensoring with ${\cal O}_{{\cal S}}(f)$, we get
\begin{equation}
\pi_*{\cal O}_{\pi^* {\cal S}}(A\sigma|_{\pi^* {\cal S}}+fF) \simeq
{\cal O}_{{\cal S}}(f) \oplus \bigoplus_{i=2}^{A} {\cal O}_{{\cal S}}
(f+i (r-2))
\label{B24}
\end{equation}
for arbitrary $A \geq 1$.
\newpage

\section{Appendix D: The Polynomial $\tilde{\cal Q}$}


Here, we give the expression for the polynomial
$\tilde{\cal Q}$ that arises in Example 2 of Section 5. We found that
\begin{eqnarray}
&&\tilde{\cal Q}=
48 \mu_1^2 \lambda_2 \rho_2^8-
48 \mu_5^2 \rho_1^8 \lambda_2-2 \mu_2 \rho_1^6 \lambda_3^4+
2 \mu_4 \rho_2^6 \lambda_1^4-
h_2 \rho_2^8 \lambda_1^3+h_4 \rho_1^8 \lambda_3^3-
24 \mu_3 \mu_2 \rho_1^3 \lambda_2 \rho_2^5-
\nonumber \\
&&72 \mu_3 \mu_2 \rho_1^2 \rho_2^6 \lambda_1+
120 \mu_3 \mu_2 \rho_1^4 \lambda_3 \rho_2^4+
96 \mu_5 \mu_1 \rho_1^3 \rho_2^5 \lambda_1-
96 \mu_5 \mu_1 \rho_1^5 \lambda_3 \rho_2^3+
48 \mu_3 \mu_1 \rho_1 \rho_2^7 \lambda_1+
\nonumber \\
&&48 \mu_3 \mu_1 \rho_1^2 \lambda_2 \rho_2^6-
144 \mu_3 \mu_1 \rho_1^3 \lambda_3 \rho_2^5-
72 \mu_4 \mu_1 \rho_1^2 \rho_2^6 \lambda_1-
24 \mu_4 \mu_1 \rho_1^3 \lambda_2 \rho_2^5+
120 \mu_4 \mu_1 \rho_1^4 \lambda_3 \rho_2^4-
\nonumber \\
&&120 \mu_4 \mu_3 \rho_1^4 \rho_2^4 \lambda_1+
24 \mu_4 \mu_3 \rho_1^5 \lambda_2 \rho_2^3+
72 \mu_4 \mu_3 \rho_1^6 \lambda_3 \rho_2^2-
168 \mu_5 \mu_4 \rho_1^6 \rho_2^2 \lambda_1+
72 \mu_5 \mu_4 \rho_1^7 \lambda_2 \rho_2+
\nonumber \\
&&24 \mu_5 \mu_4 \rho_1^8 \lambda_3-
48 \mu_5 \mu_3 \rho_1^7 \lambda_3 \rho_2-
48 \mu_5 \mu_3 \rho_1^6 \lambda_2 \rho_2^2+
144 \mu_5 \mu_3 \rho_1^5 \rho_2^3 \lambda_1-
2 h_5 \rho_1^6 \lambda_2^3 \rho_2^2+
\nonumber \\
&&72 \mu_5 \mu_2 \rho_1^6 \lambda_3 \rho_2^2-
12 h_5 \lambda_1 \rho_1^6 \lambda_2 \lambda_3 \rho_2^2+
4 h_5 \rho_1^3 \rho_2^5 \lambda_1^3+
24 \mu_5 \mu_2 \rho_1^5 \lambda_2 \rho_2^3-
24 \mu_2 \mu_1 \rho_2^8 \lambda_1-
\nonumber \\
&&120 \mu_5 \mu_2 \rho_1^4 \rho_2^4 \lambda_1-
2 h_5 \rho_1^8 \lambda_3^2 \lambda_2-
3 h_4 \rho_1^2 \rho_2^6 \lambda_1^3-
72 \mu_2 \mu_1 \lambda_2 \rho_1 \rho_2^7+
8 h_5 \lambda_1^2 \rho_1^5 \lambda_3 \rho_2^3-
\nonumber \\
&&10 h_5 \lambda_1^2 \rho_1^4 \lambda_2 \rho_2^4-
96 \mu_4 \mu_2 \rho_1^5 \lambda_3 \rho_2^3+
96 \mu_4 \mu_2 \rho_1^3 \rho_2^5 \lambda_1+
168 \mu_2 \mu_1 \rho_1^2 \lambda_3 \rho_2^6+
h_4 \rho_1^5 \lambda_2^3 \rho_2^3+
\nonumber \\
&&2 h_3 \rho_1 \rho_2^7 \lambda_1^3-
2 h_3 \rho_1^7 \lambda_3^3 \rho_2-
5 h_4 \lambda_1^2 \rho_1^4 \lambda_3 \rho_2^4+
7 h_4 \lambda_1^2 \rho_1^3 \lambda_2 \rho_2^5+
4 h_5 \rho_1^7 \lambda_3 \lambda_2^2 \rho_2+
\nonumber \\
&&4 h_5 \lambda_1 \rho_1^7 \lambda_3^2 \rho_2+
8 h_5 \lambda_1 \rho_1^5 \lambda_2^2 \rho_2^3-
2 h_3 \rho_1^5 \lambda_3 \lambda_2^2 \rho_2^3-
2 h_3 \lambda_1 \rho_1^5 \lambda_3^2 \rho_2^3+
6 h_4 \lambda_1 \rho_1^5 \lambda_2 \lambda_3 \rho_2^3+
\nonumber \\
&&2 h_3 \lambda_1 \rho_1^3 \lambda_2^2 \rho_2^5+
2 h_3 \lambda_1^2 \rho_1^3 \lambda_3 \rho_2^5-
4 h_3 \lambda_1^2 \rho_1^2 \lambda_2 \rho_2^6-
h_4 \rho_1^7 \lambda_3^2 \lambda_2 \rho_2-
h_4 \rho_1^6 \lambda_3 \lambda_2^2 \rho_2^2-
\nonumber \\
&&h_4 \lambda_1 \rho_1^6 \lambda_3^2 \rho_2^2-
5 h_4 \lambda_1 \rho_1^4 \lambda_2^2 \rho_2^4+
2 h_1 \rho_1^2 \lambda_2^3 \rho_2^6+
2 h_1 \lambda_1^2 \lambda_2 \rho_2^8+
3 h_2 \rho_1^6 \lambda_3^3 \rho_2^2-
\nonumber \\
&&h_2 \rho_1^3 \lambda_2^3 \rho_2^5+
4 h_3 \rho_1^6 \lambda_3^2 \lambda_2 \rho_2^2-
4 h_1 \rho_1^5 \lambda_3^3 \rho_2^3+
5 h_2 \lambda_1 \rho_1^4 \lambda_3^2 \rho_2^4+
h_2 \lambda_1 \rho_1^2 \lambda_2^2 \rho_2^6+
\nonumber \\
&&h_2 \lambda_1^2 \rho_1^2 \lambda_3 \rho_2^6+
h_2 \lambda_1^2 \lambda_2 \rho_1 \rho_2^7-
6 h_2 \lambda_1 \rho_1^3 \lambda_2 \lambda_3 \rho_2^5-
2 \mu_5 \rho_1^5 \lambda_2^4 \rho_2-
2 \mu_3 \lambda_1^3 \lambda_2 \rho_2^6-
\nonumber \\
&&2 \mu_4 \rho_1^6 \lambda_3^2 \lambda_2^2+
2 \mu_5 \rho_1^6 \lambda_3 \lambda_2^3-
8 \mu_5 \rho_1 \rho_2^5 \lambda_1^4+
5 h_2 \rho_1^4 \lambda_3 \lambda_2^2 \rho_2^4+
2 \mu_3 \rho_1^6 \lambda_3^3 \lambda_2-
\nonumber \\
&&8 h_1 \rho_1^3 \lambda_3 \lambda_2^2 \rho_2^5-
8 h_1 \lambda_1 \rho_1^3 \lambda_3^2 \rho_2^5-
4 \mu_5 \lambda_1 \rho_1^5 \lambda_3 \lambda_2^2 \rho_2-
4 h_1 \lambda_1 \lambda_2^2 \rho_1 \rho_2^7-
4 h_1 \lambda_1^2 \rho_1 \lambda_3 \rho_2^7+
\nonumber \\
&&12 h_1 \lambda_1 \rho_1^2 \lambda_3 \lambda_2 \rho_2^6-
7 h_2 \rho_1^5 \lambda_3^2 \lambda_2 \rho_2^3+
2 \mu_4 \lambda_1 \rho_1^6 \lambda_3^3-
10 \mu_4 \lambda_1 \rho_1^4 \lambda_3 \lambda_2^2 \rho_2^2+
20 \mu_5 \rho_1^2 \lambda_2 \rho_2^4 \lambda_1^3+
\nonumber \\
&&20 \mu_4 \lambda_1^2 \rho_1^3 \lambda_2 \lambda_3 \rho_2^3+
10 h_1 \rho_1^4 \lambda_3^2 \lambda_2 \rho_2^4-
2 \mu_2 \lambda_3 \rho_2^6 \lambda_1^3+
2 \mu_2 \lambda_1^2 \lambda_2^2 \rho_2^6+
2 \mu_3 \lambda_1^2 \lambda_2^2 \rho_1 \rho_2^5+
\nonumber \\
&&8 \mu_3 \lambda_1^3 \rho_1 \lambda_3 \rho_2^5-
2 \mu_1 \lambda_1 \lambda_2^3 \rho_2^6+
10 \mu_3 \lambda_1 \rho_1^4 \lambda_3^2 \lambda_2 \rho_2^2+
2 \mu_4 \rho_1^5 \lambda_3 \lambda_2^3 \rho_2-
10 \mu_4 \lambda_1^2 \rho_1^4 \lambda_3^2 \rho_2^2-
\nonumber \\
&&10 \mu_4 \lambda_1^3 \rho_1^2 \lambda_3 \rho_2^4-
10 \mu_3 \lambda_1^2 \rho_1^2 \lambda_3 \lambda_2 \rho_2^4-
2 \mu_4 \lambda_1^3 \lambda_2 \rho_1 \rho_2^5-
4 \mu_5 \lambda_1 \rho_1^6 \lambda_3^2 \lambda_2+
6 \mu_4 \lambda_1 \rho_1^5 \lambda_3^2 \lambda_2 \rho_2+
\nonumber \\
&&10 \mu_5 \lambda_1 \rho_1^4 \lambda_2^3 \rho_2^2+
8 \mu_5 \lambda_1^2 \rho_1^5 \lambda_3^2 \rho_2-
20 \mu_5 \lambda_1^2 \rho_1^3 \lambda_2^2 \rho_2^3+
2 \mu_1 \rho_1 \lambda_2^4 \rho_2^5+
8 \mu_1 \rho_1^5 \lambda_3^4 \rho_2-
\nonumber \\
&&6 \mu_2 \lambda_1^2 \rho_1 \lambda_2 \lambda_3 \rho_2^5-
2 \mu_3 \rho_1^5 \lambda_3^2 \lambda_2^2 \rho_2-
8 \mu_3 \lambda_1 \rho_1^5 \lambda_3^3 \rho_2+
96 \mu_5^2 \rho_1^7 \rho_2 \lambda_1+
10 \mu_2 \lambda_1^2 \rho_1^2 \lambda_3^2 \rho_2^4+
\nonumber \\
&&72 \mu_4^2 \rho_1^5 \rho_2^3 \lambda_1-
24 \mu_4^2 \rho_1^7 \lambda_3 \rho_2+
4 \mu_1 \lambda_1^2 \lambda_2 \lambda_3 \rho_2^6+
2 \mu_2 \rho_1^5 \lambda_2 \lambda_3^3 \rho_2+
10 \mu_2 \lambda_1 \rho_1^2 \lambda_2^2 \lambda_3 \rho_2^4+
\nonumber \\
&&10 \mu_2 \lambda_1 \rho_1^4 \lambda_3^3 \rho_2^2-
2 \mu_2 \lambda_1 \rho_1 \lambda_2^3 \rho_2^5-
8 \mu_1 \lambda_1^2 \rho_1 \lambda_3^2 \rho_2^5-
20 \mu_2 \lambda_1 \rho_1^3 \lambda_3^2 \lambda_2 \rho_2^3+
24 \mu_2^2 \rho_1 \rho_2^7 \lambda_1+
\nonumber \\
&&20 \mu_1 \rho_1^3 \lambda_2^2 \lambda_3^2 \rho_2^3-
10 \mu_1 \rho_1^2 \lambda_2^3 \lambda_3 \rho_2^4+
24 \mu_2^2 \rho_1^2 \lambda_2 \rho_2^6-
72 \mu_2^2 \rho_1^3 \lambda_3 \rho_2^5+
48 \mu_3^2 \rho_1^3 \rho_2^5 \lambda_1-
\nonumber \\
&&24 \mu_4^2 \rho_1^6 \lambda_2 \rho_2^2-
96 \mu_1^2 \rho_1 \lambda_3 \rho_2^7-
20 \mu_1 \rho_1^4 \lambda_2 \lambda_3^3 \rho_2^2-
48 \mu_3^2 \rho_1^5 \lambda_3 \rho_2^3+
4 \mu_1 \lambda_1 \rho_1 \lambda_2^2 \lambda_3 \rho_2^5.
\label{C1}
\end{eqnarray}


\subsection*{Acknowledgements:}

We would like to thank Dan Freed and Tony Pantev for helpful
conversations. Evgeny Buchbinder and Burt Ovrut are supported in part
by the DOE under contract No. DE-AC02-76-ER-03071. Ron Donagi is supported
in part by an NSF grant DMS-0104354.


\end{document}